\documentclass[aps,prx,prl,twocolumn,showpacs,superscriptaddress,preprintnumbers,longbibliography]{revtex4-2} 
\usepackage{graphicx}  % needed for figures
\usepackage{dcolumn}   % needed for some tables
\usepackage{bm}        % for math
\usepackage{amssymb}   % for math
\usepackage{amsmath}
\usepackage{comment}
\usepackage{graphicx}
\usepackage{color}
\usepackage{slashed}
\usepackage{braket}
\usepackage{amssymb,amsmath,graphicx,color}
\usepackage[tight]{subfigure}
\usepackage[export]{adjustbox}
\usepackage[colorlinks=true,citecolor=blue,linkcolor=red,breaklinks=true]{hyperref}
\usepackage{multirow}
\usepackage{rotating,booktabs}
\usepackage[verbose]{placeins}
\usepackage{mathrsfs}
\usepackage{tikz}
\usetikzlibrary{quantikz}
\usepackage{algorithm}
\usepackage{algpseudocode}
\usepackage{ulem}

\newcommand{\Eq}[1]{Eq.~(\ref{#1})}

\newcommand{\Fig}[1]{Fig.~\ref{#1}}

\newcommand{\coolant}{$^{172}\text{Yb}^{+}$\hspace{1mm}}
\newcommand{\qubit}{$^{171}\text{Yb}^{+}$\hspace{1mm}}

\begin{document}

\dimen\footins=5\baselineskip\relax

\title{
Dual-Isotope Sympathetic Cooling in a Long Ion Chain 
}

\author{Tianyi Wang}
\affiliation{Department of Physics, Duke University, Durham, NC 27708}
\affiliation{Duke Quantum Center, Duke University, Durham, NC 27701}
\author{Marko Cetina}
\affiliation{Department of Physics, Duke University, Durham, NC 27708}
\affiliation{Duke Quantum Center, Duke University, Durham, NC 27701}
\affiliation{Department of Electrical and Computer Engineering, Duke University, Durham, NC 27708}

\begin{abstract}
Owing to their high controllability and connectivity, one-dimensional ion chains are attractive building blocks for near-term quantum computers. However, long ion chains are susceptible to motional heating caused by fluctuating external electric fields. To enable deep computations, this source of noise must be suppressed without inducing qubit decoherence or degrading qubit connectivity. We demonstrate steady-state mid-circuit sympathetic cooling of all computationally-relevant motional modes of a 23-ion chain consisting of \qubit qubit and \coolant coolant ions without using any qubit operations. In a room-temperature system, our cooling scheme preserves the mean phonon occupations of the long-wavelength axial and radial modes near their values following state preparation, while keeping the modes used to implement entangling gates near their ground states throughout a 56-ms circuit. Crucially, we show that our cooling sequence preserves qubit coherence, and evaluate its impact on single- and two-qubit gate operations. Additionally, we demonstrate a parallel qubit reset protocol leveraging the shared radial mode coupling between species. Our sympathetic cooling scheme represents a critical step toward reducing gate errors in individually-addressed long ion chains and establishes a versatile platform for simulating open quantum systems.
\end{abstract}

\maketitle

%%%%%%%%%%%%%%%%%%%%%%%%%%%%%%%%%%%%%%%%%%%%%%%%%%%%%%%%%%%%%%%%%%%%%%%%%%%%%%%%%%%%%%%%%%%%%%%%%%%%%%%%%%%%%%%%%%%%%%%%%%%%
\section{Introduction
\label{sec:intro}}
\noindent

Trapped-ion systems are a leading platform for quantum computation and quantum simulation \cite{wineland_experimental_1998}. They have demonstrated long coherence times \cite{wang_single_2021, pi_beyond-ten-hour_2026}, high-fidelity single- and two-qubit digital gates \cite{smith_single-qubit_2025, loschnauer_scalable_2025, hughes_trapped-ion_2025}, and low state-preparation-and-measurement (SPAM) errors \cite{noek_high_2013, an_high_2022, sotirova_high-fidelity_2024}. Current efforts focus on increasing system size without compromising this performance. A system comprising nearly a hundred physical ion qubits has recently been realized within the quantum charge-coupled device (QCCD) architecture \cite{kielpinski_architecture_2002, ransford_98-qubit_2026}, demonstrating that high-fidelity ion-based computing can be scaled by shuttling ions among distinct trapping zones on the same chip. Beyond a single chip, photonic interconnects are envisioned to link multiple modules across separate chips \cite{monroe_large-scale_2014, monroe_scaling_2013, oreilly_fast_2024, saha_high-fidelity_2025}, providing an outward route to scalability. 

A complementary, inward route is to scale up the basic unit on each chip: a monolithic ion chain. Individually-addressed long ion chains provide intrinsic qubit connectivity that can facilitate scaling \cite{landsman_two-qubit_2019, huang_comparing_2024, kranzl_controlling_2022, peleg_fast_2026, wu_tilt_2020}. This connectivity greatly reduces the in-circuit overhead that is required to couple distant qubits on the same chip. The large set of motional modes of a single chain can also be harnessed to implement versatile nonunitary dynamics through native spin-motion coupling \cite{sun_quantum_2025, so_trapped-ion_2024}. 

Heating of low-frequency, long-wavelength motional modes has been recognized as a primary limitation of long ion chains \cite{Jurcevicthesis, cetina_control_2022, miao_probing_2025, katz_hybrid_2025}. In nearly equidistant chains subject to $1/f$ electric-field noise, axial heating leads to an entangling-gate error that scales as $N^6$, where $N$ is the number of ions \cite{cetina_control_2022}. Sympathetic cooling \cite{Larson_sympa_1986} is an effective technique for mitigating motional heating, but has mostly been restricted to pairs of ions \cite{rohde_sympathetic_2001, blinov_sympathetic_2002, barrett_sympathetic_2003, home_memory_2009, guggemos_sympathetic_2015, smith_microwave-driven_2026,brudney_mid-circuit_2026} or small ion crystals \cite{pino_demonstration_2021, ransford_98-qubit_2026}. Mass imbalance of different atomic species limits qubit connectivity in longer chains and impedes cooling of motional modes \cite{carter_design_2021, hou_indirect_2024}. Single-species sympathetic cooling either using a tightly focused cooling beam addressing a dipole transition \cite{mao_experimental_2021} or a global cooling beam addressing a quadrupole transition \cite{ranawat_sympathetic_2026} suffers from qubit crosstalk at high dissipation rates. Shelving qubits into metastable states \cite{allcock_omg_2021} to perform sympathetic cooling in the ground states provides high cooling power with low qubit crosstalk \cite{goldman_integration_2025, brudney_mid-circuit_2026}. However, this approach requires pre-cooling shelving and post-cooling deshelving of qubits, which introduces errors. These operations also consume valuable in-circuit time and extend the overall circuit duration. 

In this work, we demonstrate mid-circuit sympathetic cooling of all computationally relevant motional modes of an ion chain containing fourteen qubits without any qubit operations. Our sympathetic cooling scheme maintains the temperatures of the relevant long-wavelength axial and radial modes near their values after state preparation, while keeping the modes used for performing entangling gates near the motional ground states over a total experimental duration of up to 56 ms,  with 25\% gate duty factor. Qubit coherence is preserved in the presence of the sympathetic-cooling sequences. To demonstrate the applicability of our sympathetic-cooling scheme to nonunitary operations, we adapt it to implement a parallel qubit reset protocol.

This article is organized as follows. Section~\ref{sec:exp_setup} describes the general experimental setup. Section~\ref{sec:axial} characterizes heating and sympathetic cooling of the lowest-frequency axial (\textit{LA}) mode. Sections~\ref{sec:radial} and \ref{sec:radial_symp_cool} respectively present measurements of recoil heating and introduces our sympathetic cooling scheme for the lower-frequency radial (\textit{LR}) modes. Section~\ref{sec:complete_cooling} reports thermometry
%of the LA mode, the 14 lower-frequency radial (\textit{LR}) modes in the middle of the mode spectrum, and the longest-wavelength LR mode 
during prototypical experimental sequences lasting up to 56~ms that include sympathetic cooling of all relevant motional modes. 
Section~\ref{sec:coherence_and_gates} reports on qubit coherence during symapthetic cooling, and assesses the improvements in single- and two-qubit operations enabled by sympathetic cooling.
%through measurements of Ramsey fringes, sequences of sympathetic-cooling cycles interleaved with SK1-$\pi/2$ gates, and the fidelity of a single maximally entangling XX gate. 
Section~\ref{sec:reset} presents a scalable qubit-reset protocol based on shared coupling through radial motional modes. Finally, Sec.~\ref{sec:conclusions} summarizes our findings and discusses future prospects for the sympathetic-cooling scheme. 
Appendices \ref{app:loading}, \ref{app:sorting}, and \ref{app:coolant_Raman} present details of our experimental implementation, including
isotope-selective loading, ion sorting and Raman control of the coolant ions. 
Appendix~\ref{app:spectral_sel} theoretically evaluates %spectral isolation required to implement 
our dual-isotope sympathetic-cooling scheme with Yb$^+$, Ba$^+$, and Ca$^+$ ions. 
Appendix~\ref{app:recoil_cal} presents a parameter-free resolved-sideband model for axial cooling and applies this model to estimate radial recoil heating.
Finally, Appendix~\ref{app:cryo_estimate} considers the prospects for implementing this scheme in a cryogenic system containing a 60-ion chain.
%%%%%%%%%%%%%%%%%%%%%%%%%%%%%%%%%%%%%%%%%%%%%%%%%%%%%%%%%%%%%%%%%%%%%%%%%%%%%%%%%%%%%%%%%%%%%%%%%%%%%%%%%%%%%%%%%%%%%%%%%%%%

\section{Experimental Setup \label{sec:exp_setup}}

\begin{figure}[H]
  \centering
	\includegraphics[width=0.48\textwidth]{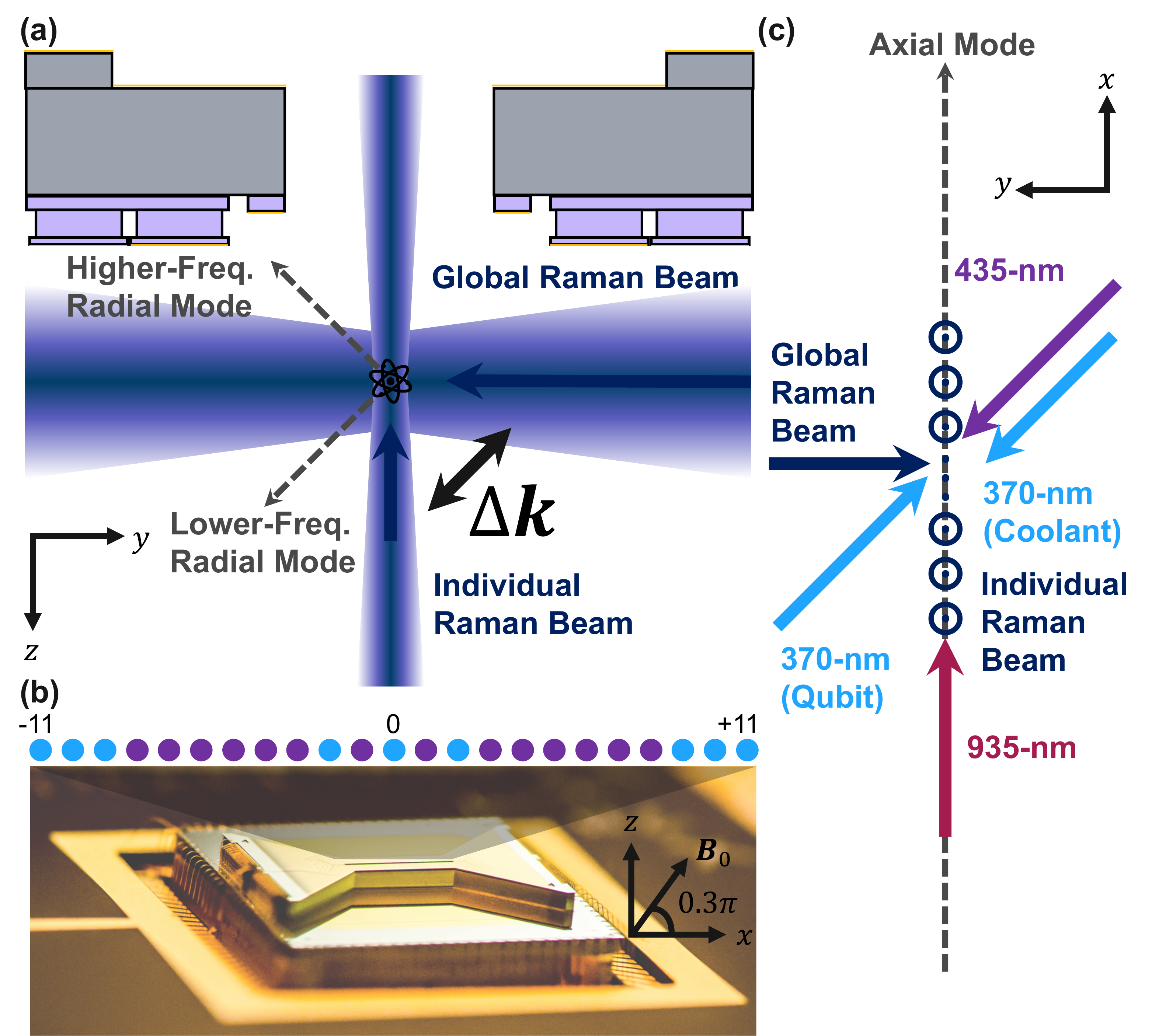}
	\caption{Experimental setup. Experiments are performed at room temperature using a microfabricated surface-electrode linear rf ion trap. (a) Side view. The dashed gray arrows indicate the radial trap principal axes associated with the higher- and lower-frequency radial modes. Two beams derived from a 355-nm pulsed Raman laser address the ions. A global Raman beam uniformly illuminates the entire ion chain, while the other Raman beam is split into separate individual beams that are tightly focused onto individual ions \cite{debnath_demonstration_2016}. For Raman operations, stimulated absorption and emission from the global and individual beams imparts the net momentum $\hbar \Delta \mathbf{k}$. (b) Front view. The ion chain, consisting of nine $^{172}$Yb$^{+}$ coolant ions (blue) and fourteen $^{171}$Yb$^{+}$ qubit ions (purple) is confined approximately $68~\mathrm{\mu m}$ away from the trap surface. Center ion is indexed as 0, and ions to the left and right are indexed by negative and positive integers, respectively. The quantization field $\mathbf{B}_0$ lies predominantly in the $xz$ plane, tilted by $54^\circ$ from $\hat{x}$ toward $\hat{z}$. (c) Top view. The ion chain is aligned along the trap axis (gray dashed arrow). The 435-nm beam (purple), which drives the quadrupole transition in coolant ions, is oriented at $45^\circ$ to the axial direction. The 370-nm coolant Doppler-cooling beam (light blue) is combined with the 435-nm beam. The 370-nm qubit Doppler-cooling beam (light blue) is oriented at $45^\circ$ to the axial direction, opposite to the 435-nm beam. The 935-nm repump beam (red) propagates along the axial direction. \label{fig:exp_setup}}
\end{figure}

We perform the experiments in a room-temperatue microfabricated surface-electrode linear rf ion trap (HOA 2.1.1, Sandia National Laboratories \cite{maunz_high_2016}). We trap \qubit ions with radial secular frequency of $\omega_{\text{sec}} = 2\pi\times 3.193~\mathrm{MHz}$. A static radial quadrupole field is applied to define the principle axes as shown in \Fig{fig:exp_setup}(a), and to split the radial mode frequencies by about $2\pi\times 285~\mathrm{kHz}$. 
We demonstrate our sympathetic cooling scheme using a dual-isotope 23-ion chain consisting of \qubit qubit and \coolant coolant ions arranged as shown in \Fig{fig:exp_setup}(b). In this mixed-species 23-ion chain,  the lower-frequency radial ({\it LR}) modes have frequencies spanning from $2\pi\times 2.761~\mathrm{MHz}$ to $2\pi\times 3.038~\mathrm{MHz}$. 

We use two 355-nm laser beams derived from the same pulsed laser for the Raman addressing of ions. The two beams are orthogonal to each other and to the ion chain axis, as shown in \Fig{fig:exp_setup}(a). One beam has a large waist to address the entire ion chain (\textit{global beam}); the second beam is split by a diffractive optical element (DOE) into separate beams that are routed through a 32-channel acousto-optic modulator (AOM) \cite{debnath_demonstration_2016} and tightly focused to address individual ions (\textit{individual beams}). The global and individual Raman beams are linearly polarized along the trap $\hat{z}$ and $\hat{y}$ axes, respectively. The $1/e^2$-intensity radius ({\it waist}) of the individual beams is measured to be $w=616(12)~\mathrm{nm}$ by scanning the location of a tightly confined single ion across the beam \cite{cetina_control_2022}. 

Raman beams drive transitions between states in the ground electronic manifold in both species via absorption and emission of photons between the global and individual beams, transferring momentum to the ions along LR mode direction, as shown in \Fig{fig:exp_setup}. On qubits, this Raman process drives the transition between the clock states $|F=0\rangle$ ($\ket{1})$) and $|F=1, m_F=0\rangle$ ($\ket{0}$) in the ground electronic manifold. On coolants, the Raman process drives the transition between the $|F=1/2, m_F=-1/2\rangle$ and $|F=1/2, m_F=+1/2\rangle$ states (see Appendix~\ref{app:coolant_Raman} for details). The quantization field $\mathbf{B}_0$ (magnitude $5.53~\mathrm{G}$) lies predominantly in the $xz$ plane, and is tilted by $54^\circ$ from $\hat{x}$ toward $\hat{z}$ as shown in \Fig{fig:exp_setup}(b). Under this field configuration, the Zeeman splitting between the two coolant ground states is $15.48~\mathrm{MHz}$. 

A global 435-nm beam is used exclusively for sympathetic cooling on the $^{2}S_{1/2}-^{2}D_{3/2}$ quadrupole transition. The 435-nm beam is aligned at $45^\circ$ relative to the chain axis as shown in \Fig{fig:exp_setup}(c). Also aligned at $45^\circ$ to the trap axis are separate 369-nm beams for Doppler cooling of qubit and coolant ions on the
D1 transition. A global 935-nm repumping beam is aligned along the axial direction. 

Each experiment begins with Doppler cooling of both isotopes. The relative ordering of the isotopes can change because of collisions with background gas molecules. For a dual-isotope chain containing 23 ions, the reordering rate is approximately one event per minute under our experimental conditions. The isotope ordering is monitored during the Doppler cooling stage. If a reordering event is detected, we briefly reduce the trap radio-frequency (RF) drive and apply shuttling voltage waveforms to restore the target isotope configuration (see Appendix~\ref{app:sorting} for details). After ensuring the isotope ordering, Raman sideband cooling is performed on the \qubit qubit ions, reducing the occupation of the middle radial modes to below 0.2 quanta while optically pumping the qubits into $\ket{1}$. Following this state preparation, the quantum circuit is executed.

%%%%%%%%%%%%%%%%%%%%%%%%%%%%%%%%%%%%%%%%%%%%%%%%%%%%%%%%%%%%%%%%%%%%%%%%%%%%%%%%%%%%%%%%%%%%%%%%%%%%%%%%%%%%%%%%%%%%%%%%%%%%
\section{Axial sympathetic cooling\label{sec:axial}}
\noindent

\begin{figure}[H]
  \centering
	\includegraphics[width=0.48\textwidth]{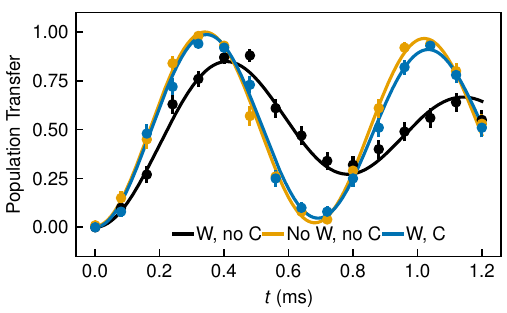}
	\caption{Carrier Raman Rabi oscillations of qubit ion +4. The probability of population transfer to the excited state $P$ is plotted as a function of the Rabi pulse duration $t$. Error bars denote $1\sigma$ binomial uncertainties. Black: after a 20-ms idle wait with no axial sympathetic cooling (W, no C). Yellow: immediately after state preparation with no idle wait and no axial sympathetic cooling (no W, no C). Blue: after a 20-ms idle wait followed by 8 ms of axial sympathetic cooling (W, C). Cooling reduces the fitted carrier-decay parameter from $0.27(2)$ after the idle wait (black) to $0.069(8)$ (blue), approaching the post-state-preparation value of $0.042(10)$ (yellow).}
	\label{fig:carrier_compare}
\end{figure}

To determine the axial temperature of the ion chain, we apply square Raman pulses to simultaneously drive carrier Rabi oscillations on all $^{171}$Yb$^{+}$ qubits between the $|0\rangle$ and $|1\rangle$ states. Figure~\ref{fig:carrier_compare} (yellow) shows the resulting oscillation in excitation probability of ion +4 immediately following state preparation. 

To model the data, we consider only motion in the weakly-confined LA mode. The measured $|0\rangle$-state probability, $p_{b,i}$, for ion $i$ is fit to the model in Ref.~\cite{cetina_control_2022},

\begin{align}\label{eq:axialmodel}
    P = p_{b, i} = \frac{1-(1+\theta_i^2 \Omega^2_{i, 0} t^2)^{-1/2} \cos(\Omega_{i, 0} t - \phi_i)}{2}.
\end{align}
Here, $\Omega_{i,0}$ denotes the carrier Rabi frequency of ion $i$ in a cold chain, and $\phi_i=\arctan(\Omega_{i,0}\theta_i t)$ (yellow curve in \Fig{fig:carrier_compare}). The fitted decay parameter $\theta_i$ is then related to the mean occupation number, $\bar{n}$, of the LA mode using the relation from \cite{cetina_control_2022}
\begin{align}\label{eq:theta}
%    \theta_i = -b_i^2 \xi^2 \frac{\Omega_{i, 0}^{\prime\prime}}{\Omega_{i, 0}} \bar{n}=\frac{2b_i^2\xi^2}{w^2}\bar{n}.
    \theta_i = \frac{2b_i^2\xi^2}{w^2}\bar{n}.
\end{align}
Here, $b_i$ the participation factor of ion $i$ in the LA mode with angular frequency $\omega_0$ (see calculation details in Appendix~\ref{app:recoil_cal}), and $\xi=\sqrt{\hbar/(2M\omega_0)}$ is the root-mean-square spatial extent of the ground-state wavefunction of a single ion with mass $M$ confined in a trap with angular frequency $\omega_0$. Here, we take $M=171~\mathrm{a.m.u.}$, corresponding to the mass of the qubit ions.

\begin{figure}[H]
  \centering
	\includegraphics[width=0.48\textwidth]{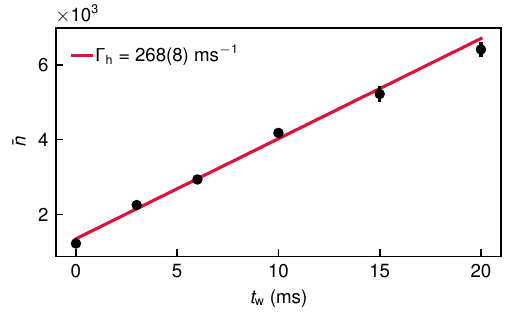}
	\caption{Idle heating rate of the LA mode. The average mode occupation $\bar{n}$ is plotted as a function of idling duration $t_{\mathrm{w}}$. Error bars denote the standard error of the ion mean with weights inversely proportional to the individual binomial variances. A linear fit to the resulting data (red line) yields an idle heating rate $\Gamma_{\mathrm{h}}=268(8)~\mathrm{ms}^{-1}$.}
	\label{fig:axial_heating}
\end{figure}

We measure the heating rate $\Gamma_{\mathrm{h}}$ of the LA mode by driving simultaneous carrier Rabi oscillations on all qubits after a variable wait time $t_{\mathrm{w}}$ following state preparation. Figure~\ref{fig:carrier_compare} (black) shows a representative carrier Rabi oscillation measured with $t_{\mathrm{w}}=20~\mathrm{ms}$. For each wait time, we fit the Rabi oscillation data to the model described by Eq.~\ref{eq:axialmodel} and extract the mean occupation number $\bar{n}$ using Eq.~\ref{eq:theta}. The resulting values of $\bar{n}$ are plotted in \Fig{fig:axial_heating}. A linear fit to the data  yields the  heating rate $\Gamma_{\mathrm{h}}=268(8)~\mathrm{ms}^{-1}$.

%

% Axial cooling scheme 
We suppress axial heating of the chain by sympathetic cooling using the co-trapped \coolant coolant ions. Continuous sideband cooling is implemented on the $^{2}S_{1/2}$--$^{2}D_{3/2}$ 435-nm quadrupole transition, following the method described in Ref.~\cite{cetina_control_2022}. During cooling, the 435-nm beam contains two frequency components that simultaneously address both $\Delta m_F=0$ transitions with resonant Rabi frequencies $\Omega_{\mathrm{SD},m_J=+1/2}=2\pi\times283~\mathrm{kHz}$ and $\Omega_{\mathrm{SD},m_J=-1/2}=2\pi\times311~\mathrm{kHz}$ on the central ions. Both 435-nm tones are detuned from the respective resonances by $\Delta_{\mathrm{SD},m_J=-1/2}=\Delta_{\mathrm{SD},m_J=+1/2}=-2\pi\times680~\mathrm{kHz}$. The 435-nm beam is expanded along the chain, with the measured carrier quadrupole Rabi frequencies on the edge coolant ions at positions $-10$ and $+10$ equal to approximately $0.91$ $\Omega_{\rm{SD},\pm1/2}$ and $0.74$ $\Omega_{\rm{SD},\pm1/2}$ respectively. The state of the coolant is reset using a 935-nm beam that resonantly drives the $^{2}D_{3/2}$ -- $^{3}[3/2]_{1/2}$ transition. With the parameters used for axial cooling, we measure the repumping rate from the %$^{2}D_{3/2}, 
$m_J=\pm1/2$ states to be $\gamma_{+1/2}=2\pi\times90~\mathrm{kHz}$ and $\gamma_{-1/2}=2\pi\times225~\mathrm{kHz}$, respectively.

We measure the axial temperature of the chain after axial sympathetic cooling by driving simultaneous carrier Rabi oscillations on all qubits. Figure~\ref{fig:carrier_compare} (blue) shows the carrier Rabi oscillation measured on ion +4 after 8 ms of axial sympathetic cooling following a wait time of $t_{\mathrm{w}}=20~\mathrm{ms}$. Compared with the measurement without sympathetic cooling, the oscillation exhibits substantially reduced decay, indicating that the axial temperature has nearly returned to its value immediately after state preparation.

\begin{figure}[H]
  \centering
	\includegraphics[width=0.48\textwidth]{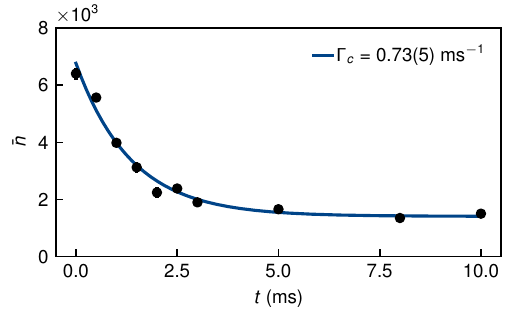}
	\caption{Evolution of the mean occupation of the LA mode $\bar{n}$ during sympathetic cooling. Carrier Raman Rabi oscillations are driven on all qubits following a 20-ms idle delay and subsequent axial sympathetic cooling of duration $t$. Error bars denote the standard error of the ion mean with weights inversely proportional to the individual binomial variances. A fit of the data to Eq.~\ref{eq:axial_cooling_model} (blue curve) yields the equilibrium mean occupation $\bar{n}_{\mathrm{eq}}=1.41(9)\times10^3$ and the cooling rate constant of $\Gamma_{\mathrm{c}}=0.73(5)~\mathrm{ms}^{-1}$.}
	\label{fig:axial_cooling}
\end{figure}

We characterize the axial sympathetic cooling dynamics by measuring carrier Rabi oscillations on all qubits after varying the cooling duration. Each measurement is performed following state preparation, a 20-ms idle period to allow the axial mode to heat, and axial sympathetic cooling of various duration. The measured Rabi oscillations data, shown in \Fig{fig:axial_cooling}, are fit to the model described in \cite{leibfried_quantum_2003}, 
\begin{align}\label{eq:axial_cooling_model}
    \dot{\bar{n}} = -\Gamma_{\mathrm{c}} \bar{n} + \bar{n}_{\mathrm{eq}}.
\end{align}
The fit yields the cooling rate constant $\Gamma_{\mathrm{c}} = 0.73(5)~\mathrm{ms}^{-1}$ and the equilibrium average occupation $\bar{n}_{\mathrm{eq}}=1.41(9)\times10^3$. Based on the analytic parameter-free cooling model presented in Appendix~\ref{app:recoil_cal}, we believe that the performance of our axial cooling is limited by the available 435 nm laser power; increasing the 435 nm laser power would enable a higher axial cooling rate constant $\Gamma_{\mathrm{c}}$ and a lower equilibrium occupation $\bar{n}_{\mathrm{eq}}$.

%%%%%%%%%%%%%%%%%%%%%%%%%%%%%%%%%%%%%%%%%%%%%%%%%%%%%%%%%%%%%%%%%%%%%%%%%%%%%%%%%%%%%%%%%%%%%%%%%%%%%%%%%%%%%%%%%%%%%%%%%%%%
\section{Radial recoil heating}% and sympathetic cooling
\label{sec:radial}
\noindent
Entangling gates are affected by motion of radial modes with projection along the Raman recoil direction (see \Fig{fig:exp_setup}a). In our system, Raman beams primarily couple to the LR modes; we drive the middle-wavelength LR modes with spin-dependent forces to implement Mølmer--Sørensen (MS) gates. While these modes are cooled to near their motional ground state at the beginning of a quantum circuit, during axial sympathetic cooling, recoil associated with the absorption of 435-nm photons and the spontaneous emission of 297-nm photons can heat them, leading to subsequent gate errors.

\begin{figure}[H]
  \centering
	\includegraphics[width=0.48\textwidth]{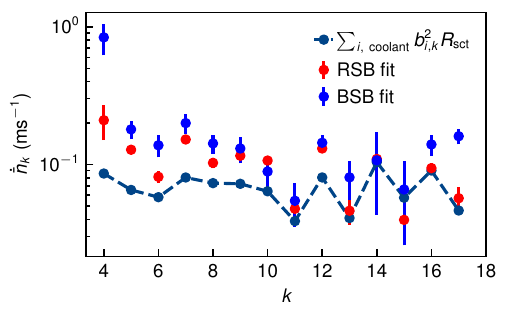}
	\caption{Average heating rates $\dot{\bar{n}}_k$ of 14 middle LR modes during 3 ms of axial sympathetic cooling. Mode index $k=0$ corresponds to the longest-wavelength LR mode. Blue points indicate the heating rates obtained from linear fits to the mean mode occupations, $\bar{n}_k$, measured at different cooling times, using blue-sideband Rabi oscillations. Red points show the corresponding rates obtained from red-sideband Rabi oscillations. Dark blue dashed points show the average recoil-induced heating rates during the 3-ms axial cooling sequence obtained from a zero-parameter resolved sideband model described in Appendix.~\ref{app:recoil_cal}.}
	\label{fig:radial_heating}
\end{figure}

We characterize this radial heating by recording first-order radial sideband Rabi oscillations on the qubit ions. Each qubit ion is associated with a distinct radial mode, and first-order, either red or blue, sideband Raman transitions are driven simultaneously on all qubits using Stark-shift-compensated Gaussian pulses.
We determine the radial-mode occupation by fitting the first-order blue- and red-sideband Rabi oscillations to the model described in Ref.~\cite{wineland_experimental_1998}. For the blue-sideband fits, the detuning, mean phonon occupation $\bar{n}_k$, and the carrier Rabi frequency are treated as free parameters. The weak excitation of the red sideband precludes independent fitting of the detunings. Therefore, for the red-sideband Rabi oscillations, we set the detuning to zero and use the carrier Rabi frequency and $\bar{n}_k$ as the free parameters. 

We measure the heating rates of 14 LR modes during a 3-ms interval of axial sympathetic cooling. The axial cooling conditions are identical to those described in Section~\ref{sec:axial}. For each radial mode $k$, we determine the mean phonon number, $\bar{n}_k$, as a function of the cooling time and perform a linear fit to extract the heating rate, $\dot{\bar{n}}_k$, from the fitted slope. The heating rates determined from the blue and red sideband measurements are shown in \Fig{fig:radial_heating}. The axial coupling described by Eq.~\ref{eq:axialmodel} is not captured by the fitting model; consequently, the blue-sideband fits systematically overestimate $\bar{n}_k$. At the same time, residual detuning not accounted for in our red-sideband fitting model leads to a systematic underestimate of the true $\bar{n}_k$. We therefore expect actual heating rates to lie between the values obtained from the blue-sideband and the red-sideband measurements.

To interpret our measurements, we calculate the recoil-induced heating based on the total photon scattering rate during axial cooling. We obtain the photon scattering rate using a semiclassical resolved-sideband model (see Appendix~\ref{app:recoil_cal} for details). The predicted recoil heating is in good agreement with the observed heating of the short-wavelength radial modes. We attribute the increased heating of the long-wavelength radial modes to residual idle heating.

\section{Radial Sympathetic Cooling}
\label{sec:radial_symp_cool}
\begin{figure}[H]
  \centering
	\includegraphics[width=0.48\textwidth]{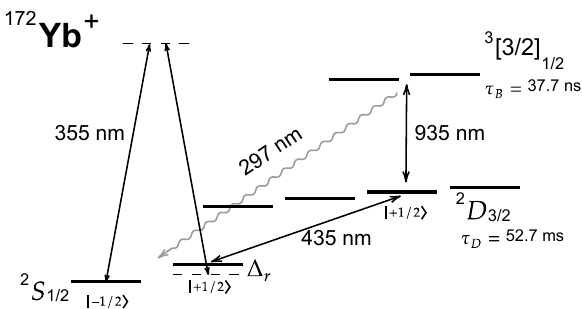}
	\caption{Raman pulsed sideband cooling cycle for radial sympathetic cooling. For each addressed coolant ion, the cooling sequence begins with a first-order red-sideband Raman transition on the associated radial mode on coolant ions, with the Raman detuning from the carrier given by $\Delta_r$. This is followed by a 435-nm carrier pulse and a subsequent 935-nm repumping pulse.}
	\label{fig:radial_cooling_scheme}
\end{figure}

To recool the radial modes following axial cooling, we perform sympathetic radial cooling. Because the radial sidebands are significantly weaker than the axial sidebands, we employ a pulsed sideband cooling scheme. To maximize momentum transfer, the sideband transitions are driven using the Raman beams, while the 435-nm and 935-nm lasers are used for optical pumping (\Fig{fig:radial_cooling_scheme}).

\begin{figure*}[t]
  \centering
	\includegraphics[width=\textwidth]{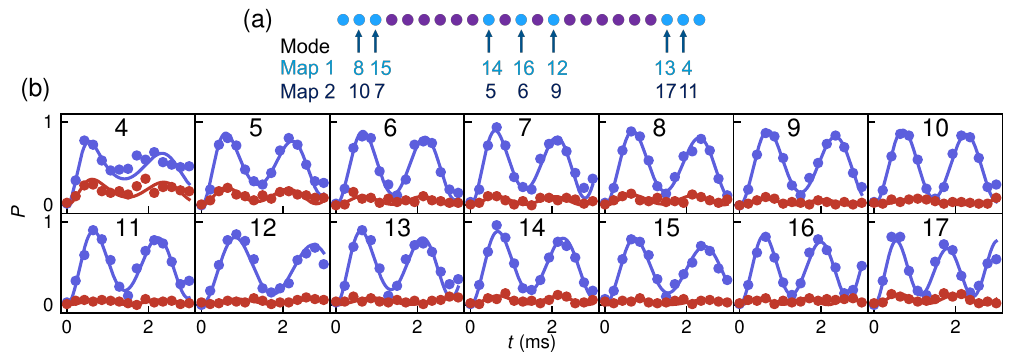}
	\caption{Sympathetic cooling of the gate radial modes. (a) Mode-to-coolant mapping for both steps used for radial sympathetic cooling (blue: \coolant coolant ions; purple: \qubit qubit ions). The cooling sequence alternates between the two steps. (b) First-order blue- and red-sideband Raman Rabi oscillations following a single sympathetic cooling cycle. Following the sympathetic cooling cycle, Raman sideband pulses are applied in parallel to qubits {-5, 7, 8, -6, 3, 1, 5, -1, 6, -3, -4, 4, -8, -7} to probe LR modes 4–17 respectively. Blue-sideband fitted $\bar{n}_k$ are 1.91(55), 0.49(10), 0.20(5), 0.32(8), 0.27(5), 0.24(4), 0.12(5), 0.28(9), 0.23(8), 0.31(10), 0.27(5), 0.26(5), 0.20(4), 0.22(10). Red-sideband fitted $\bar{n}_k$ are 0.57(9), 0.30(3), 0.13(2), 0.13(2), 0.15(2), 0.08(1), 0.09(1), 0.07(1), 0.08(1), 0.10(2), 0.13(2), 0.07(1), 0.09(2), 0.16(2). }
	\label{fig:radial_cooling_bsbrsb_first_round}
\end{figure*}

During each cooling step, first-order red-sideband Raman transitions are driven simultaneously on the seven coolant ions. Each coolant ion is used to cool two distinct radial modes, with the cooling sequence alternating between the two modes in successive cooling steps. The coolant--mode assignments are shown in \Fig{fig:radial_cooling_bsbrsb_first_round}(a). We apply square Raman pulses to coolants during the radial cooling with pulse duration of $30~\mathrm{\mu s}$ with no dynamical decoupling. After each sideband pulse, the coolant ions are reset using an optical pumping sequence consisting of a 435-nm carrier pulse with the 935-nm beam turned off, followed by a 935-nm repump pulse with the 435-nm beam turned off. This pumping sequence is repeated five times during each cooling cycle for a total duration of $12~\mathrm{\mu s}$. Each mode undergoes 20 cooling cycles.

%Radial cooling scheme 

After cooling the middle LR modes, we sideband-cool the longest-wavelength LR mode. Since this mode couples strongly to the noisy electric fields near the trap surface, it requires much higher cooling power. During each cooling cycle, all seven coolant ions are driven simultaneously on the first-order red sideband of this radial mode. The optical pumping sequence is identical to that used for cooling the middle LR modes. A total of 30 cooling cycles is used, each employing a Raman sideband pulse with a fixed duration of $20~\mathrm{\mu s}$. 

Figure~\ref{fig:radial_cooling_bsbrsb_first_round}(b) shows the first-order blue- and red-sideband Raman Rabi oscillations measured after one full round of sympathetic cooling. A single cooling round consists of 3-ms axial sympathetic cooling, two 0.96-ms radial cooling steps addressing the middle LR modes (modes 4--17), and a 1.14-ms cooling step for the longest-wavelength LR mode. Following the sympathetic cooling sequence, Raman sideband pulses are applied in parallel to qubits ${-5, 7, 8, -6, 3, 1, 5, -1, 6, -3, -4, 4, -8, -7}$, corresponding to LR modes 4--17 respectively to measure the temperature.

We assess the cooling of the longest-wavelength LR mode by comparing the first-order Raman blue-sideband Rabi oscillation measured immediately after state preparation with that after a single round of sympathetic cooling. The blue-sideband Rabi oscillations are measured on qubit ion $-4$ with the carrier Rabi frequency set to $\Omega_{\mathrm{c,hr}}=2\pi\times70~\mathrm{kHz}$. Both data sets are fit to the resonant first-order blue-sideband Rabi oscillation model described in Ref.~\cite{wineland_experimental_1998}, with the mean phonon number, $\bar{n}_{\mathrm{hr}}$, as the only fit parameter.

\begin{figure}[H]
  \centering
	\includegraphics[width=0.48\textwidth]{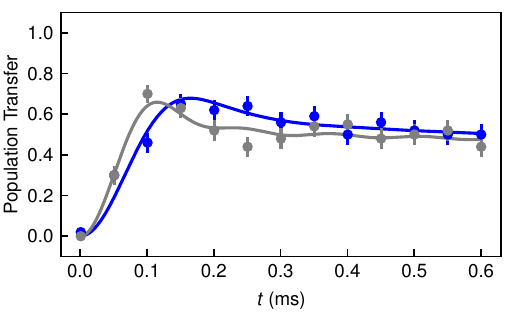}
	\caption{First-order blue-sideband Raman Rabi oscillations of the longest-wavelength LR mode. The population transfer probability to the excited state is plotted as a function of the Rabi pulse duration $t$. The blue-sideband drive is on qubit ion $-4$ with a carrier Rabi frequency of $\Omega_{\mathrm{c,hr}}=2\pi\times70~\mathrm{kHz}$, either immediately after state preparation (gray) or following single round of the full sympathetic cooling sequence (blue). Error bars denote $1\sigma$ binomial uncertainties. The mean occupation $\bar{n}_{\mathrm{hr}}$ is extracted as the only fitting parameter, yielding $\bar{n}_{\mathrm{hr}}=16(2)$ (gray) and $8(4)$ (blue). }
	\label{fig:radial_com_bsb_compare}
\end{figure}

The measurement immediately after state preparation yields $\bar{n}_{\mathrm{hr}}=16(2)$ (gray curve in \Fig{fig:radial_com_bsb_compare}). After one round of sympathetic cooling, the fitted occupation decreases to $\bar{n}_{\mathrm{hr}}=8(4)$ (blue curve in \Fig{fig:radial_com_bsb_compare}). This suggests that sympathetic cooling effectively recools the longest-wavelength LR mode, compensating for the idle and recoil heating accrued during the previous sequence.

%%%%%%%%%%%%%%%%%%%%%%%%%%%%%%%%%%%%%%%%%%%%%%%%%%%%%%%%%%%%%%%%%%%%%%%%%%%%%%%%%%%%%%%%%%%%%%%%%%%%%%%%%%%%%%%%%%%%%%%%%%%%
\section{Steady-state sympathetic cooling\label{sec:complete_cooling}}
\noindent
For sympathetic cooling to support long quantum circuits, it must be applied periodically to maintain all computationally relevant modes at low temperature throughout  the circuit. We benchmark this capability using a sequence comprising $N_r$ rounds of a prototypical sequence. Each round begins with a 2-ms idle period that is allocated for circuit operations. This is followed by a sympathetic cooling sequence consisting of 3 ms of axial cooling, two 0.96-ms radial cooling steps addressing the 14 middle LR modes (modes 4--17), and a 1.14-ms cooling step for the longest-wavelength LR mode. In this prototypical sequence, approximately $25\%$ of the total time is reserved for qubit operations, while the remaining time is divided approximately equally between axial and radial sympathetic cooling. We perform up to seven rounds of this sequence, corresponding to a total experimental duration of around $56~\mathrm{ms}$, including $14~\mathrm{ms}$ of idle time.

\begin{figure}[H]
  \centering
	\includegraphics[width=0.5\textwidth]{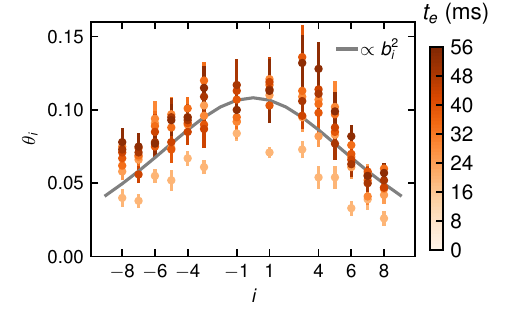}
	\caption{Spatial profile of the fitted carrier Raman Rabi decay parameter $\theta_i$. The fitted carrier Rabi decay parameter is measured for each qubit ion $i$ in the chain. The color scale indicates the total circuit duration $t_{\mathrm{e}}$ preceding the carrier Raman Rabi oscillation (see color bar). The gray curve shows the scaled squared participation factor $b_i^2$ of the LA mode.}
	\label{fig:acrc_theta_profile}
\end{figure}

After the final round, we characterize the axial and radial mode temperature using the thermometry methods described in Sections~\ref{sec:axial} and \ref{sec:radial}. Figure.~\ref{fig:acrc_theta_profile} shows the fitted decay parameters for the individual qubit ions. The decay parameters are consistent across all rounds and their spatial profile closely follows that predicted by \Eq{eq:theta}, indicating that motion in the LA mode dominates the noise in the carrier Rabi oscillations. %the squared participation factors of the LA mode.

\begin{figure}[H]
  \centering
	\includegraphics[width=0.48\textwidth]{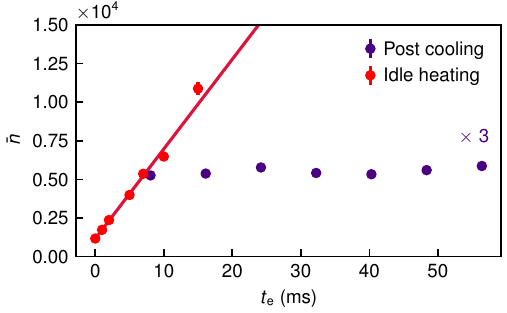}
	\caption{Measured mean occupation $\bar{n}$ of the LA mode following circuit execution of total duration $t_e$. The circuits consist either of idle evolution (red) or repeated rounds of a prototypical sequence with sympathetic cooling (purple). Error bars denote the standard error of the ion mean with weights inversely proportional to the individual binomial variances. The displayed post-cooling $\bar{n}$ values are multiplied 3. The red line is a linear fit to the idle-heating data.}
	\label{fig:acrc_axial}
\end{figure}

Figure~\ref{fig:acrc_axial} shows the average phonon number of the LA mode determined from the fitted decay parameters using \Eq{eq:theta} after different number of rounds of the prototypical sequence with a total duration of $t_e$ (see Appendix~\ref{app:cooling_supp_data} for the detailed data). For comparison, we also plot the occupation measured after an idle period of duration $t_e$. The idle data correspond to a heating rate of $581(13)~\mathrm{ms}^{-1}$, which is higher than the $268(8)~\mathrm{ms}^{-1}$ reported in Section~\ref{sec:axial}, probably due to increased technical electric-field noise at the time. With sympathetic axial cooling included in the sequence, the occupation of the LA mode remains below 2000 quanta after up to seven rounds.

\begin{figure}[H]
  \centering
	\includegraphics[width=0.48\textwidth]{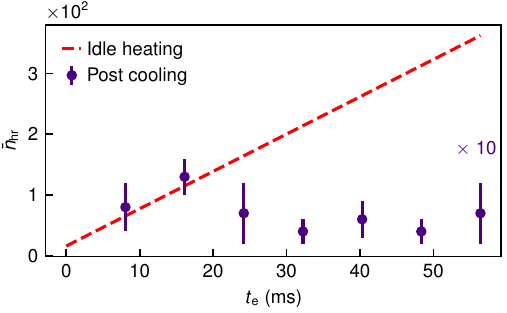}
	\caption{Measured mean occupation $\bar{n}$ of the longest-wavelength LR mode following circuit execution of total duration $t_e$. The circuits consist repeated rounds of a prototypical sequence with sympathetic cooling (purple). Error bars denote the $1\sigma$-binomial-uncertainty-weighted fit errors. The displayed post-cooling $\bar{n}_{\mathrm{hr}}$ values are multiplied 10. The red line is the estimate based on the idle heating rate.}
	\label{fig:acrc_lw_lr}
\end{figure}

Figure~\ref{fig:acrc_lw_lr} compares the average phonon number $\bar{n}_{\mathrm{hr}}$ of the longest-wavelength LR mode following idling with that measured after repeated rounds of the prototypical sequence. For the LR mode of a single tightly confined ion, whose frequency is close to that of the longest-wavelength LR mode of the chain, we measure a heating rate of $0.267~\mathrm{ms}^{-1}$. Assuming that the heating rate scales linearly with the number of ions, we estimate a rate of $6.141~\mathrm{ms}^{-1}$ for the corresponding chain mode. Together with the initial temperature from Fig.~\ref{fig:radial_com_bsb_compare}, this estimate yields the idle-heating prediction shown by the dashed red line. After repeated rounds of the prototypical sequence with total duration $t_{\mathrm{e}}$, we determine $\bar{n}_{\mathrm{hr}}$ by fitting blue-sideband Rabi oscillations on qubit ion $-4$ using the model described in Sec.~\ref{sec:radial_symp_cool}. The resulting measurements, shown in purple, demonstrate that sympathetic cooling maintains the mode occupation close to its value after state preparation (see Appendix~\ref{app:cooling_supp_data} for the corresponding measurements). 

We measure the occupations of the 14 LR modes (modes 4--17) with first-order sideband Rabi oscillations. We show the fitted $\bar{n}_k$ from the blue-sideband data in Fig.~\ref{fig:acrc_radial} (see Appendix~\ref{app:cooling_supp_data} for the detailed data). The radial sympathetic cooling sequence maintains the mean phonon number, $\bar{n}_k$, below approximately $0.55$ for all measured modes except mode 4, the longest-wavelength mode, which is presumably more strongly affected by idle heating.

\begin{figure}[H]
  \centering
	\includegraphics[width=0.48\textwidth]{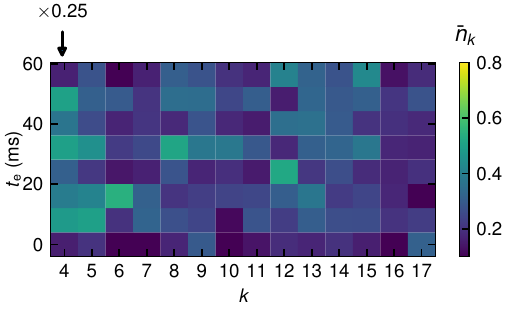}
	\caption{Mean occupations $\bar{n}_k$ of the middle 14 LR modes determined from blue-sideband Rabi oscillations following repeated rounds of the prototypical sequence with sympathetic cooling, with total circuit duration $t_e$. The data for mode 4 are scaled by 0.25.}
	\label{fig:acrc_radial}
\end{figure}
%

%%%%%%%%%%%%%%%%%%%%%%%%%%%%%%%%%%%%%%%%%%%%%%%%%%%%%%%%%%%%%%%%%%%%%%%%%%%%%%%%%%%%%%%%%%%%%%%%%%%%%%%%%%%%%%%%%%%%%%%%%%%%

\section{Qubit coherence and digital gates with sympathetic cooling 
\label{sec:coherence_and_gates}}
\noindent

We evaluate the impact of sympathetic cooling on qubit coherence by inserting multiple rounds of the prototypical sequence with full sympathetic cooling described in Section~\ref{sec:complete_cooling} between the two $\pi/2$ pulses of a Ramsey sequence. Square $\pi/2$ pulses are applied sequentially to the qubits, maintaining a constant sequence duration on each qubit. Simultaneously scanning the phase of the final pulse for all qubits yields Ramsey fringes (see Appendix~\ref{app:cooling_supp_data} for the fringe data). 

\begin{figure}[H]
  \centering
	\includegraphics[width=0.48\textwidth]{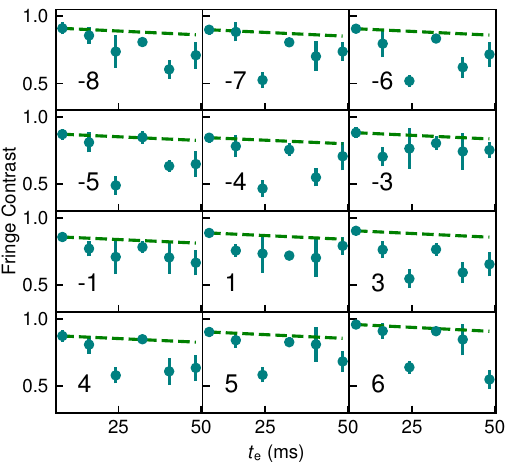}
	\caption{Ramsey phase scan fringe contrast. Square $\pi/2$ pulses are applied sequentially to qubits before and after repeated rounds of a prototypical sequence with complete sympathetic cooling of total circuit duration $t_e$. The fringe contrast is extracted as a fit parameter (dark green). Error bars denote the fit errors. The light green dashed curve shows the Ramsey contrast decay measured on a single tightly confined ion using microwave control, scaled to match the initial data point.}
	\label{fig:acrc_ramsey_contrast}
\end{figure}

The fitted fringe contrasts are shown in \Fig{fig:acrc_ramsey_contrast} as a function of total circuit duration $t_{\mathrm{e}}$. As reference, the green dashed line shows the Ramsey contrast decay measured using microwave pulses on a single tightly confined ion. The Ramsey contrast decay in the chain generally follows the reference with dips at certain points. Fluctuations in the 435-nm beam power produce variations in the AC Stark shift during sympathetic cooling, leading to additional qubit dephasing. This effect can be mitigated by further optimizing the active power stabilization of the 435-nm beam. Pulse-area errors arising from Raman beam pointing drifts contribute to the observed coherence loss. In current Ramsey experiments, compiling long pulse sequences results in significant experimental downtime. Reducing the compilation time can enable more frequent ion-based calibrations, allowing slow experimental drifts to be tracked and compensated more effectively. Reducing compilation overhead also enables the implementation of composite pulse sequences, such as SK1 and BB1 \cite{brown_arbitrarily_2004}, which can improve robustness against Raman beam power fluctuations.

We test a sample circuit comprising consecutive sympathetic cooling cycles interleaved with SK1-$\pi/2$ single-qubit gates. We apply axial sympathetic cooling during the sequence considering the single-qubit gates do not directly couple to the radial modes. Each gate block consists of sequentially applying a single Gaussian SK1-$\pi/2$ pulse to each qubit ion in the chain. A $100~\mathrm{\mu s}$ axial sympathetic-cooling cycle is inserted before every gate block. 

\begin{figure}[H]
  \centering
	\includegraphics[width=0.48\textwidth]{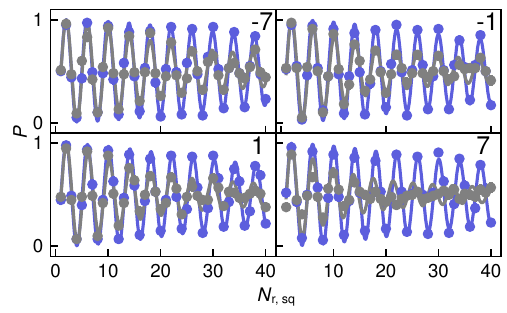}
    \caption{Performance of single-qubit gates. A total of $N_{\mathrm{r,sq}}$ blocks of SK1 $\pi/2$ gates applied sequentially to each qubit are performed either consecutively (gray) or interleaved with $100~\mathrm{\mu s}$ of axial sympathetic cooling (blue). The measured state populations of qubits with selected indices (indicated in the upper-right corner of each panel) are shown as function of $N_{\mathrm{r,sq}}$. Error bars denote $1\sigma$ binomial uncertainties.} 
	\label{fig:acrc_sk1_compare_main}
\end{figure}

We show the population transfer of qubits following this circuit in \Fig{fig:acrc_sk1_compare_main}. The accumulation of SK1 pulse errors is substantially reduced when axial sympathetic cooling is incorporated (blue) compared with the case without sympathetic cooling (gray) (data for remaining qubits are shown in Appendix~\ref{app:cooling_supp_data}). These results further demonstrate that the interspersed sympathetic cooling sequence preserves coherent qubit evolution while mitigating motional heating accumulated during the circuit.

\begin{figure}[H]
  \centering
	\includegraphics[width=0.3\textwidth]{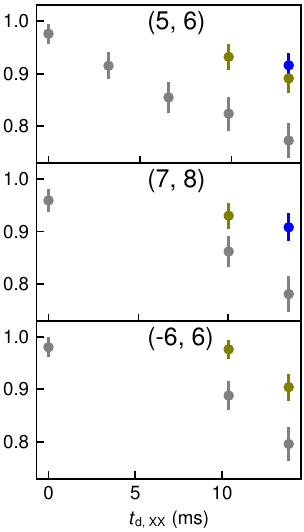}
	\caption{Performance of entangling gates.  The fidelity $F$ of a single XX-$\pi/4$ gate is measured for the ion pairs $(5,6)$, $(7,8)$, and $(-6,6)$ following an idle delay of duration $t_{\mathrm{d,XX}}$. The gate is performed immediately after the idle delay (gray), following 3 ms of axial sympathetic cooling (green), or following 4 ms of axial sympathetic cooling (blue). Error bars are obtained by propagating the uncertainties in the parity-contrast fits and the $1\sigma$ binomial uncertainties in the even-state populations.}
	\label{fig:acrc_xx_delay}
\end{figure}

For entangling gates, we evaluate the performance of individual entangling gate in the presence of sympathetic cooling. We investigate three representative ion pairs located at different positions in the chain. In each experiment, we characterize a single maximally entangling XX gate. The sequence begins with a 10-ms idle period, implemented by applying XX gates with the individual Raman beams turned off to the addressed ion pair. Following the idle period, axial sympathetic cooling is applied when enabled. A single maximally entangling XX gate is then performed on the selected ion pair, followed by BB1-$\pi/2$ analysis pulses with the phase of the final pulse scanned to measure the parity fringe.

The gate fidelity is calculated as $F=(p_{00}+p_{11}+C)/2$ \cite{sackett_experimental_2000}, where $p_{00}$ and $p_{11}$ are the populations of the $\ket{00}$ and $\ket{11}$ states respectively, and $C$ is the parity fringe contrast. We show the measured single gate fidelity following delay of $t_{\mathrm{d, XX}}$ in \Fig{fig:acrc_xx_delay}. For the ion pair $(-6,6)$, the single-gate fidelity improves from $0.89(3)$ to $0.976(18)$ after 3-ms axial sympathetic cooling following an approximately 10-ms idle period, approaching the reference fidelity of $0.980(19)$ measured immediately after state preparation. Under similar idle conditions, the fidelity increases from $0.86(3)$ to $0.93(3)$ for pair $(7,8)$ and from $0.82(3)$ to $0.93(3)$ for pair $(5,6)$ after 3-ms axial sympathetic cooling. These results demonstrate that axial sympathetic cooling effectively recovers the entangling-gate fidelity lost because of axial motional heating during the idle period.

%%%%%%%%%%%%%%%%%%%%%%%%%%%%%%%%%%%%%%%%%%%%%%%%%%%%%%%%%%%%%%%%%%%%%%%%%%%%%%%%%%%%%%%%%%%%%%%%%%%%%%%%%%%%%%%%%%%%%%%%%%%%
\section{Qubit reset 
\label{sec:reset}}
\noindent

We develop a qubit-reset protocol that exploits the shared coupling of the qubit and coolant ions to the LR modes. The protocol assumes that the selected radial mode is initially cooled close to its motional ground state, either by direct sideband cooling of the qubit ions or by radial sympathetic cooling. We first identify a radial mode for which the target qubit has a large participation factor and then select a coolant ion that also exhibits strong participation in the same mode.

\begin{figure}[H]
  \centering
	\includegraphics[width=0.24\textwidth]{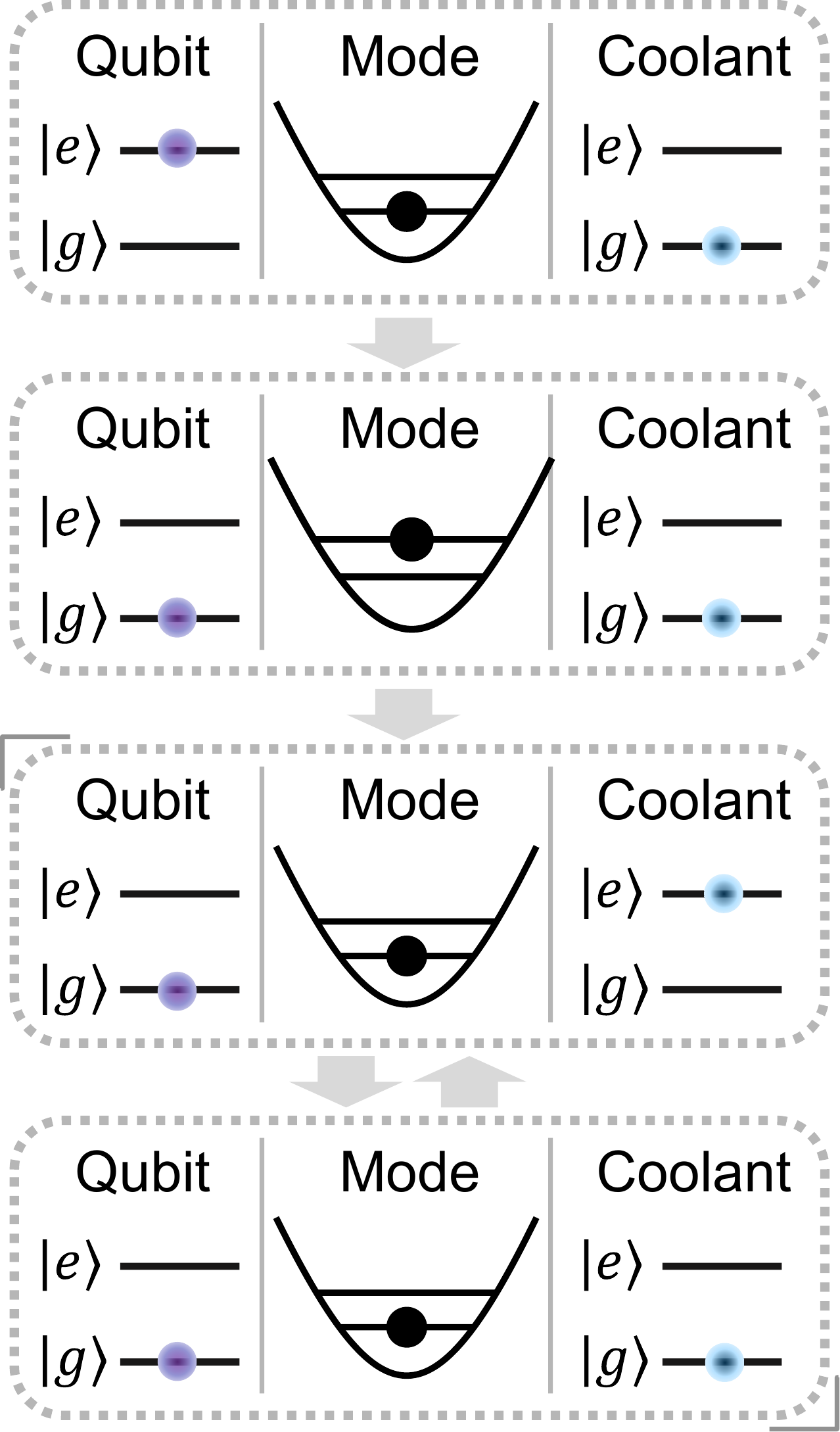}
	\caption{Qubit reset protocol. The reset protocol exploits the shared coupling of the coolant and qubit ions to the LR modes. The protocol begins by preparing the selected radial mode close to its ground state and initializing the coolant ion in the $^2S_{1/2},m_J=-1/2$ state (g). A first-order red-sideband $\pi$ pulse is then applied to transfer the qubit ion population from the excited state (e) to the ground state (g). For an initially excited qubit ion, this process increases the mean occupation of the radial mode by one phonon. Then, the radial cooling sequence is applied to the selected coolant ion, returning the LR mode close to its ground state.}
	\label{fig:qubit_reset_protocol}
\end{figure}

\begin{figure*}[t]
  \centering
	\includegraphics[width=\textwidth]{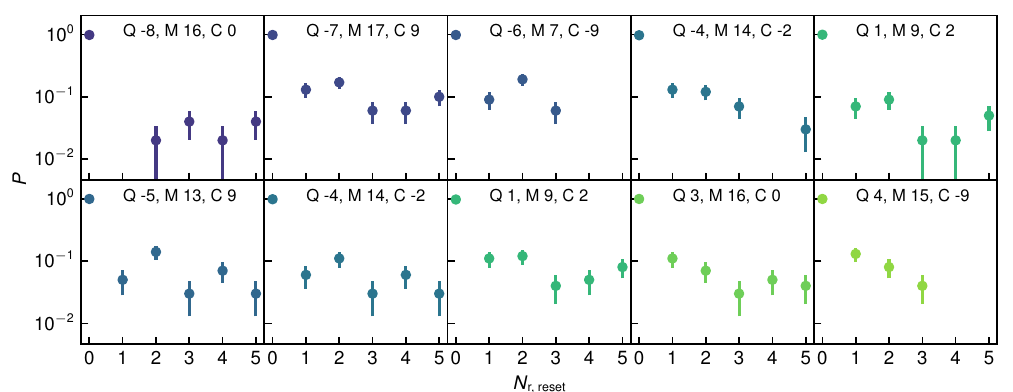}
	\caption{Collective reset of five qubits. Results using qubit-mode-coolant mapping 1 (top) and mapping 2 (bottom). All qubits are initially prepared in the excited state, and the reset sequence is applied for $N_{\text{r, reset}}$ rounds. The configuration for each qubit ion is denoted as Q i, M j, C k, indicating that qubit ion $i$ is reset using radial mode $j$ and coolant ion $k$. Error bars denote the $1\sigma$ binomial uncertainties.
    }
	\label{fig:collective_reset}
\end{figure*}

The protocol is illustrated in \Fig{fig:qubit_reset_protocol}. First, the target qubit is coupled to the selected radial mode by applying a first-order red-sideband $\pi$ pulse. We then perform $N_{\mathrm{c}}$ radial sympathetic cooling cycles on the associated coolant ion, which resets the coolant to ground state and recool the mode. The reset protocol can be executed in parallel on multiple qubits, provided that each target qubit is associated with a sufficiently cold radial mode and a coolant ion with appreciable participation in that mode.

We demonstrate this parallel protocol by simultaneously resetting five qubits using two distinct qubit--mode--coolant mappings, as shown in \Fig{fig:collective_reset}. The experimental sequence consists of a $220~\mathrm{\mu s}$ qubit sideband pulse followed by $N_c=20$ coolant cooling cycles ($960~\mathrm{\mu s}$), resulting in a total duration of $1.18~\mathrm{ms}$ for a single collective reset.

The qubits are initialized in the excited state before the reset sequence. After a single reset cycle, the excited-state population of all selected qubits is reduced to approximately $0.1$ (without correcting for SPAM errors). The reset is limited primarily by residual excitation of the motional mode above its ground state. Nevertheless, because the protocol naturally incorporates radial sympathetic cooling and resets multiple qubits in parallel, the additional experimental overhead is limited to the duration of the single initial qubit sideband pulse.

%%%%%%%%%%%%%%%%%%%%%%%%%%%%%%%%%%%%%%%%%%%%%%%%%%%%%%%%%%%%%%%%%%%%%%%%%%%%%%%%%%%%%%%%%%%%%%%%%%%%%%%%%%%%%%%%%%%%%%%%%%%%
\section{Conclusions and outlook 
\label{sec:conclusions}}
\noindent
We demonstrated a dual-isotope sympathetic-cooling scheme in a long Yb$^+$ ion chain. Our scheme maintains the long-wavelength axial and radial modes of the chain near their post-state-preparation temperature, thereby addressing the heating problem identified in Ref.~\cite{cetina_control_2022}; it also efficiently recools the modes used for entangling gates to near their ground states. Our sympathetic-cooling scheme preserves qubit coherence and can potentially be extended to other ion species such as Ba$^+$ and Ca$^+$ (see Appendix~\ref{app:spectral_sel}).

Our approach provides a practical route to high-fidelity quantum gates in medium- to large-scale trapped-ion processors based on long ion-chain units. It effectively suppresses the two dominant sources of motional error, axial heating during circuit execution and radial recoil heating induced by axial sympathetic cooling, while leaving a substantial fraction of the experimental sequence available for coherent quantum operations. By maintaining low and stable motional occupations over tens of milliseconds, the demonstrated protocol represents an important step toward executing deep quantum circuits with high fidelity in long ion chains.

Beyond enhancing unitary operations, the components of our sympathetic-cooling scheme provide a versatile toolbox for engineering nonunitary dynamics. In addition to the parallel qubit reset, these tools enable simulations of dissipative spin-boson dynamics, with selected radial modes serving as a bosonic bath whose dissipation is engineered through radial sympathetic cooling. Photon recoil generated during axial sympathetic cooling provides an additional control knob for tuning the excitation of the radial modes.

Our sympathetic cooling scheme is inherently scalable to longer ion chains. The relative ordering of the two isotopes is maintained through infrequent shuttling operations performed outside the quantum circuit, introducing no additional runtime overhead during circuit execution. Furthermore, because the two isotopes have similar masses, the mode participation remains well balanced between the qubit and coolant ions, even in long chains. The overhead associated with axial cooling does not increase significantly with system size, provided that a similar coolant-ion fraction is maintained. In contrast, the efficiency of parallel radial sympathetic cooling is expected to decrease as the mode participation factors scale approximately as $1/\sqrt{N}$, leading to longer sideband pulse durations in larger systems. This scaling can be mitigated by optimizing the cooling of each radial mode using multiple coolant ions within a single cooling round and by increasing the Raman sideband drive strength where experimentally feasible \cite{vybornyi_collective_2026}.
 
We expect our sympathetic cooling scheme to yield substantially greater improvements in long-chain scalability when implemented in a well-engineered cryogenic system. The present work is demonstrated in a room-temperature system, where the comparatively large heating rate of the LA mode sets the dominant contribution to the sympathetic cooling time budget. In state-of-art cryogenic systems, however, the heating rate of this mode can be reduced by more than two orders of magnitude. Such a reduction would substantially decrease the axial cooling time required during circuit execution and, consequently, the radial cooling needed to remove the recoil heating generated by axial cooling. To quantify this improvement, we perform a rough estimate for a 60-ion mixed Yb$^{+}$ chain based on the cryogenic system parameters reported in Ref.~\cite{brandl_cryogenic_2016}. To suppress the axial-heating associated single maximally entangling XX amplitude-modulated MS gate infidelity to below $10^{-3}$, the total sympathetic cooling overhead is estimated to account for only 21\% of circuit time (see Appendix~\ref{app:cryo_estimate} for details). 

\section*{Acknowledgment}
\noindent
T.W. is grateful to L. Feng for instructional help on the experimental setup, and thanks Y. Yu for valuable discussions. M.C. and T.W. are supported by 
the National Science Foundation's Quantum Leap Challenge Institute for Robust Quantum Simulation under Award NSF (OMA-2120757) and the NSF STAQ project (Phy-2325080). M.C. also acknowledges support from the U.S. Department of Energy, Office of Science, National Quantum Information Science Research Centers, Quantum Systems Accelerator.

%%%%%%%%%%%%%%%%%%%%%%%%%%%%%%%%%%%%%%%%%%%%%%%%%%%%%%%%%%%%%%%%%%%%%%%%%%%%%%%%%%%%%%%%%%%%%%%%%%%%%%%%%%%%%%%%%%%%%%%%%%%%
\section*{Competing Interests}
\noindent
M.C. is a co-inventor on 
%U.S. patents US11710061B2, US11152756B2, US11083075B2, and US11573477B2, 
patents assigned to University of Maryland that were licensed to IonQ Inc. T.W. declares no competing interests.

%%%%%%%%%%%%%%%%%%%%%%%%%%%%%%%%%%%%%%%%%%%%%%%%%%%%%%%%%%%%%%%%%%%%%%%%%%%%%%%%%%%%%%%%%%%%%%%%%%%%%%%%%%%%%%%%%%%%%%%%%%%%
\section*{Data availability}
\noindent
The data that support the findings of this work is openly available in \cite{RDM}.

%%%%%%%%%%%%%%%%%%%%%%%%%%%%%%%%%%%%%%%%%%%%%%%%%%%%%%%%%%%%%%%%%%%%%%%%%%%%%%%%%%%%%%%%%%%%%%%%%%%%%%%%%%%%%%%%%%%%%%%%%%%%
\appendix
%%%%%%%%%%%%%%%%%%%%%%%%%%%%%%%%%%%%%%%%%%%%%%%%%%%%%%%%%%%%%%%%%%%%%%%%%%%%%%%%%%%%%%%%%%%%%%%%%%%%%%%%%%%%%%%%%%%%%%%%%%%%
\section{Isotope-Selective Loading 
\label{app:loading}}
\noindent
The dual-isotope chain is built up isotope-selectively one ion at a time following the target spatial isotope sequence. We load both the \qubit qubit and the \coolant coolant ions from a thermal source. A key technical difficulty is to load the coolants efficiently while maintaining isotope selectivity. Our neutral ytterbium sample is enriched in $^{171}\text{Yb}$\hspace{1mm} and contains only a small natural abundance of $^{172}\text{Yb}$\hspace{1mm} (2.5\%) \cite{eganthesis}. In the two-step photoionization process used to load $\text{Yb}^+$ \cite{eganthesis}, we increase the optical power of the 394-nm beam used in the second ionization step to around 80 mW with a $1/e^2$-intensity radius of around $44~\mathrm{\mu m}$ during the coolant-loading pulse. For the first-step of ionization, the isotope shift of the $^1S_0$ -- $^1P_1$ transition for $^{172}\text{Yb}$\hspace{1mm} relative to $^{171}\text{Yb}$\hspace{1mm} is $\Delta_l=-2\pi\times304~\mathrm{MHz}$ \cite{kleinert_measurement_2016}, while the natural linewidth of the $^1P_1$ state is $2\pi\times29~\mathrm{MHz}$ \cite{bowers_experimental_1996}. To preserve isotope selectivity, we reduce the 399-nm beam power to approximately the saturation intensity during the coolant loading pulse.

Following the loading pulse, we apply a Doppler-cooling pulse and then shuttle the trapping well from the load slot to the trap center. To determine the ion presence and species, we sequentially turn on the 369-nm qubit and coolant beams while monitoring the fluorescence. Once we confirm the desired species has been trapped, the new ion is merged with the existing chain.

%%%%%%%%%%%%%%%%%%%%%%%%%%%%%%%%%%%%%%%%%%%%%%%%%%%%%%%%%%%%%%%%%%%%%%%%%%%%%%%%%%%%%%%%%%%%%%%%%%%%%%%%%%%%%%%%%%%%%%%%%%%%
\section{Deterministic Isotope Sorting 
\label{app:sorting}}
\noindent
In each experimental shot, we determine the isotope configuration of the ion chain by collecting fluorescence for $500~\mathrm{\mu s}$ immediately after the nominal Doppler-cooling sequence using a near-resonant 369-nm beam that drives the coolant D1 transition. The measured fluorescence counts are compared with pre-calibrated thresholds to generate a binary array representing the detected isotope configuration. If the detected configuration differs from the target configuration, the measurement is repeated up to two additional times. If the configuration remains incorrect after three consecutive checks and no ion loss is detected, the sorting subroutine is executed. Otherwise, if the target configuration is confirmed to be correct, the sorting step is skipped and the experiment proceeds directly to Raman sideband cooling. If ion loss is detected at any point, the sorting subroutine is also bypassed, and the corresponding experimental shot is flagged as invalid.

In the sorting subroutine, the ion chain is rearranged into the target isotope configuration through a sequence of nearest-neighbor swaps using the algorithm described in Ref.~\cite{andrew_tianyi_sorting}. At the start of the subroutine, the trap rf amplitude is taken out of the lock and ramped down using the rf source (Rohde \& Schwarz SMC100A). The reduced rf amplitude is calibrated such that the radial secular frequency of the \qubit qubit ions is approximately $2\pi\times2~\mathrm{MHz}$.

The sorting subroutine consists of an iterative sorting loop. At each iteration, a sorting solution is generated in real time from the detected configuration and the target configuration using the sorting algorithm. The solution comprises a sequence of nearest-neighbor swaps that transforms the current configuration into the target configuration. Only the first swap in the sequence is executed during each iteration, after which the isotope configuration is remeasured. The loop terminates when the detected configuration matches the target configuration or when a maximum of 90 swap attempts has been reached. At the end of the sorting loop, the trap rf amplitude is ramped back to its nominal value and the rf-amplitude lock is re-engaged. 

Each nearest-neighbor swap consists of three stages: splitting, swapping, and merging. During the splitting stage, the target ion pair is isolated from the rest of the chain through a sequence of up to three approximately equal chain-splitting operations implemented with DC-voltage waveforms. The remaining chain segments are then transported away from the trap center to prevent them from being perturbed during the subsequent swap operation.

During the swapping stage, DC-voltage waveforms are applied to compress the isolated ion pair and tune the axial mode close to degeneracy with one of the radial modes. As the waveform progresses, the principal axes of the trapping potential rotate, causing the ion pair to undergo a $180^\circ$-rotation and thereby exchange positions.

During the merging stage, the splitting sequence is played in reverse to recombine the swapped ion pair with the previously separated chain segments, thereby restoring the original chain with the selected ion pair swapped.

The axial shuttling region extends from approximately $-750~\mathrm{\mu m}$ to $+750~\mathrm{\mu m}$ relative to trap center. For a 23-ion chain, the four possible subwell configurations are ${\mathrm{C1R3},\mathrm{L1C1R2},\mathrm{L2C1R1},\mathrm{L3C1}}$, where $\mathrm{Ln}$ and $\mathrm{Rn}$ denote $n$ subwells on the negative and positive sides of the trap axis respectively, and $\mathrm{C1}$ denotes a single subwell at the trap center. For example, the subwell positions are $(0,+250,+470,+750)~\mathrm{\mu m}$ for the $\mathrm{C1R3}$ configuration and $(-250,0,+470,+750)~\mathrm{\mu m}$ for the $\mathrm{L1C1R2}$ configuration.

This shuttling range is chosen primarily to transport the outer, isolated subwells sufficiently far from the central subwells, thereby protecting the isolated ions while the central subwells undergo the splitting and swapping operations. The shuttling range is not extended further because axial stray-field compensation becomes infeasible in regions where axially symmetric DC electrode pairs are electrically connected.

The voltage waveform for the swapping stage is constructed by superposing two components. The first is the final voltage configuration of the splitting sequence, which preserves the trapping potential experienced by the isolated ions at the end of the splitting stage. The second drives the exchange of the central ion pair. This component is largely confined to the trap center region and has a substantially larger amplitude than the shuttling waveforms. It is generated analytically such that the principal axes of the two lowest-frequency motional modes rotate continuously by $180^\circ$ as the waveform progresses from its initial to final configuration. 

%%%%%%%%%%%%%%%%%%%%%%%%%%%%%%%%%%%%%%%%%%%%%%%%%%%%%%%%%%%%%%%%%%%%%%%%%%%%%%%%%%%%%%%%%%%%%%%%%%%%%%%%%%%%%%%%%%%%%%%%%%%%
\section{Raman Control of Coolant Ions
\label{app:coolant_Raman}}
\noindent
Our Raman system employs a lin $\perp$ lin polarization configuration. The optical fields of the global beam, $\mathbf{E}_{\mathrm{g}}(\mathbf{r},t)$, and the individual beam, $\mathbf{E}_{\mathrm{i}}(\mathbf{r},t)$, generate an effective magnetic field, $\mathbf{B}_{\mathrm{eff}}$, through the vector light shift. $\mathbf{B}_{\mathrm{eff}}$ is proportional to the vector product of the two optical fields,$-i \mathbf{E_{\mathrm{i}}}^*(\mathbf{r}, t) \times \mathbf{E_{\mathrm{g}}}(\mathbf{r}, t)$. 

The qubit states of \qubit, expressed in the $\ket{J,m_J;I,m_I}$ basis (hereafter abbreviated as $\ket{m_J;m_I}$), are 
\begin{align}
   \ket{1}=1/\sqrt{2} (\ket{+1/2;-1/2} - \ket{-1/2;+1/2}),
\end{align}
\begin{align}
   \ket{0}=1/\sqrt{2} (\ket{+1/2;-1/2} + \ket{-1/2;+1/2}).
\end{align}
The \coolant ion has zero nuclear spin. The coolant basis is therefore simply the Zeeman basis of the $^2S_{1/2}$ manifold, $\ket{+1/2}$ and $\ket{-1/2}$, expressed in the $\ket{m_J}$ basis. These states are defined with respect to the quantization magnetic field, $\mathbf{B}_0$. When the effective Raman field, $\mathbf{B}_{\mathrm{eff}}$, is parallel to $\mathbf{B}_0$, the Raman interaction flips the \qubit singlet-triplet qubit-state pair while leaving the \coolant Zeeman-state pair unchanged up to a global phase. When the effective Raman field, $\mathbf{B}_{\mathrm{eff}}$, is perpendicular to $\mathbf{B}_0$, the interaction instead flips the \coolant Zeeman-state pair while leaving the \qubit singlet–triplet pair unchanged up to a global phase.

In our experimental setup, the Raman beams are aligned such that the effective field, $\mathbf{B}_{\mathrm{eff}}$, is directed along the $\hat{x}$. The laser repetition rate is stabilized at $f_{\mathrm{rep}}=118.307201~\mathrm{MHz}$, while the AOMs in the path of the global and individual beams are driven at $f_g=180.950439~\mathrm{MHz}$ and $f_i=197~\mathrm{MHz}$ respectively.

The transition between the \qubit qubit states is resonantly driven by pairs of comb teeth separated by $107f_{\mathrm{rep}}$, corresponding to the absorption of a photon from the global beam followed by stimulated emission into the individual beam. In contrast, the transition between the \coolant coolant states is driven by the absorption of a photon from the individual beam followed by stimulated emission into the global beam using comb-tooth pairs with zero tooth-index separation, with the remaining carrier-frequency offset provided by the global-beam AOM.

Under the same power-optimized global- and individual-beam drive conditions, the measured carrier Raman periods are $4.04~\mathrm{\mu s}$ for the \coolant transition and $2.78~\mathrm{\mu s}$ for the \qubit transition.

Magnetic-field fluctuations and vector light shifts limit the coherence of Raman operations on the \coolant states. In the absence of magnetic shielding, ambient magnetic-field noise fundamentally limits the coherence time of the coolant states. In addition, imperfect polarization purification and optical Magnus effects \cite{stafeev_circular_2022} leave residual circular polarization components, predominantly in the individual beams. Through the vector coupling, these residual circular components generate an additional effective magnetic field, $\mathbf{B}_{\mathrm{circ}}(\mathbf{r},t)$, directed along $-i\boldsymbol{\varepsilon}_{\mathrm{i}}^*\times\boldsymbol{\varepsilon}_{\mathrm{i}}$, where $\boldsymbol{\varepsilon}_{\mathrm{i}}(\mathbf{r},t)$ denotes the polarization vector of the residual circularly polarized component of the individual beam. Although the transverse component of $\mathbf{B}_{\mathrm{circ}}$ is negligible because $B_{\mathrm{circ}}\ll B_{\mathrm{eff}}$, its projection along the quantization axis modifies the state splitting and gives rise to drive-dependent shifts and dephasing. As a result, the magnetically sensitive coolant states experience vector Stark shifts approximately three orders of magnitude larger than those experienced by the clock qubit states of the \qubit qubit ions.

During each experimental run, we calibrate $\mathbf{B}_{\mathrm{circ}}$ for every coolant ion in the chain using a Ramsey frequency scan. The resulting frequency shifts vary among ions because they are dominated by the individual addressing beams, whose local optical conditions differ across the chain. Accurate calibration of these frequency shifts is essential because their magnitudes are generally comparable to the first-order sideband Raman Rabi frequencies coupling the \coolant ions to the LR modes. These calibrated shifts are highly sensitive to changes in beam pointing, presumably because variations in the beam–ion alignment alter the polarization ellipticity induced by the optical Magnus effect.

To reduce this sensitivity, we optimize the telecentricity of the individual-beam array to bring the beam foci onto a common plane. The lateral alignment of each individual beam is then precalibrated in both directions within the trap plane by optimizing the intensity on corresponding ion. Finally, the axial tilt of each beam is optimized by minimizing the axial displacement at the ion when driving bichromatically at axial mode frequency\cite{huang_comparing_2024}.

%%%%%%%%%%%%%%%%%%%%%%%%%%%%%%%%%%%%%%%%%%%%%%%%%%%%%%%%%%%%%%%%%%%%%%%%%%%%%%%%%%%%%%%%%%%%%%%%%%%%%%%%%%%%%%%%%%%%%%%%%%%%
\section{Spectral Isolation
\label{app:spectral_sel}}
\noindent
Here, we estimate the spectral isolation between the qubit ions and coolant ions during dual-isotope sympathetic cooling through the narrow-line quadrupole transition. We extend the analysis from $\mathrm{Yb}^{+}$ to $\mathrm{Ba}^{+}$ and $\mathrm{Ca}^{+}$, which are the other two species commonly used for experiments with long trapped ion chains. 

We assume the quadrupole laser has two frequency components, each resonantly addressing a $\Delta m_J=0$ transition of the coolant ions with a Rabi frequency of $\Omega_{\mathrm{SD}}$. The quadrupole couplings are based on the experimental setup described in Sec.~\ref{sec:exp_setup}. 

Because the motional-mode frequencies and Zeeman splittings are generally more than an order of magnitude smaller than the detunings associated with the isotope shifts and hyperfine splittings, we evaluate the off-resonant quadrupole coupling to the qubit ions by approximating the quadrupole drive as resonant with the coolant transition. Finally, we assume an ion spacing of $d=3.83~\mathrm{\mu m}$, corresponding to that in our experimental system and representative of typical spacings in individually addressed trapped-ion systems.

\subsection{Framework}
Far-off-resonant photon scattering from the quadrupole beam via the D1 and D2 transitions can induce decoherence in the qubits. For hyperfine qubits in the ground-state manifold, this process is dominated by Raman scattering. Following Refs.~\cite{boguslawski_raman_2023, moore_photon_2023}, we derive the corresponding Raman scattering rate, given in Eq.~\ref{eq:raman_d1_d2_raman_full}.
\begin{equation}
\begin{aligned}
   &\Gamma_{\text{Raman}}=\frac{\omega_{l}^3c^2\pi I_l}{3\hbar}\left(\frac{2\omega_{l}}{\omega_{e^\prime}^2-\omega_{l}^2}\frac{A_{e^\prime}}{\omega_{e^\prime}^3}-\frac{2\omega_{l}}{\omega_{e}^2-\omega_l^2}\frac{A_{e}}{\omega_{e}^3}\right)\\
   &+\Theta(\omega_l-\omega_{g,g^\prime})\frac{2(\omega_{l}-\omega_{g, g^\prime})^3c^2\pi I_l}{\hbar}
   \left(\frac{2\omega_{l}}{\omega_{e^\prime}^2-\omega_{l}^2}\frac{A_{e^\prime,g^\prime}}{\omega_{e^\prime,g^\prime}^3}\right)^2 \\
   &+\Theta(\omega_l-\omega_{g,g^\prime}) \frac{(\omega_{l}-\omega_{g, g^\prime})^3c^2\pi I_l}{\hbar}\left(\frac{2\omega_{l}}{\omega_{e}^2-\omega_{l}^2}\frac{A_{e,g^\prime}}{\omega_{e,g^\prime}^3}\right)^2 \\
   &+\Theta(\omega_l-\omega_{g,g^{\prime\prime}}) \frac{2(\omega_{l}-\omega_{g, g^{\prime\prime}})^3c^2\pi I_l}{\hbar}\left(\frac{2\omega_{l}}{\omega_{e^\prime}^2-\omega_{l}^2}\frac{A_{e^\prime,g^{\prime\prime}}}{\omega_{e^\prime,g^{\prime\prime}}^3}\right)^2
   \label{eq:raman_d1_d2_raman_full}, 
\end{aligned}
\end{equation}
Here, $\omega_{l}$ is the drive angular frequency. $I_l$ is the beam intensity at the ion. The labels $e^\prime$ and $e$ denote the excited manifolds $^2P_{3/2}$ and $^2P_{1/2}$ respectively, while $g^\prime$ and $g^{\prime\prime}$ denote the metastable manifolds $^2D_{3/2}$ and $^2D_{5/2}$. The terms $\omega_{e^\prime}$ and $\omega_{e}$ are the transition angular frequencies between the ground and the corresponding excited manifolds, whereas $\omega_{g,g^{\prime}}$ and $\omega_{g,g^{\prime\prime}}$ are the transition angular frequencies between the ground and the corresponding metastable manifolds. The parameters $A_{e^\prime}$ and $A_{e}$ represent the Einstein $A$ coefficients for transitions between the excited and ground manifolds, with $A_{e^\prime, g^\prime}$, $A_{e, g^{\prime\prime}}$, and $A_{e, g^\prime}$ representing the coefficients between the corresponding excited and metastable manifolds. $\Theta(x)$ is the Heaviside step function used to enforce energy conservation. We neglect hyperfine splittings and isotope shifts, as the detunings between the drive and each excited-state manifold are considerably larger.

Another source of decoherence arises from the direct off-resonant coupling to the quadrupole transition. The fluctuating Stark shift induced by this coupling can be significant under certain experimental conditions as pointed out in Ref.~\cite{cetina_control_2022}. In addition, photon scattering induced by the quadrupole drive contributes to qubit decoherence. Here, we develop a general framework for evaluating the quadrupole photon scattering rate during sympathetic cooling. The same framework can be adapted to estimate the associated Stark shift for a specific experimental implementation.

For each driving tone, the scattering rate induced by the 435-nm beam is
\begin{align}
   \Gamma_{\text{SD}}=\frac{\Gamma_{\text{D}}}{2}\sum_i \frac{s_i}{1+s_i+\left(\frac{\Delta_{\text{SD, i}}}{\Gamma_{\text{D}}/2}\right)^2}\label{eq:qubit_sd}.
\end{align}
Here, the summation is over all qubit transitions $i$. $\Gamma_{\text{D}}$ is the natural linewidth of the selected metastable states. The saturation parameter is calculated as $s_i=2(\frac{\Omega_i}{\Gamma_{\text{D}}})^2$, where the resonant Rabi frequency for transition $i$ between $^{2}S_{1/2} (F, m_F)$ and $^{2}D_{J^\prime} (F', m_{F'})$ is $\Omega_i=C_{F, m_F}^{F', m_{F'}}(q)\hspace{1mm} \Omega_{\text{SD}}$. $C_{F, m_F}^{F', m_{F'}}(q)$ is the ratio of the Rabi frequency magnitude for the considered qubit $\Delta m_F=q$ coupling to that for the coolant $\Delta m_J=0$ coupling. Following \cite{edmonds1957angular}, we calculated it as 
\begin{equation}
\begin{aligned}
   C_{F, m_F}^{F', m_{F'}}(q)&=|\frac{\langle F', m_{F}'|T_q^{(2)}|F, m_F\rangle}{\langle J', m_{J}'|T_0^{(2)}|J, m_J\rangle}|\\
   &=|\frac{g^{(q)}(\gamma, \phi)}{g^{(0)}(\gamma, \phi)}|\sqrt{(2F'+1)(2F+1)}\\
   &\times |\frac{\begin{pmatrix}
    F' & 2 & F \\
    -m_F' & q & m_F
    \end{pmatrix} 
    \begin{Bmatrix}
    J' & F' & I \\
    F & J & 2
    \end{Bmatrix}
    }{\begin{pmatrix}
    J' & 2 & J \\
    -m_J' & 0 & m_J
    \end{pmatrix}}|\label{eq:quad_coupling_ratio}.
\end{aligned}
\end{equation}
Here, $T_q^{(2)}$ is the rank-2 spherical tensor coupling the electronic degrees of freedom of the qubit ion. Qubit ion has nuclear angular quantum number $I$. $\langle J',m_{J}'|T_q^{(2)}|J,m_J\rangle$ is the quadrupole-transition matrix element for the coolant transition, where $J$ and $J^\prime$ are the total electronic angular momentum of ground and metastable manifolds. $m_{J}'=m_J=\pm1/2$, corresponding to the driven $\Delta m_J=0$ transitions. $g^{(q)}$ denotes the geometric coupling coefficient of $T_q^{(2)}$, determined by the geometry of the quantization axis and the quadrupole beam, as given by Eq.~\ref{eq:SD_geo} \cite{james_quantum_1998}. For the experimental parameters described earlier, $\gamma=0.651\pi, \phi=0.636\pi$.
\begin{equation}
\begin{aligned}
g^{(0)} &=-\frac{1}{2}\cos \gamma \sin (2\phi ),\\
g^{( \pm 1)} &=\mp \frac{1}{\sqrt{6}} [\cos \gamma \cos (2\phi )\pm i\sin \gamma \cos \phi ], \\
g^{( \pm 2)} &=\frac{1}{\sqrt{6}}\left[\frac{1}{2}\cos \gamma \sin (2\phi )\pm i\sin \gamma \sin \phi \right].
\end{aligned}\label{eq:SD_geo}
\end{equation}

During sympathetic cooling, spontaneously emitted photons from the coolant ions can induce crosstalk on the neighboring qubit ions. From the dipole-transition matrix elements between the excited and ground states of the coolant ions considered here, we find equal probabilities for the emission of $\pi$-, $\sigma_-$-, and $\sigma_+$-polarized photons. In general, the polarization of a photon emitted by a coolant ion differs from that observed by a qubit ion. We denote by $D_{\epsilon' \rightarrow \epsilon}$ the probability density per unit solid angle that a photon emitted by the coolant with polarization $\epsilon'$ is observed at the qubit with polarization $\epsilon$, and thus drives the corresponding qubit transition. This probability depends on the angle $\theta$ between the axial direction and the common quantization axis.

\begin{equation}
\begin{aligned}
   D_{\pi\rightarrow\pi}&=\frac{3}{2}\sin^4\theta,\\
   D_{\pi\rightarrow\sigma^{\pm}}&=\frac{3}{4}\sin^2\theta\cos^2\theta,\\
   D_{\sigma^{\pm}\rightarrow\pi}&=\frac{3}{4}\sin^2\theta\cos^2\theta,\\
   D_{\sigma^{\pm}\rightarrow\sigma^{\pm}}&=\frac{3}{8}(1+\cos^2\theta)^2,\\
   D_{\sigma^{\pm}\rightarrow\sigma^{\mp}}&=\frac{3}{8}\sin^4\theta.\\
   \label{eq:polarization}
\end{aligned}
\end{equation}
We evaluate the scattering probability $P_{\mathrm{sp}}$ for a qubit in state $\ket{F,m_F}$ due to a single photon spontaneously emitted by a neighboring coolant ion. The total probability that a photon received by the qubit has polarization $\epsilon$ is obtained by incoherently summing the contributions from all emitted polarizations $\epsilon'$. This incoherent sum is justified because the polarization of the emitted photon is correlated with the final internal state of the coolant, making the different $\epsilon'$-photon emission pathways distinguishable. We also incoherently sum the contributions for the qubit ion driven by an $\epsilon$-photon.

\begin{equation}
\begin{aligned}
   P_{\mathrm{sp}}&= \sum_{\epsilon, \epsilon', F', m_{F^\prime}}\frac{D_{\epsilon'\rightarrow\epsilon}}{3} \frac{\sigma_0 \frac{(\Gamma/2)^2}{(\Gamma/2)^2+(\Delta_{\mathrm{sp}})^2}}{4\pi d^2} \xi \\&\times (2J'+1)\frac{|\bra{F', m_F'}r_\epsilon\ket{F, m_F}|^2}{|\langle J'||\mathbf{r}||J\rangle|^2}
   \label{eq:rel_crosstalk_R}.
\end{aligned}
\end{equation}

In Eq.~\ref{eq:rel_crosstalk_R}, $\frac{1}{3}$ accounts for the probability of the emitted photon being of $\epsilon'$ polarization. $\sigma_0=3 \lambda^2/(2\pi)$ is the resonant cross section. $\Gamma$ is the natural linewidth of the selected excited state. $\xi$ is the branching ratio from excited state manifold to the ground state manifold. The relative transition strength for the coupled transition in the qubit is given by Ref. \cite{edmonds1957angular} as
\begin{equation}
\begin{aligned}
  \frac{|\bra{F', m_F'}r_\epsilon\ket{F, m_F}|^2}{|\langle J'||\mathbf{r}||J\rangle|^2}&=(2F'+1)(2F+1)\\
   &\begin{pmatrix}
    F' & 1 & F \\
    -m_F' & \epsilon & m_F
    \end{pmatrix}^2
    \begin{Bmatrix}
    J' & J & 1 \\
    F & F' & I
    \end{Bmatrix}^2
   \label{eq:dipole_rel}.
\end{aligned}
\end{equation}
Using $P_{\mathrm{sp}}$, we can then calculate the scattering rate of a qubit ion as $R_{\mathrm{c}}P_{\mathrm{sp}}$, where $R_{\mathrm{c}}$ is the photon scattering rate of the neighboring coolant ion.

\subsection{$\text{Yb}^{+}$}
We consider \qubit as the qubit ion and \coolant as the coolant ion. We use Eq.~\ref{eq:raman_d1_d2_raman_full} to estimate Raman scattering via the D1 and D2 transitions induced by a single addressing tone of the 435-nm beam. With $\omega_{e^\prime}=911.227~\mathrm{THz}$ \cite{meggers_second_1967}, $\omega_{e}=811.323~\mathrm{THz}$ \cite{zalivako_improved_2019}, $\omega_{l}=688.349~\mathrm{THz}$ \cite{stenger_absolute_2001}, $A_{e^\prime}=1.62\times10^8~\mathrm{s^{-1}}$, and $A_{e}=1.23\times10^8~\mathrm{s^{-1}}$ \cite{morton_atomic_2000}, using Eq.~\ref{eq:raman_d1_d2_raman_full}, the Raman scattering rate induced by a single quadrupole-drive tone is given by $\Gamma_{\text{Raman}}=\Omega_{\mathrm{SD}}^2\times1.1\times10^{-16}~\mathrm{s}$, as shown in Table.~\ref{tab:yb_scat_sum}. 

Using the measured isotope shifts of the even Yb isotopes \cite{counts_evidence_2020} together with the isotope shift of the 411-nm transition between \qubit and \coolant \cite{roberts_measurement_1999}, we infer the isotope shift of the 435-nm transition from a King plot relating its modified isotope shift to that of the 411-nm transition. This analysis yields a centroid shift of $2\pi\times1.336~\mathrm{GHz}$ for \qubit relative to \coolant. 

The hyperfine splitting of $^{2}S_{1/2}, ^{2}D_{3/2},^{3}[3/2]_{1/2}$ in \qubit are respectively $2\pi\times 12.6428~\mathrm{GHz}$ \cite{fisk_paper},  $2\pi\times 0.86~\mathrm{GHz}$ \cite{olmschenk_manipulation_2007}, and $2\pi\times 2.2095~\mathrm{GHz}$ \cite{olmschenk_manipulation_2007}. We can then determine the detunings $\Delta_{\mathrm{SD}}$ and coefficients $C_{F,m_F}^{F',m_{F'}}(q)$ for all allowed 435-nm transitions from the two qubit states, $\ket{F=0,m_F=0}$ and $\ket{F=1,m_F=0}$, to the different excited-state hyperfine manifolds in \qubit, as summarized in Table.~\ref{tab:yb_sd}. The natural linewidth of the $^{3}D_{3/2}$ state is $\Gamma_{\mathrm{D}}=2\pi\times3~\mathrm{Hz}$ \cite{Yu_yb_linewidth}. Using Eq.~\ref{eq:qubit_sd}, the scattering rates induced by a single driving tone of the 435-nm beam for the two qubit states are respectively $\Gamma_{\mathrm{SD}}(\ket{F=0, m_F=0})=\Omega_{\mathrm{SD}}^2\times1.1\times10^{-20}~\mathrm{s}$, $\Gamma_{\mathrm{SD}}(\ket{F=1, m_F=0})=\Omega_{\mathrm{SD}}^2\times5.0\times10^{-19}~\mathrm{s}$ as shown in Table.~\ref{tab:yb_scat_sum}.

\begin{table}[htbp]
    \centering
    \renewcommand{\arraystretch}{1.3}
    \caption{435-nm transition relative qubit Rabi frequency magnitude $C_{F, m_F}^{F', m_{F'}}(q)$ and detuning $\Delta_{\mathrm{SD}}$.}
    \label{tab:yb_sd}
    \begin{tabular*}{\linewidth}{@{\extracolsep{\fill}}cccccc}
    \hline\hline 
     $F$ & $q$ & $F^\prime$ & $m_{F^\prime}$ & $ C_{F, m_F}^{F', m_{F'}}(q)$& $\displaystyle \Delta _{\mathrm{SD}} /(2\pi\times\mathrm{GHz})$  \\
    \hline 
    0 & 0 & 2 & 0 & 1 & 11.14\\
    0 & $\pm$1 & 2 & $\pm$1 & 1.126 & 11.14\\
    0 & $\pm$2 & 2 & $\pm$2 & 1.962 & 11.14\\
    1 & 0 & 1 & 0 & 1 & -2.36\\
    1 & $\pm$1 & 1 & $\pm$1 & 0.975 & -2.36\\
    1 & $\pm$1 & 2 & $\pm$1 & 0.563 & -1.50\\
    1 & $\pm$2 & 2 & $\pm$2 & 1.962 & -1.50\\
    \hline\hline
    \end{tabular*}
\end{table}

The natural linewidth of the $^{3}[3/2]_{1/2}$ state is $\Gamma=2\pi\times4.22~\mathrm{MHz}$\cite{Yu_yb_linewidth}. The branching ratio $\xi$ is $0.982$ \cite{Le_bracket_branching}. With the measured shift of $-2\pi\times110~\mathrm{MHz}$ \qubit $^{2}D_{3/2}, F=1$ -- $^{3}[3/2]_{1/2}, F=0$ relative to \coolant \cite{McLoughlin_935}, together with the hyperfine splittings, we calculate the detuning $\Delta_{\mathrm{sp}}$ for all non-forbidden transitions of qubit and the corresponding contribution to the sum (without interference terms), as listed in Table~\ref{tab:yb_db_rel_scattering rate}. The qubit scattering probability for both qubit states are then $P_{\text{sp}}(\ket{F=0, m_F=0})=5.0\times10^{-12}$, and $P_{\text{sp}}(\ket{F=1, m_F=0})=3.3\times10^{-11}$ as shown in Table.~\ref{tab:yb_scat_sum}.

\begin{table}[htbp]
    \centering
    \renewcommand{\arraystretch}{1.3}
    \caption{297-nm photon scattering probability on \qubit qubit for all non-forbidden transitions.}
    \label{tab:yb_db_rel_scattering rate}
    \begin{tabular*}{\linewidth}{@{\extracolsep{\fill}}cccc}
    \hline\hline 
     $F$ & $F'$ & $\displaystyle \Delta _{\text{sp}} /\text{GHz}$ & $P_{\text{sp}}$ \\
    \hline 
    1 & 0 & -2.25 & $2.2\times 10^{-11}$ \\
    1 & 1 & -4.46 & $1.1\times 10^{-11}$ \\
    0 & 1 & 8.18 & $5.0\times 10^{-12}$ \\
    \hline\hline
    \end{tabular*}
\end{table}

\begin{table}[H]
    \centering
    \renewcommand{\arraystretch}{1.3}
    \caption{\qubit scattering rate in processes exclusively present in sympathetic cooling.}
    \label{tab:yb_scat_sum}
    \begin{tabular*}{\linewidth}{@{\extracolsep{\fill}}ccc}
    \hline\hline 
     Qubit State & Process & Scattering Rate \\
    \hline 
    $\ket{F=0,1, m_F=0}$ & 435-nm D1 D2 & $\Omega_{\mathrm{SD}}^2\times1.1\times 10^{-16}$ s \\
    $\ket{F=0, m_F=0}$ & 435-nm SD & $\Omega_{\mathrm{SD}}^2\times1.1\times 10^{-20}$ s \\
    $\ket{F=0, m_F=0}$ & 297-nm Crosstalk & $R_{\mathrm{c}}\times5.0\times 10^{-12}$ \\
    $\ket{F=1, m_F=0}$ & 435-nm SD & $\Omega_{\mathrm{SD}}^2\times5.0\times 10^{-19}$ s \\
    $\ket{F=1, m_F=0}$ & 297-nm Crosstalk & $R_{\mathrm{c}}\times3.3\times 10^{-11}$ \\
    \hline\hline
    \end{tabular*}
\end{table}

\subsection{$\text{Ba}^{+}$}
To assess the applicability of our sympathetic cooling scheme to $\mathrm{Ba}^{+}$, we consider $^{138}\mathrm{Ba}^{+}$ as the coolant ion and evaluate both $^{137}\mathrm{Ba}^{+}$ and $^{133}\mathrm{Ba}^{+}$ as candidate qubit ions. Unlike $\mathrm{Yb}^{+}$, $\mathrm{Ba}^{+}$ has no low-lying $F$ states. We consider cooling schemes using both the $^{2}D_{3/2}$ and $^{2}D_{5/2}$ states. In either case, repumping proceeds through one of the $P$ manifolds.

\begin{figure}[H]
  \centering
	\includegraphics[width=0.5\textwidth]{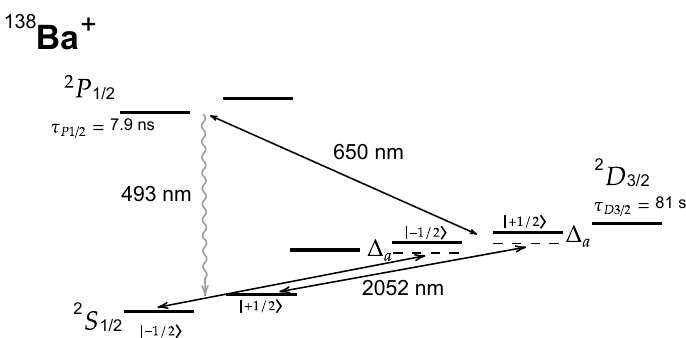}
	\caption{Axial sympathetic cooling scheme through $^{2}D_{3/2}$ on $^{138}\text{Ba}^{+}$ coolants. A dual-tone 2052-nm beam, with each tone detuned from the carrier by $\Delta_a$, drives the $\Delta m_J =0$ quadrupole transition between the $^2S_{1/2}$ and $^2D_{3/2}$ states. The cooling cycle is completed by spontaneous emission of a 493-nm photon, followed by carrier repumping with a 650-nm beam to reinitialize the coolant ion.}
	\label{fig:ba_2052_axial}
\end{figure}

\begin{figure}[H]
  \centering
	\includegraphics[width=0.5\textwidth]{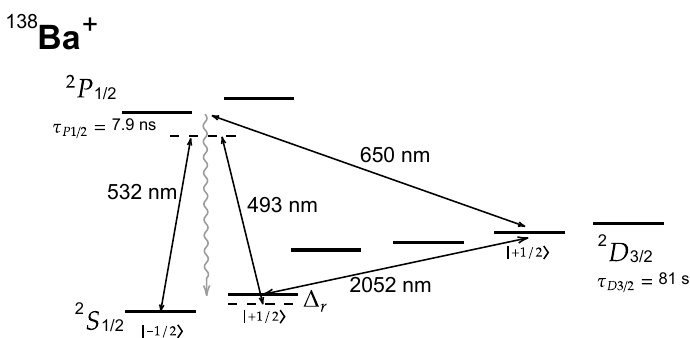}
	\caption{Radial sympathetic cooling scheme through $^{2}D_{3/2}$ on $^{138}\text{Ba}^{+}$ coolants. The coolant ion is initialized in the $^{2}S_{1/2}, m_J=-1/2$ state. A first-order red-sideband transition is then driven with the 532-nm Raman beam. The population is subsequently transferred by a resonant single-tone 2052-nm beam driving the $\Delta m_J=0$ transition between $^{2}S_{1/2}, m_J=+1/2$ and $^{2}D_{3/2}, m_J=+1/2$. The cooling cycle is completed by spontaneous emission of a 493-nm photon, followed by a carrier 650-nm pulse to reinitialize the coolant ion.}
	\label{fig:ba_2052_radial}
\end{figure}

We first consider sympathetic cooling through the $^{2}D_{3/2}$ state. For axial cooling (\Fig{fig:ba_2052_axial}), we propose driving the two $\Delta m_J=0$ transitions with a dual-tone 2052-nm beam, with each tone detuned from the carrier by $\Delta_a$. Repumping is performed on the 650-nm transition through the $^{2}P_{1/2}$ state. For radial cooling (\Fig{fig:ba_2052_radial}), we initialize the coolant ion in the $^{2}S_{1/2},m_J=-1/2$ state and implement pulsed Raman sideband cooling using 532-nm Raman beams. The optical pumping sequence consists of a single-tone carrier pulse on the 2052-nm transition followed by a carrier repump on the 650-nm transition.
For cooling through $^2D_{3/2}$, we use Eq.~\ref{eq:raman_d1_d2_raman_full} to estimate Raman scattering via the D1 and D2 transitions induced by a single addressing tone of the 2052-nm beam. With $\omega_{e^\prime}=658.108~\mathrm{THz}$, $\omega_{e}=607.423~\mathrm{THz}$ \cite{karlsson_revised_1999}, $\omega_{l}=146.114~\mathrm{THz}$ \cite{dijck_determination_2015}, $A_{e^\prime}=1.19\times10^8~\mathrm{s^{-1}}$, and $A_{e}=9.69\times10^7~\mathrm{s^{-1}}$ \cite{arnold_measurements_2019, kelly_6p2p_1978, zhang_branching_2020}, the Raman scattering rate via D1 and D2 transitions is $\Gamma_{\text{Raman}}=\Omega_{\mathrm{SD}}^2\times8.3\times10^{-20}~\mathrm{s}$ as shown in Table.~\ref{tab:ba_133_2052_scat_sum}.

We first consider the $^{133}\mathrm{Ba}^{+}$ qubits. The isotope shifts of the 2052-nm and 493-nm transitions relative to $^{138}\mathrm{Ba}^{+}$ are $2\pi\times157~\mathrm{MHz}$ and $2\pi\times373~\mathrm{MHz}$ respectively \cite{Hucul_spec}. The hyperfine splittings of the $^{2}S_{1/2}$, $^{2}D_{3/2}$, and $^{2}P_{1/2}$ states are $2\pi\times9.925~\mathrm{GHz}$ \cite{knab_precision_1987}, $2\pi\times0.937~\mathrm{GHz}$ \cite{christensen_high-fidelity_2020}, and $2\pi\times1.840~\mathrm{GHz}$ \cite{christensen_high-fidelity_2020}. The resulting detunings for the allowed quadrupole transitions are shown in Table.~\ref{tab:ba_133_2052}. The natural linewidth of $^{2}D_{3/2}$ is $\Gamma_{\mathrm{D}}=2\pi\times0.0020~\mathrm{Hz}$ \cite{yu_radiative_1997}. Using Eq.~\ref{eq:qubit_sd}, the quadrupole-drive-induced scattering rates for a single driving tone applied to both qubit states are respectively $\Gamma_{\mathrm{SD}}(\ket{F=0, m_F=0})=\Omega_{\mathrm{SD}}^2\times1.5\times10^{-23}~\mathrm{s}$, $\Gamma_{\mathrm{SD}}(\ket{F=1, m_F=0})=\Omega_{\mathrm{SD}}^2\times1.5\times10^{-22}~\mathrm{s}$ as shown in Table.~\ref{tab:ba_133_2052_scat_sum}.

\begin{table}[htbp]
    \centering
    \renewcommand{\arraystretch}{1.3}
    \caption{2052-nm transition relative qubit Rabi frequency magnitude $C_{F, m_F}^{F', m_{F'}}(q)$ and detuning $\Delta_{\mathrm{SD}}$ on $^{133}\text{Ba}^{+}$ qubit.}
    \label{tab:ba_133_2052}
    \begin{tabular*}{\linewidth}{@{\extracolsep{\fill}}cccccc}
    \hline\hline 
     $F$ & $q$ & $F^\prime$ & $m_{F^\prime}$ & $ C_{F, m_F}^{F', m_{F'}}(q)$& $\displaystyle \Delta _{\mathrm{SD}} /(2\pi\times\mathrm{GHz})$  \\
    \hline 
    0 & 0 & 2 & 0 & 1 & -7.638\\
    0 & $\pm$1 & 2 & $\pm$1 & 1.126 & -7.638\\
    0 & $\pm$2 & 2 & $\pm$2 & 1.962 & -7.638\\
    1 & 0 & 1 & 0 & 1 & 3.224\\
    1 & $\pm$1 & 1 & $\pm$1 & 0.975 & 3.224\\
    1 & $\pm$1 & 2 & $\pm$1 & 0.563 & 2.287\\
    1 & $\pm$2 & 2 & $\pm$2 & 1.962 & 2.287\\
    \hline\hline
    \end{tabular*}
\end{table}
We evaluate the 493-nm crosstalk scattering using Eq.~\ref{eq:rel_crosstalk_R}. The branching ratio for spontaneous decay from $^2P_{1/2}$ to $^2S_{1/2}$ is 0.7318 \cite{Ba_493_branching_ratio}. The natural linewidth of the $^{2}P_{1/2}$ state is $\Gamma=2\pi\times20.1~\mathrm{MHz}$ \cite{arnesen_Ba_P_linewidth}. We calculate the detunings $\Delta_{\mathrm{sp}}$ and the corresponding qubit scattering probabilities $P_{\mathrm{sp}}$ for all allowed transitions of the $^{133}\mathrm{Ba}^{+}$ qubit, as summarized in Table~\ref{tab:ba_133_493_rel_scattering rate}. The qubit scattering probability for both qubit states are then $P_{\text{sp}}(\ket{F=0, m_F=0})=2.7\times10^{-10}$, and $P_{\text{sp}}(\ket{F=1, m_F=0})=2.1\times10^{-9}$ as shown in Table.~\ref{tab:ba_133_2052_scat_sum}.

\begin{table}[htbp]
    \centering
    \renewcommand{\arraystretch}{1.3}
    \caption{493-nm photon scattering probability $P_{\mathrm{sp}}$ on $^{133}\text{Ba}^{+}$ qubit for all non-forbidden transitions.}
    \label{tab:ba_133_493_rel_scattering rate}
    \begin{tabular*}{\linewidth}{@{\extracolsep{\fill}}cccc}
    \hline\hline 
     $F$ & $F'$ & $\displaystyle \Delta _{\text{sp}} /\text{GHz}$ & $P_{\mathrm{sp}}$ \\
    \hline 
    1 & 1 & 2.394 & $1.8\times10^{-9}$ \\
    1 & 0 & 4.234 & $2.8\times10^{-10}$ \\
    0 & 1 & -7.531 & $2.7\times10^{-10}$ \\
    \hline\hline
    \end{tabular*}
\end{table}

\begin{table}[htbp]
    \centering
    \renewcommand{\arraystretch}{1.3}
    \caption{$^{133}\text{Ba}^{+}$ scattering rate in processes exclusively present in sympathetic cooling through $^2D_{3/2}$.}
    \label{tab:ba_133_2052_scat_sum}
    \begin{tabular*}{\linewidth}{@{\extracolsep{\fill}}ccc}
    \hline\hline 
     Qubit State & Process & Scattering Rate \\
    \hline 
    $\ket{F=0,1, m_F=0}$ & 2052-nm D1 D2 & $\Omega_{\mathrm{SD}}^2\times8.3\times 10^{-20}$ s \\
    $\ket{F=0, m_F=0}$ & 2052-nm SD & $\Omega_{\mathrm{SD}}^2\times1.5\times 10^{-23}$ s \\
    $\ket{F=0, m_F=0}$ & 493-nm Crosstalk & $R_{\mathrm{c}}\times2.7\times 10^{-10}$ \\
    $\ket{F=1, m_F=0}$ & 2052-nm SD & $\Omega_{\mathrm{SD}}^2\times1.5\times 10^{-22}$ s \\
    $\ket{F=1, m_F=0}$ & 493-nm Crosstalk & $R_{\mathrm{c}}\times2.1\times 10^{-9}$ \\
    \hline\hline
    \end{tabular*}
\end{table}

For $^{137}\mathrm{Ba}^{+}$ qubits, the isotope shifts of the 2052-nm and 493-nm transitions relative to $^{138}\mathrm{Ba}^{+}$ are $2\pi\times284~\mathrm{MHz}$ \cite{villemoes_isotope_1993} and $2\pi\times271~\mathrm{MHz}$ \cite{Hucul_spec}. The hyperfine splittings of the $^{2}S_{1/2}$ and $^{2}P_{1/2}$ states are $2\pi\times8.038~\mathrm{GHz}$ \cite{Blatt_ba137} and $2\pi\times1.49~\mathrm{GHz}$ \cite{low_practical}. We calculate the hyperfine structure of the $^{2}D_{3/2}$ state using the Casimir formula together with the hyperfine constants $A=189.73~\mathrm{MHz}$ and $B=44.54~\mathrm{MHz}$ from Refs.~\cite{lewty_spectroscopy_2012,lewty_spectroscopy_2013}. The resulting detunings for all allowed quadrupole transitions are shown in Table.~\ref{tab:ba_137_2052}. Using Eq.~\ref{eq:qubit_sd}, the quadrupole-drive-induced scattering rates for a single driving tone applied to both qubit states are respectively $\Gamma_{\mathrm{SD}}(\ket{F=1, m_F=0})=\Omega_{\mathrm{SD}}^2\times3.0\times10^{-23}~\mathrm{s}$, $\Gamma_{\mathrm{SD}}(\ket{F=2, m_F=0})=\Omega_{\mathrm{SD}}^2\times1.3\times10^{-22}~\mathrm{s}$ as shown in Table.~\ref{tab:ba_137_2052_scat_sum}.

\begin{table}[htbp]
    \centering
    \renewcommand{\arraystretch}{1.3}
    \caption{2052-nm transition relative qubit Rabi frequency magnitude $C_{F, m_F}^{F', m_{F'}}(q)$ and detuning $\Delta_{\mathrm{SD}}$ on $^{137}\text{Ba}^{+}$ qubit.}
    \label{tab:ba_137_2052}
    \begin{tabular*}{\linewidth}{@{\extracolsep{\fill}}cccccc}
    \hline\hline 
     $F$ & $q$ & $F^\prime$ & $m_{F^\prime}$ & $ C_{F, m_F}^{F', m_{F'}}(q)$& $\displaystyle \Delta _{\mathrm{SD}} /(2\pi\times\mathrm{GHz})$  \\
     \hline
    1 & 0 & 1 & 0 & 0.316 & 4.797\\
    1 & 0 & 3 & 0 & 0.949 & 5.746\\
    1 & $\pm$1 & 1 & $\pm$1 & 0.308 & 4.797\\
    1 & $\pm$1 & 2 & $\pm$1 & 0.398 & 5.132\\
    1 & $\pm$1 & 3 & $\pm$1 & 1.007 & 5.746\\
    1 & $\pm$2 & 2 & $\pm$2 & 1.387 & 5.132\\
    1 & $\pm$2 & 3 & $\pm$2 & 1.387 & 5.746\\
    2 & 0 & 0 & 0 & 0.707 & -3.386\\
    2 & 0 & 2 & 0 & 0.707 & -2.906\\
    2 & $\pm$1 & 1 & $\pm$1 & 0.925 & -3.241\\
    2 & $\pm$1 & 2 & $\pm$1 & 0.398 & -2.906\\
    2 & $\pm$1 & 3 & $\pm$1 & 0.504 & -2.292\\
    2 & $\pm$2 & 2 & $\pm$2 & 1.387 & -2.906\\
    2 & $\pm$2 & 3 & $\pm$2 & 1.387 & -2.292\\
    \hline\hline
    \end{tabular*}
\end{table}

We calculate $\Delta_{\mathrm{sp}}$ and qubit scattering probability $P_{\mathrm{sp}}$ for all non-forbidden 493-nm transitions on $^{137}\text{Ba}^{+}$ qubit as shown in Table~\ref{tab:ba_137_493_rel_scattering rate}. The qubit scattering probability for both qubit states are then $P_{\text{sp}}(\ket{F=2, m_F=0})=2.2\times10^{-9}$, and $P_{\text{sp}}(\ket{F=1, m_F=0})=3.9\times10^{-10}$ as shown in Table.~\ref{tab:ba_137_2052_scat_sum}.

\begin{table}[htbp]
    \centering
    \renewcommand{\arraystretch}{1.3}
    \caption{493-nm photon scattering probability $P_{\mathrm{sp}}$ on $^{137}\text{Ba}^{+}$ qubit for all non-forbidden transitions.}
    \label{tab:ba_137_493_rel_scattering rate}
    \begin{tabular*}{\linewidth}{@{\extracolsep{\fill}}cccc}
    \hline \hline 
     $ F$ & $ F'$ & $\displaystyle \Delta _{\text{sp}} /\text{GHz}$ & $P_{\mathrm{sp}}$ \\
     \hline 
    2 & 1 & -3.673& $5.7\times10^{-10}$\\
    2 & 2 & -2.183& $1.6\times10^{-9}$\\
    1 & 1& 4.365& $1.4\times10^{-11}$\\
    1 & 2 & 5.855& $3.8\times10^{-10}$\\
    \hline\hline
    \end{tabular*}
\end{table}

\begin{table}[H]
    \centering
    \renewcommand{\arraystretch}{1.3}
    \caption{$^{137}\text{Ba}^{+}$ scattering rate in processes exclusively present in sympathetic cooling through $^2D_{3/2}$.}
    \label{tab:ba_137_2052_scat_sum}
    \begin{tabular*}{\linewidth}{@{\extracolsep{\fill}}ccc}
    \hline\hline 
     Qubit State & Process & Scattering Rate \\
    \hline 
    $\ket{F=1,2, m_F=0}$ & 2052-nm D1 D2 & $\Omega_{\mathrm{SD}}^2\times8.3\times 10^{-20}$ s \\
    $\ket{F=1, m_F=0}$ & 2052-nm SD & $\Omega_{\mathrm{SD}}^2\times3.0\times 10^{-23}$ s\\
    $\ket{F=1, m_F=0}$ & 493-nm Crosstalk & $R_{\mathrm{c}}\times3.9\times 10^{-10}$ \\
    $\ket{F=2, m_F=0}$ & 2052-nm SD & $\Omega_{\mathrm{SD}}^2\times1.3\times 10^{-22}$ s\\
    $\ket{F=2, m_F=0}$ & 493-nm Crosstalk & $R_{\mathrm{c}}\times2.2\times 10^{-9}$ \\
    \hline\hline
    \end{tabular*}
\end{table}

\begin{figure}[H]
  \centering
	\includegraphics[width=0.5\textwidth]{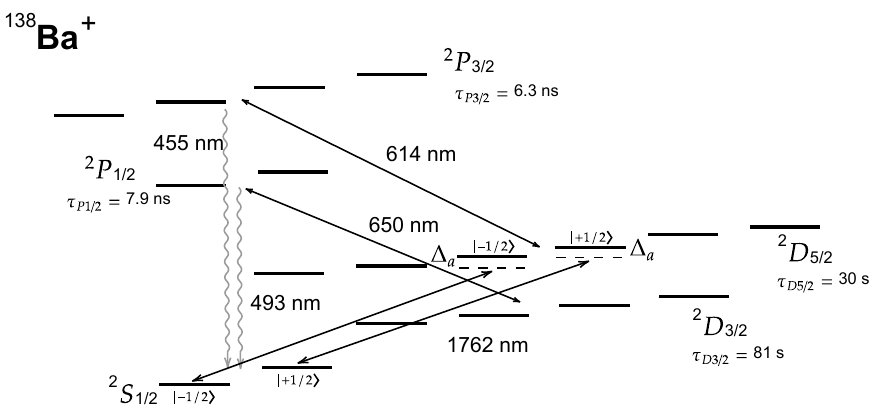}
	\caption{Axial sympathetic cooling scheme through $^{2}D_{5/2}$ on $^{138}\text{Ba}^{+}$ coolants. A dual-tone 1762-nm beam, with each tone detuned from the carrier by $\Delta_a$, drives the $\Delta m_J=0$ quadrupole transition between the $^2S_{1/2}$ and $^2D_{5/2}$ states. The cooling cycle is completed by spontaneous emission of a 455-nm photon, followed by carrier repumping with a 614-nm beam to reinitialize the coolant ion. An additional 650-nm repumping pulse is required to depopulate the $^2D_{3/2}$ state, transferring population back to the $^2S_{1/2}$ ground state via spontaneous emission of 493-nm photons.}
	\label{fig:ba_1762_axial}
\end{figure}

\begin{figure}[H]
  \centering
	\includegraphics[width=0.5\textwidth]{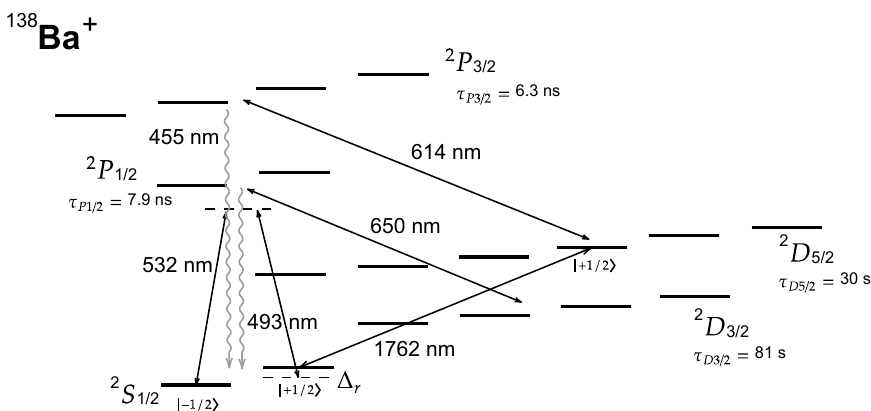}
	\caption{Radial sympathetic cooling scheme through $^{2}D_{5/2}$ on $^{138}\text{Ba}^{+}$ coolants. The coolant ion is initialized in the $^{2}S_{1/2}, m_J=-1/2$ state. A first-order red-sideband transition is then driven with the 532-nm Raman beam. The population is subsequently transferred by a resonant single-tone 2052-nm beam driving the $\Delta m_J=0$ transition between $^{2}S_{1/2}, m_J=+1/2$ and $^{2}D_{5/2}, m_J=+1/2$. The cooling cycle is completed by spontaneous emission of a 455-nm photon, followed by carrier repumping with a 614-nm pulse to reinitialize the coolant ion. An additional 650-nm repumping pulse is required to depopulate the $^2D_{3/2}$ state, transferring population back to the $^2S_{1/2}$ ground state via spontaneous emission of 493-nm photons.
    }
	\label{fig:ba_1762_radial}
\end{figure}

For sympathetic cooling through the $^{2}D_{5/2}$ state, we propose the scheme shown in \Fig{fig:ba_1762_axial}. During axial cooling, a dual-tone 1762-nm beam drives the two $\pi$ transitions, with each tone detuned from the carrier by $\Delta_a$. The primary repump is provided by the 614-nm transition to the $^{2}P_{3/2}$ state. Because $2.8\%$ of the population decays from $^{2}P_{3/2}$ to $^{2}D_{3/2}$ \cite{zhang_branching_2020}, an additional 650-nm repump is required to depopulate the $^{2}D_{3/2}$ manifold. The dominant spontaneous-emission channel is the 455-nm transition from $^{2}P_{3/2}$ to $^{2}S_{1/2}$, with only a minor contribution from 493-nm photons emitted via the $^{2}P_{1/2}$ state. The radial cooling scheme (Fig.~\ref{fig:ba_1762_radial}) differs from the 2052-nm implementation only in the optical pumping sequence, which uses the 1762-nm quadrupole transition together with the additional repump laser.


We use Eq.~\ref{eq:raman_d1_d2_raman_full} to estimate Raman scattering via the D1 and D2 transitions induced by a single addressing tone of the 1762-nm beam. With $\omega_{g, g^\prime}=146.114~\mathrm{THz}$ \cite{dijck_determination_2015}, $A_{e^\prime, g^\prime}=4.51\times10^6~\mathrm{s^{-1}}$, $A_{e, g^\prime}=3.55\times10^7~\mathrm{s^{-1}}$ \cite{arnold_measurements_2019, kelly_6p2p_1978, zhang_branching_2020}, and other quantities being the same as aformentioned values, the Raman scattering from single 1762-nm beam is $\Gamma_{\text{Raman}}=\Omega_{\mathrm{SD}}^2\times1.1\times10^{-19}~\mathrm{s}$ as shown in Table.~\ref{tab:ba_133_1762_scat_sum}.

For $^{133}\mathrm{Ba}^{+}$ qubits, the isotope shifts of the 1762-nm and 455-nm transitions relative to $^{138}\mathrm{Ba}^{+}$ are $2\pi\times142~\mathrm{MHz}$ (calculated from the measured isotope shifts of the 615-nm and 455-nm transitions \cite{christensen_high-fidelity_2020}) and $2\pi\times358~\mathrm{MHz}$ \cite{christensen_high-fidelity_2020}. The hyperfine splittings of the $^{2}P_{3/2}$ and $^{2}D_{5/2}$ states are $2\pi\times0.623~\mathrm{GHz}$ and $2\pi\times83~\mathrm{MHz}$ \cite{christensen_high-fidelity_2020}. The resulting detunings for the allowed quadrupole transitions are shown in Table.~\ref{tab:ba_133_2052}. The natural linewidth of the $^{2}D_{5/2}$ state is $\Gamma_{D}=2\pi\times0.0053~\mathrm{Hz}$ \cite{zhang_branching_2020}. Using Eq.~\ref{eq} with $J'=5/2$, the quadrupole-drive-induced scattering rates for a single driving tone applied to both qubit states are respectively $\Gamma_{\mathrm{SD}}(\ket{F=0, m_F=0})=\Omega_{\mathrm{SD}}^2\times4.4\times10^{-23}~\mathrm{s}$, $\Gamma_{\mathrm{SD}}(\ket{F=1, m_F=0})=\Omega_{\mathrm{SD}}^2\times3.4\times10^{-22}~\mathrm{s}$ as shown in Table.~\ref{tab:ba_133_1762_scat_sum}.

\begin{table}[htbp]
    \centering
    \renewcommand{\arraystretch}{1.3}
    \caption{1762-nm transition relative qubit Rabi frequency magnitude $C_{F, m_F}^{F', m_{F'}}(q)$ and detuning $\Delta_{\mathrm{SD}}$ on $^{133}\text{Ba}^{+}$ qubit.}
    \label{tab:ba_133_1762}
    \begin{tabular*}{\linewidth}{@{\extracolsep{\fill}}cccccc}
    \hline\hline 
     $F$ & $q$ & $F^\prime$ & $m_{F^\prime}$ & $ C_{F, m_F}^{F', m_{F'}}(q)$& $\displaystyle \Delta _{\mathrm{SD}} /(2\pi\times\mathrm{GHz})$  \\
    \hline 
    0 & 0 & 2 & 0 & 1 & -7.350\\
    0 & $\pm$1 & 2 & $\pm$1 & 1.126 & -7.350\\
    0 & $\pm$2 & 2 & $\pm$2 & 1.962 & -7.350\\
    1 & 0 & 3 & 0 & 1 &  2.658\\
    1 & $\pm$1 & 2 & $\pm$1 & 0.375 & 2.575\\
    1 & $\pm$1 & 3 & $\pm$1 & 1.062 &  2.658\\
    1 & $\pm$2 & 2 & $\pm$2 & 1.308 & 2.575\\
    1 & $\pm$2 & 3 & $\pm$2 & 1.462 &  2.658\\
    \hline\hline
    \end{tabular*}
\end{table}

For 455-nm crosstalk scattering, the polarization ratios of spontaneously emitted photons from the coolant ion are equal. The branching ratio for spontaneous decay from $^2P_{3/2}$ to $^2S_{1/2}$ is 0.7417 \cite{zhang_branching_2020}. The natural linewidth of the $^{2}P_{3/2}$ state is $\Gamma=2\pi\times25.4~\mathrm{MHz}$ \cite{arnesen_Ba_P_linewidth}. We calculate the detunings $\Delta_{\mathrm{sp}}$ and qubit scattering probability $P_{\mathrm{sp}}$ for all non-forbidden transitions of the $^{133}\mathrm{Ba}^{+}$ qubit, as listed in Table~\ref{tab:ba_133_455_rel_scattering rate}. The qubit scattering probability for both qubit states are respectively $P_{\text{sp}}(\ket{F=0, m_F=0})=9.5\times10^{-10}$, and $P_{\text{sp}}(\ket{F=1, m_F=0})=5.9\times10^{-9}$ as shown in Table.~\ref{tab:ba_133_1762_scat_sum}.

\begin{table}[htbp]
    \centering
    \renewcommand{\arraystretch}{1.3}
    \caption{455-nm photon scattering probability $P_{\mathrm{sp}}$ on $^{133}\text{Ba}^{+}$ qubit for all non-forbidden transitions.}
    \label{tab:ba_133_455_rel_scattering rate}
    \begin{tabular*}{\linewidth}{@{\extracolsep{\fill}}cccc}
    \hline \hline 
     $ F$ & $ F'$ & $\displaystyle \Delta _{\text{sp}} /\text{GHz}$ & $P_{\text{sp}}$ \\
    \hline
    1 & 2 & 2.606 & $5.2\times10^{-9}$ \\
    1 & 1 & 3.229 & $6.9\times10^{-10}$ \\
    0 & 1 & -6.697 & $9.5\times10^{-10}$ \\
     \hline\hline
    \end{tabular*}
\end{table}

\begin{table}[H]
    \centering
    \renewcommand{\arraystretch}{1.3}
    \caption{$^{133}\text{Ba}^{+}$ scattering rate in processes exclusively present in sympathetic cooling through $^2D_{5/2}$.}
    \label{tab:ba_133_1762_scat_sum}
    \begin{tabular*}{\linewidth}{@{\extracolsep{\fill}}ccc}
    \hline\hline 
     Qubit State & Process & Scattering Rate \\
    \hline 
    $\ket{F=0,1, m_F=0}$ & 1762-nm D1 D2 & $\Omega_{\mathrm{SD}}^2\times9.0\times 10^{-20}$ s\\
    $\ket{F=0, m_F=0}$ & 1762-nm SD & $\Omega_{\mathrm{SD}}^2\times4.4\times 10^{-23}$ s\\
    $\ket{F=0, m_F=0}$ & 455-nm Crosstalk & $R_{\mathrm{c}}\times9.5\times 10^{-10}$ \\
    $\ket{F=1, m_F=0}$ & 1762-nm SD & $\Omega_{\mathrm{SD}}^2\times3.4\times 10^{-22}$ s\\
    $\ket{F=1, m_F=0}$ & 455-nm Crosstalk & $R_{\mathrm{c}}\times5.9\times 10^{-9}$ \\
    \hline\hline
    \end{tabular*}
\end{table}

For $^{137}\mathrm{Ba}^{+}$ qubits, the isotope shifts of the 1762-nm and 455-nm transitions relative to $^{138}\mathrm{Ba}^{+}$ are $2\pi\times277~\mathrm{MHz}$ (calculated from the measured isotope shifts of the 615-nm and 455-nm transitions \cite{villemoes_isotope_1993,wendt_relativistic_1984}) and $2\pi\times279~\mathrm{MHz}$ \cite{wendt_relativistic_1984}. We calculate the hyperfine structure of the $^{2}D_{5/2}$ state using the Casimir formula with hyperfine constants $A=-12.03~\mathrm{MHz}$ and $B=59.53~\mathrm{MHz}$, and that of the $^{2}P_{3/2}$ state with $A=127.2~\mathrm{MHz}$ and $B=44.5~\mathrm{MHz}$ \cite{wendt_relativistic_1984}. The resulting detunings for the allowed quadrupole transitions are as shown in Table.~\ref{tab:ba_137_1762}. Using Eq.~\ref{eq} with $J'=5/2$, the quadrupole-drive-induced scattering rates for a single driving tone applied to both qubit states are respectively $\Gamma_{\mathrm{SD}}(\ket{F=1, m_F=0})=\Omega_{\mathrm{SD}}^2\times8.4\times10^{-23}~\mathrm{s}$, $\Gamma_{\mathrm{SD}}(\ket{F=2, m_F=0})=\Omega_{\mathrm{SD}}^2\times3.1\times10^{-22}~\mathrm{s}$ as shown in Table.~\ref{tab:ba_137_1762_scat_sum}.

\begin{table}[htbp]
    \centering
    \renewcommand{\arraystretch}{1.3}
    \caption{1762-nm transition relative qubit Rabi frequency magnitude $C_{F, m_F}^{F', m_{F'}}(q)$ and detuning $\Delta_{\mathrm{SD}}$ on $^{137}\text{Ba}^{+}$ qubit.}
    \label{tab:ba_137_1762}
    \begin{tabular*}{\linewidth}{@{\extracolsep{\fill}}cccccc}
    \hline\hline 
     $F$ & $q$ & $F^\prime$ & $m_{F^\prime}$ & $ C_{F, m_F}^{F', m_{F'}}(q)$& $\displaystyle \Delta _{\mathrm{SD}} /(2\pi\times\mathrm{GHz})$  \\
     \hline
    1 & 0 & 1 & 0 & 0.775 & 5.405\\
    1 & 0 & 3 & 0 & 0.632 & 5.271\\
    1 & $\pm$1 & 1 & $\pm$1 & 0.755 & 5.405\\
    1 & $\pm$1 & 2 & $\pm$1 & 0.497 & 5.333\\
    1 & $\pm$1 & 3 & $\pm$1 & 0.672 & 5.271\\
    1 & $\pm$2 & 2 & $\pm$2 & 1.730 & 5.333\\
    1 & $\pm$2 & 3 & $\pm$2 & 0.925 & 5.271\\
    2 & 0 & 2 & 0 & 0.378 & -2.704\\
    2 & 0 & 4 & 0 & 0.926 & -2.768\\
    2 & $\pm$1 & 1 & $\pm$1 & 0.252 & -2.633\\
    2 & $\pm$1 & 2 & $\pm$1 & 0.213 & -2.704\\
    2 & $\pm$1 & 3 & $\pm$1 & 0.504 & -2.767\\
    2 & $\pm$1 & 4 & $\pm$1 & 0.952 & -2.768\\
    2 & $\pm$2 & 2 & $\pm$2 & 0.741 & -2.704\\
    2 & $\pm$2 & 3 & $\pm$2 & 1.387 & -2.767\\
    2 & $\pm$2 & 4 & $\pm$2 & 1.172 & -2.768\\
    \hline\hline
    \end{tabular*}
\end{table}

We calculate $\Delta_{\mathrm{sp}}$ and qubit scattering probability $P_{\mathrm{sp}}$ for all non-forbidden 455-nm transitions on $^{137}\text{Ba}^{+}$ qubit as listed in Table~\ref{tab:ba_137_455_rel_scattering rate}. The qubit scattering probability for both qubit states are then $P_{\text{sp}}(\ket{F=2, m_F=0})=7.1\times10^{-9}$, and $P_{\text{sp}}(\ket{F=1, m_F=0})=1.7\times10^{-9}$ as shown in Table.~\ref{tab:ba_137_1762_scat_sum}.

\begin{table}[htbp]
    \centering
    \renewcommand{\arraystretch}{1.3}
    \caption{455-nm photon scattering probability $P_{\mathrm{sp}}$ on $^{137}\text{Ba}^{+}$ qubit for all non-forbidden transitions.}
    \label{tab:ba_137_455_rel_scattering rate}
    \begin{tabular*}{\linewidth}{@{\extracolsep{\fill}}cccc}
    \hline \hline 
     $ F$ & $ F'$ & $\displaystyle \Delta _{\text{sp}} /\text{GHz}$ & $P_{\mathrm{sp}}$ \\
    \hline
    2 & 1 & -3.062 & $2.3\times10^{-10}$ \\
    2 & 2 & -2.900 & $1.8\times10^{-9}$ \\
    2 & 3 & -2.426 & $5.1\times10^{-9}$ \\
    1 & 2 & 5.138 & $6.7\times10^{-10}$ \\
    1 & 1 & 4.976 & $7.3\times10^{-10}$ \\
    1 & 0 & 4.942 & $2.9\times10^{-10}$ \\
     \hline\hline
    \end{tabular*}
\end{table}

\begin{table}[H]
    \centering
    \renewcommand{\arraystretch}{1.3}
    \caption{$^{137}\text{Ba}^{+}$ scattering rate in processes exclusively present in sympathetic cooling through $^2D_{5/2}$.}
    \label{tab:ba_137_1762_scat_sum}
    \begin{tabular*}{\linewidth}{@{\extracolsep{\fill}}ccc}
    \hline\hline 
     Qubit State & Process & Scattering Rate \\
    \hline 
    $\ket{F=1, 2, m_F=0}$ & 1762-nm D1 D2 & $\Omega_{\mathrm{SD}}^2\times9.0\times 10^{-20}$ s\\
    $\ket{F=1, m_F=0}$ & 1762-nm SD & $\Omega_{\mathrm{SD}}^2\times8.4\times10^{-23}$ s\\
    $\ket{F=1, m_F=0}$ & 455-nm Crosstalk & $R_{\mathrm{c}}\times1.7\times 10^{-9}$ \\
    $\ket{F=2, m_F=0}$ & 1762-nm SD & $\Omega_{\mathrm{SD}}^2\times3.1\times 10^{-22}$ s\\
    $\ket{F=2, m_F=0}$ & 455-nm Crosstalk & $R_{\mathrm{c}}\times7.1\times 10^{-9}$ \\
    \hline\hline
    \end{tabular*}
\end{table}

\subsection{$\text{Ca}^{+}$}
Current $\mathrm{Ca}^{+}$ quantum-computing architectures include optical qubits encoded in the $^{2}S_{1/2}$ and $^{2}D_{5/2}$ states of $^{40}\mathrm{Ca}^{+}$ \cite{haffner_ca_rev,schmidt-kaler_realization_2003}, as well as hyperfine qubits encoded in the clock states of the $^{2}S_{1/2}$ manifold of $^{43}\mathrm{Ca}^{+}$ \cite{Myerson_ca43,ballance_ca_43,harty_ca_43,smith_microwave-driven_2026}. Here, we consider the implementation of our dual-isotope sympathetic-cooling scheme for both architectures.

\subsubsection{Optical qubits}

For optical qubits, we consider $^{48}\mathrm{Ca}^{+}$ as the coolant species and $^{40}\mathrm{Ca}^{+}$ as the qubit species. The optical qubits are encoded in $^2S_{1/2}, m_{J, Q}=-1/2$ and $^2D_{5/2}, m_{J^\prime, Q}=-1/2$ states. 

\begin{figure}[H]
  \centering
	\includegraphics[width=0.5\textwidth]{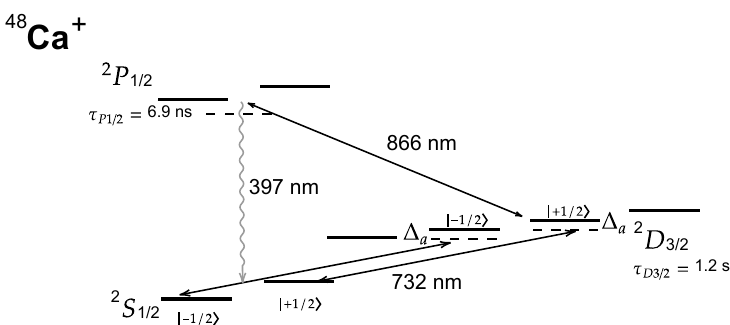}
	\caption{Axial sympathetic cooling scheme for $^{40}\mathrm{Ca}^{+}$ optical qubits with $^{48}\mathrm{Ca}^{+}$ coolant ions. A dual-tone 732-nm beam, with each tone detuned from the carrier by $\Delta_a$, drives the $\Delta m_J=0$ quadrupole transition between the $^2S_{1/2}$ and $^2D_{3/2}$ states. The cooling cycle is completed by spontaneous emission of a 397-nm photon, followed by carrier repumping with a 866-nm beam to reinitialize the coolant ion.}
	\label{fig:ca_optical_axial}
\end{figure}
\begin{figure}[H]
  \centering
	\includegraphics[width=0.5\textwidth]{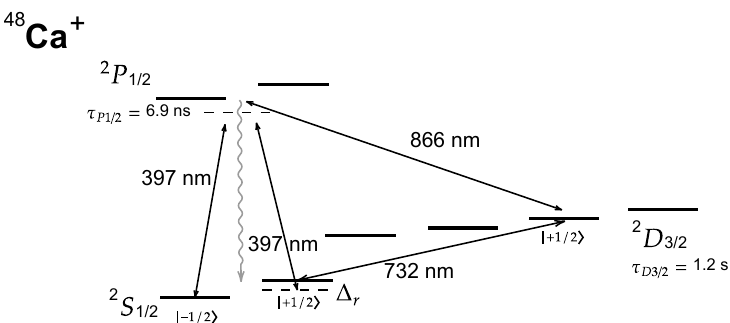}
	\caption{Radial sympathetic cooling scheme for $^{40}\mathrm{Ca}^{+}$ optical qubits with $^{48}\mathrm{Ca}^{+}$ coolant ions. The coolant ion is initialized in the $^{2}S_{1/2}, m_J=-1/2$ state. A first-order red-sideband transition is then driven with the 397-nm Raman beam. The population is subsequently transferred by a resonant single-tone 732-nm beam driving the $\Delta m_J=0$ transition between $^{2}S_{1/2}, m_J=+1/2$ and $^{2}D_{3/2},m_J=+1/2$. The cooling cycle is completed by spontaneous emission of a 397-nm photon, followed by a carrier 866-nm pulse to reinitialize the coolant ion.
    }
	\label{fig:ca_optical_radial}
\end{figure}

We propose implementing sympathetic cooling through the $^{2}D_{3/2}$ state, following a scheme analogous to the 2052-nm $\mathrm{Ba}^{+}$ implementation. As shown in \Fig{fig:ca_optical_axial}, axial cooling is achieved by driving the two $\Delta m_J=0$ transitions with a dual-tone 732-nm beam, with each tone detuned from the carrier by $\Delta_a$. The quadrupole beam is applied simultaneously with the 866-nm repump beam. During the cooling process, spontaneous decay from the $^2P_{1/2}$ state generates 397-nm photons. For radial cooling (\Fig{fig:ca_optical_radial}), the coolant ion is initialized in the $^{2}S_{1/2}, m_J=-1/2$ state. Pulsed Raman sideband cooling is then implemented using 397-nm Raman beams \cite{ballance_hybrid_2015} together with a 732-nm carrier pump and an 866-nm carrier repump.

We use Eq.~\ref{eq:raman_d1_d2_raman_full} to estimate Raman scattering via the D1 and D2 transitions induced by a single addressing tone of the 732-nm beam. With $\omega_{e^\prime}=761.827~\mathrm{THz}$, $\omega_{e}=755.299~\mathrm{THz}$ \cite{morton_atomic_2000}, $\omega_{l}=409.223~\mathrm{THz}$ \cite{chang_systematic-free_2024}, $A_{e^\prime}=1.35\times10^8~\mathrm{s^{-1}}$, and $A_{e}=1.32\times10^8~\mathrm{s^{-1}}$ \cite{jin_precision_1993, ramm_precision_2013, gerritsma_precision_2008}, the scattering rate is $\Gamma_{\text{Raman}}=\Omega_{\mathrm{SD}}^2\times3.4\times10^{-20}~\mathrm{s}$ as shown in Table.~\ref{tab:ca_40_scat_sum}.

\begin{table}[htbp]
    \centering
    \renewcommand{\arraystretch}{1.3}
    \caption{732-nm transition relative qubit Rabi frequency magnitude $C_{J, m_{J, Q}}^{J^\prime, m_{J^\prime, Q}}(q)$ and detuning $\Delta_{\mathrm{SD}}$ on $^{40}\text{Ca}^{+}$ optical qubit.}
    \label{tab:ca_40_732}
    \begin{tabular*}{\linewidth}{@{\extracolsep{\fill}}cccccc}
    \hline\hline 
     $m_{J, Q}$ & $q$ & $m_{J^\prime, Q}$ & $ C_{J, m_{J, Q}}^{J^\prime, m_{J^\prime, Q}}(q)$& $\displaystyle \Delta _{\mathrm{SD}} /(2\pi\times\mathrm{GHz})$  \\
    \hline
    -1/2 & 0 & -1/2 & 1 & -10.003\\
    -1/2 & 1 & 1/2 & 1.379 & -10.003\\
    -1/2 & -1 & -3/2 & 0.796 & -10.003\\
    -1/2 & 2 & 3/2 & 2.774 & -10.003\\
    \hline\hline
    \end{tabular*}
\end{table}

Both the $^{48}\mathrm{Ca}^{+}$ coolant ions and the $^{40}\mathrm{Ca}^{+}$ qubit ions have nuclear spin $I=0$; consequently, there is no hyperfine structure to consider. Relative to $^{48}\mathrm{Ca}^{+}$, the isotope shifts of the 732-nm and 397-nm transitions in $^{40}\mathrm{Ca}^{+}$ are $-2\pi\times10.003~\mathrm{GHz}$ \cite{chang_systematic-free_2024} and $-2\pi\times1.705~\mathrm{GHz}$ \cite{Gebert_Precision_2015}. The detuning of the quadrupole transition is therefore determined entirely by the isotope shift. The natural linewidth of the $^{2}D_{3/2}$ state is $\Gamma_{D}=2\pi\times0.135~\mathrm{Hz}$ \cite{Kreuter_lifetime_2005}. Given that there is no hyperfine structure present, we use Eq.~\ref{eq:quad_coupling_ratio} with $I=0$ to calculate $C_{J, m_{J, Q}}^{J^\prime, m_{J^\prime, Q}}(q)$. Using Eq.~\ref{eq:qubit_sd}, the quadrupole-drive-induced scattering rate is dominantly from the qubit state in the ground manifold. The scattering rate for a single driving tone is $\Gamma_{\mathrm{SD}}(\ket{^2S_{1/2}, m_{J, Q}=-1/2})=\Omega_{\mathrm{SD}}^2\times6.0\times10^{-22}~\mathrm{s}$ as shown in Table.~\ref{tab:ca_40_scat_sum}.

\begin{table}[H]
    \centering
    \renewcommand{\arraystretch}{1.3}
    \caption{$^{40}\text{Ca}^{+}$ scattering rate in processes exclusively present in sympathetic cooling.}
    \label{tab:ca_40_scat_sum}
    \begin{tabular*}{\linewidth}{@{\extracolsep{\fill}}ccc}
    \hline\hline 
     Qubit State & Process & Scattering Rate \\
    \hline 
    $\ket{^2S_{1/2}, m_J=-1/2}$ & 732-nm D1 D2 & $\Omega_{\mathrm{SD}}^2\times3.4\times 10^{-20}$ s\\
    $\ket{^2S_{1/2}, m_J=-1/2}$ & 732-nm SD & $\Omega_{\mathrm{SD}}^2\times6.0\times10^{-22}$ s\\
    $\ket{^2S_{1/2}, m_J=-1/2}$ & 397-nm Crosstalk & $R_{\mathrm{c}}\times1.8\times 10^{-8}$ \\
    \hline\hline
    \end{tabular*}
\end{table}

The detuning for the 397-nm crosstalk also originates purely from the isotope shift, yielding $\Delta_{\text{sp}}=-2\pi\times1.705~\mathrm{GHz}$. The natural linewidth of $^{2}P_{1/2}$ is $\Gamma=2\pi\times23.05~\mathrm{MHz}$ \cite{hettrich_measurement_2015}. The branching ratio for spontaneous decay from $^2P_{1/2}$ to $^2S_{1/2}$ is 0.9357 \cite{ramm_precision_2013}. The total 397-nm photon scattering probability $P_{\mathrm{sp}}$ is $1.8\times10^{-8}$ as shown in Table.~\ref{tab:ca_40_scat_sum}.

\subsubsection{Hyperfine Qubits}

For hyperfine qubits, we consider $^{40}\mathrm{Ca}^{+}$ as the coolant species and $^{43}\mathrm{Ca}^{+}$ as the qubit species. We propose implementing sympathetic cooling through the $^{2}D_{5/2}$ state, following a scheme analogous to the 1762-nm $\mathrm{Ba}^{+}$ implementation. 
\begin{figure}[H]
  \centering
	\includegraphics[width=0.5\textwidth]{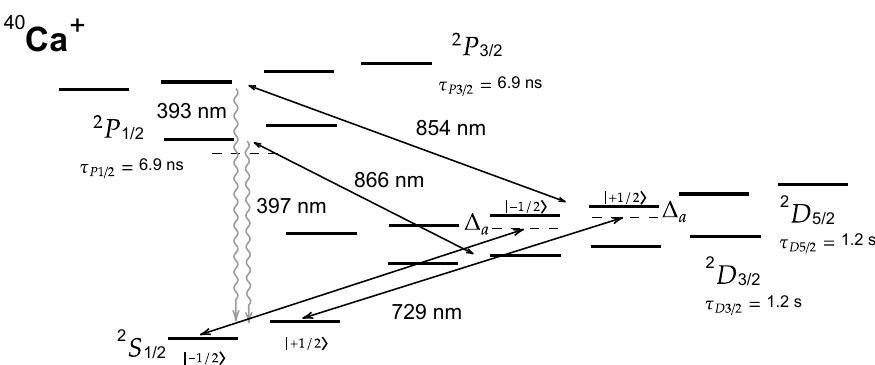}
	\caption{Axial sympathetic cooling scheme for $^{43}\mathrm{Ca}^{+}$ hyperfine qubits with $^{40}\mathrm{Ca}^{+}$ coolant ions. A dual-tone 729-nm beam, with each tone detuned from the carrier by $\Delta_a$, drives the $\Delta m_J=0$ quadrupole transition between the $^2S_{1/2}$ and $^2D_{5/2}$ states. The cooling cycle is completed by spontaneous emission of a 393-nm photon, followed by a carrier 854-nm pulse to reinitialize the coolant ion. An additional carrier 866-nm repumping pulse is required to depopulate the $^2D_{3/2}$ state, transferring population back to the $^2S_{1/2}$ ground state via spontaneous emission of 397-nm photons.
    }
	\label{fig:ca_ground_axial}
\end{figure}

\begin{figure}[H]
  \centering
	\includegraphics[width=0.5\textwidth]{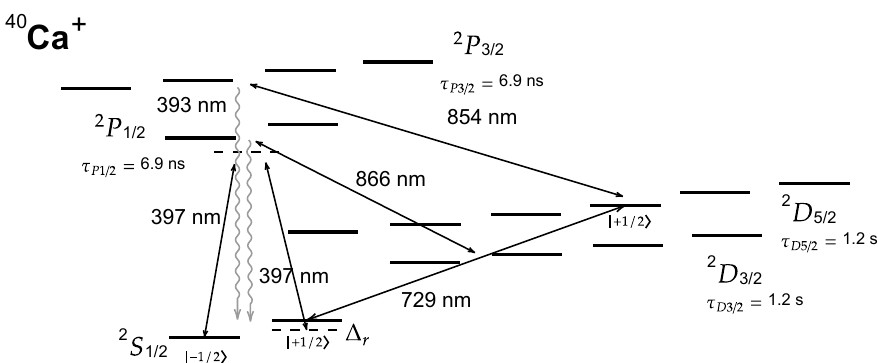}
	\caption{Radial sympathetic cooling scheme for $^{43}\mathrm{Ca}^{+}$ hyperfine qubits with $^{40}\mathrm{Ca}^{+}$ coolant ions. The coolant ion is initialized in the $^{2}S_{1/2}, m_J=-1/2$ state. A first-order red-sideband transition is then driven with the 397-nm Raman beam. The population is subsequently transferred by a resonant single-tone 729-nm beam driving the $\Delta m_J=0$ transition between $^{2}S_{1/2},m_J=+1/2$ and $^{2}D_{5/2},m_J=+1/2$. The cooling cycle is completed by spontaneous emission of a 393-nm photon, followed by a carrier 854-nm pulse to reinitialize the coolant ion. An additional carrier 866-nm repumping pulse is required to depopulate the $^2D_{3/2}$ state, transferring population back to the $^2S_{1/2}$ ground state via spontaneous emission of 397-nm photons.}
	\label{fig:ca_ground_radial}
\end{figure}

As shown in \Fig{fig:ca_ground_axial}, the axial cooling scheme differs from the optical-qubit implementation primarily in that the quadrupole transition is shifted from 732~nm to 729~nm. The primary repump is provided by the 854-nm transition to the $^{2}P_{3/2}$ state. Because $0.66\%$ of the population decays from $^{2}P_{3/2}$ to $^{2}D_{3/2}$ \cite{gerritsma_precision_2008}, an additional 866-nm repump is required to depopulate the $^{2}D_{3/2}$ manifold. The radial cooling scheme (\Fig{fig:ca_ground_radial}) employs the same Raman sideband cooling protocol as the $^{48}\mathrm{Ca}^{+}$ implementation. The carrier pump is replaced by the 729-nm quadrupole transition, and the corresponding repump is changed to 854-nm. The additional 866-nm repump is likewise required to empty the $^{2}D_{3/2}$ state.

We use Eq.~\ref{eq:raman_d1_d2_raman_full} to estimate Raman scattering via the D1 and D2 transitions induced by a single addressing tone of the 729-nm beam. With $\omega_{g, g^\prime}=409.223~\mathrm{THz}$ \cite{chang_systematic-free_2024}, $A_{e^\prime, g^\prime}=9.55\times10^6~\mathrm{s^{-1}}$, $A_{e, g^\prime}=9.07\times10^7~\mathrm{s^{-1}}$ \cite{jin_precision_1993, ramm_precision_2013, gerritsma_precision_2008}, and other quantities being the same as aformentioned values, the scattering rate is $\Gamma_{\text{Raman}}=\Omega_{\mathrm{SD}}^2\times2.4\times10^{-20}~\mathrm{s}$ as shown in Table.~\ref{tab:ca_43_scat_sum}.

\begin{table}[H]
    \centering
    \renewcommand{\arraystretch}{1.3}
    \caption{729-nm transition relative qubit Rabi frequency magnitude $C_{F, m_F}^{F', m_{F'}}(q)$ and detuning $\Delta_{\mathrm{SD}}$ on $^{43}\text{Ca}^{+}$ qubit.}
    \label{tab:ca_43_729}
    \begin{tabular*}{\linewidth}{@{\extracolsep{\fill}}cccccc}
    \hline\hline 
     $F$ & $q$ & $F^\prime$ & $m_{F^\prime}$ & $ C_{F, m_F}^{F', m_{F'}}(q)$& $\displaystyle \Delta _{\mathrm{SD}} /(2\pi\times\mathrm{GHz})$  \\
     \hline
    3 & 0 & 1 & 0 & 0.655 & 2.362\\
    3 & 0 & 3 & 0 & 0.577 & 2.345\\
    3 & 0 & 5 & 0 & 0.488 & 2.311\\
    3 & $\pm$1 & 1 & $\pm$1 & 0.426 & 2.362\\
    3 & $\pm$1 & 2 & $\pm$1 & 0.737 & 2.355\\
    3 & $\pm$1 & 3 & $\pm$1 & 0.230 & 2.345\\
    3 & $\pm$1 & 4 & $\pm$1 & 0.499 & 2.330\\
    3 & $\pm$1 & 5 & $\pm$1 & 0.492 & 2.311\\
    3 & $\pm$2 & 2 & $\pm$2 & 0.642 & 2.355\\
    3 & $\pm$2 & 3 & $\pm$2 & 1.266 & 2.345\\
    3 & $\pm$2 & 4 & $\pm$2 & 1.230 & 2.330\\
    3 & $\pm$2 & 5 & $\pm$2 & 0.566 & 2.311\\
    4 & 0 & 2 & 0 & 0.218 & 5.581\\
    4 & 0 & 4 & 0 & 0.441 & 5.556\\
    4 & 0 & 6 & 0 & 0.870 & 5.511\\
    4 & $\pm$1 & 2 & $\pm$1 & 0.164 & 5.581\\
    4 & $\pm$1 & 3 & $\pm$1 & 0.383 & 5.571\\
    4 & $\pm$1 & 4 & $\pm$1 & 0.136 & 5.556\\
    4 & $\pm$1 & 5 & $\pm$1 & 0.573 & 5.536\\
    4 & $\pm$1 & 6 & $\pm$1 & 0.864 & 5.511\\
    4 & $\pm$2 & 2 & $\pm$2 & 0.071 & 5.581\\
    4 & $\pm$2 & 3 & $\pm$2 & 0.422 & 5.571\\
    4 & $\pm$2 & 4 & $\pm$2 & 1.006 & 5.556\\
    4 & $\pm$2 & 5 & $\pm$2 & 1.321 & 5.536\\
    4 & $\pm$2 & 6 & $\pm$2 & 0.952 & 5.511\\
    \hline\hline
    \end{tabular*}
\end{table}

Relative to the $^{40}\mathrm{Ca}^{+}$ coolant ions, the isotope shifts of the 729-nm, 393-nm, and 397-nm transitions in $^{43}\mathrm{Ca}^{+}$ are $2\pi\times4.135~\mathrm{GHz}$ \cite{benhelm_measurement_2007}, $2\pi\times0.688~\mathrm{GHz}$ \cite{lucas_isotope-selective_2004}, and $2\pi\times0.688~\mathrm{GHz}$ \cite{lucas_isotope-selective_2004}. The hyperfine splittings of the $^{2}S_{1/2}$ and $^{2}P_{1/2}$ states are $2\pi\times3.2256~\mathrm{GHz}$ \cite{andl_isotope_1982} and $2\pi\times0.5816~\mathrm{GHz}$ \cite{andl_isotope_1982} respectively. We calculate the hyperfine structure of the $^{2}D_{5/2}$ state using the hyperfine constants $A=-3.89~\mathrm{MHz}$ and $B=-4.24~\mathrm{MHz}$ from Ref.~\cite{benhelm_measurement_2007}, yielding a hyperfine splitting of $2\pi\times77~\mathrm{MHz}$. Similarly, the calculated hyperfine splittings of the $^{2}D_{3/2}$ and $^{2}P_{3/2}$ states are $2\pi\times567~\mathrm{MHz}$ and $2\pi\times375~\mathrm{MHz}$, obtained using the hyperfine constants $A=-47.3~\mathrm{MHz}$, $B=-3.7~\mathrm{MHz}$ \cite{benhelm_measurement_2007} and $A=-31.4~\mathrm{MHz}$, $B=-5.8~\mathrm{MHz}$ \cite{andl_isotope_1982} respectively. The resulting detunings for the allowed 729-nm quadrupole transitions are shown in Table.~\ref{tab:ca_43_729}. The natural linewidth of the $^{2}D_{5/2}$ state is $\Gamma_{D}=2\pi\times0.136~\mathrm{Hz}$ \cite{Kreuter_lifetime_2005}. Using Eq.~\ref{eq:qubit_sd}, the quadrupole-drive-induced scattering rates for a single driving tone applied to both qubit states are respectively $\Gamma_{\mathrm{SD}}(\ket{F=3, m_F=0})=\Omega_{\mathrm{SD}}^2\times1.1\times10^{-20}~\mathrm{s}$, $\Gamma_{\mathrm{SD}}(\ket{F=4, m_F=0})=\Omega_{\mathrm{SD}}^2\times2.0\times10^{-21}~\mathrm{s}$ as shown in Table.~\ref{tab:ca_43_scat_sum}.

For 393-nm crosstalk scattering, the polarization ratios of spontaneously emitted photons from the coolant ion are equal. The branching ratio for spontaneous decay from $^2P_{3/2}$ to $^2S_{1/2}$ is 0.9347 \cite{gerritsma_precision_2008}, and the natural linewidth of the $^{2}P_{3/2}$ state is $\Gamma=2\pi\times23.97~\mathrm{MHz}$ \cite{meir_combining_2020}. Using Eq.~\ref{eq:rel_crosstalk_R}, we calculate the detunings $\Delta_{\mathrm{sp}}$ and qubit scattering probability $P_{\mathrm{sp}}$ for all allowed transitions of the $^{43}\mathrm{Ca}^{+}$ qubit, as listed in Table~\ref{tab:ca_43_393_rel_scattering_rate}. The qubit scattering probability for both qubit states are then $P_{\text{sp}}(\ket{F=3, m_F=0})=3.6\times10^{-8}$, and $P_{\text{sp}}(\ket{F=4, m_F=0})=8.9\times10^{-9}$ as shown in Table.~\ref{tab:ca_43_scat_sum}.

\begin{table}[H]
    \centering
    \renewcommand{\arraystretch}{1.3}
    \caption{393-nm photon scattering probability $P_{\mathrm{sp}}$ on $^{43}\text{Ca}^{+}$ qubit for all non-forbidden transitions.}
    \label{tab:ca_43_393_rel_scattering_rate}
    \begin{tabular*}{\linewidth}{@{\extracolsep{\fill}}cccc}
    \hline \hline 
     $ F$ & $ F'$ & $\displaystyle \Delta _{\text{sp}} /\text{GHz}$ & $\Delta_{\mathrm{sp}}$ \\
    \hline
    3 & 2 & -0.918 & $1.51\times10^{-8}$ \\
    3 & 3 & -1.008 & $1.33\times10^{-8}$ \\
    3 & 4 & -1.132 & $7.46\times10^{-9}$ \\
    4 & 3 & 2.218 & $7.03\times10^{-10}$ \\
    4 & 4 & 2.094 & $2.40\times10^{-9}$ \\
    4 & 5 & 1.933 & $5.84\times10^{-9}$ \\
    \hline\hline
    \end{tabular*}
\end{table}

\begin{table}[H]
    \centering
    \renewcommand{\arraystretch}{1.3}
    \caption{$^{43}\text{Ca}^{+}$ scattering rate in processes exclusively present in sympathetic cooling.}
    \label{tab:ca_43_scat_sum}
    \begin{tabular*}{\linewidth}{@{\extracolsep{\fill}}ccc}
    \hline\hline 
     Qubit State & Process & Scattering Rate\\
    \hline 
    $\ket{F=3, 4, m_F=0}$ & 729-nm D1 D2 & $\Omega_{\mathrm{SD}}^2\times2.4\times 10^{-20}$ s\\
    $\ket{F=3, m_F=0}$ & 729-nm SD & $\Omega_{\mathrm{SD}}^2\times1.1\times10^{-20}$ s\\
    $\ket{F=3, m_F=0}$ & 393-nm Crosstalk & $R_{\mathrm{c}}\times3.6\times 10^{-8}$ \\
    $\ket{F=4, m_F=0}$ & 729-nm SD & $\Omega_{\mathrm{SD}}^2\times2.0\times 10^{-21}$ s\\
    $\ket{F=4, m_F=0}$ & 393-nm Crosstalk & $R_{\mathrm{c}}\times8.9\times 10^{-9}$ \\
    \hline\hline
    \end{tabular*}
\end{table}

\subsection{Primary Scattering during Sympathetic Cooling}

Here, we discuss the primary sources of scattering-induced qubit errors during sympathetic cooling. Table~\ref{tab:dom_scat_summ} shows the photon scattering rates associated with these sources for all the considered qubits with $\Omega_{\mathrm{SD}}$ and $R_{\mathrm{sp, c}}$ both equal to $2\pi\times1.0~\mathrm{MHz}$. In \qubit the Raman scattering induced by the quadrupole beam via the D1 and D2 transitions (D1+D2 Raman scattering) dominates. In Ca$^{+}$ and Ba$^{+}$ qubit ions, the errors arise primarily from absorption of the photons that are spontaneously emitted from the neighboring coolant ions (SE crosstalk). The rate of off-resonant quadrupole scattering in Ba$^{+}$ is approximately three orders of magnitude lower than that in Yb$^{+}$, primarily owing to the long lifetime of the metastable $D$ state.
The relatively high crosstalk-induced error rates in Ba$^{+}$ and Ca$^{+}$ ions arise from the short lifetimes of the $P$ states that are used as the excited states for cooling. This crosstalk is considerably weaker in \qubit owing to the availability of the narrower-linewidth $^3[3/2]_{1/2}$ state, which is electric-dipole coupled to the ground-state qubit manifold. 
Conversely, the shorter-wavelength quadrupole drive used in \qubit produces stronger Raman scattering. \qubit has the lowest scattering-induced error rate among all considered species, making it well-suited for high-dissipation implementation of our cooling scheme.

\begin{table}[htbp]
    \centering
    \renewcommand{\arraystretch}{1.3}
    \caption{Primary sources of scattering-induced qubit errors for all considered qubit species. The associated scattering rates are evaluated for $\Omega_{\mathrm{SD}} = R_{\mathrm{sp, c}} = $ $2\pi\times1~\mathrm{MHz}$.} 
    \label{tab:dom_scat_summ}
    \begin{tabular*}{\linewidth}{@{\extracolsep{\fill}}ccc}
    \hline\hline 
     Qubit & Primary Source & Scattering Rate/$\mathrm{s}^{-1}$ \\
    \hline 
    $^{171}\text{Yb}^{+}$ & D1+D2 Raman & $4.3\times 10^{-3}$ \\
    $^{133}\text{Ba}^{+}$ & spont. emission (SE) & $1.3\times 10^{-2}$(493 nm) \\
        & crosstalk & $3.7\times 10^{-2}$(455 nm) \\
    $^{137}\text{Ba}^{+}$ & SE crosstalk & $1.4\times 10^{-2}$(493 nm) \\
        & & $4.5\times 10^{-2}$(455 nm) \\
    $^{40}\text{Ca}^{+}$ & SE crosstalk  & $1.1\times 10^{-1}$\\
    $^{43}\text{Ca}^{+}$ & SE crosstalk  & $2.3\times 10^{-1}$\\
    \hline\hline
    \end{tabular*}
\end{table}

\subsection{Comparison with Raman Scattering during Gates}

The Raman beams can induce photon scattering in the qubit ions through the D1 and D2 transitions. During sympathetic cooling, this error can be eliminated using fully individual Raman addressing. During qubit-gate operations, however, photon-scattering errors remain. Following Refs.~\cite{moore_photon_2023, le_kien_dynamical_2013} and assuming equal peak electric-field amplitudes for the two Raman beams, the magnitude of the carrier Raman Rabi frequency is given by 
\begin{align}
   \Omega_{\mathrm{Raman}}=\frac{\pi c^2 I_l}{\hbar}\left(\frac{2\omega_{l}}{\omega_{e^\prime}^2-\omega_{l}^2}\frac{A_{e^\prime}}{\omega_{e^\prime}^3}-\frac{2\omega_{l}}{\omega_{e}^2-\omega_{l}^2}\frac{A_{e}}{\omega_{e}^3}\right)\cos\theta_R\label{eq:raman_rabi_freq}.
\end{align}
Here, $I_l$ is the intensity of each Raman beam at the ion position, and $\theta_R$ is the angle between the quantization magnetic field $\mathbf{B}_0$ and the effective magnetic field $\mathbf{B}_{\mathrm{eff}}$ generated by the Raman coupling. The associated Raman-scattering rate due to each Raman beam is given by Eq.~\ref{eq:raman_d1_d2_raman_full}. We evaluate this rate for hyperfine ground-state qubits in all the considered qubit species. The total Raman scattering rates induced by both Raman beams for the two qubit states are listed in Table~\ref{tab:raman_d1_d2}. For implementations with $\Omega_{\mathrm{R}}\approx 2\pi\times1~\mathrm{MHz}$, the total Raman-scattering error rate from both beams is typically on the order of $1~\mathrm{s^{-1}}$ or higher.

\begin{table}[htbp]
    \centering
    \renewcommand{\arraystretch}{1.3}
    \caption{Rates  $\Gamma_R$ of qubit errors induced by Raman scattering from both Raman beams of wavelength $\lambda_{\mathrm{R}}$ implemented on evaluated qubit species.}
    \label{tab:raman_d1_d2}
    \begin{tabular*}{\linewidth}{@{\extracolsep{\fill}}ccc}
    \hline\hline 
     Qubit & $\lambda_{\mathrm{R}}/\mathrm{nm}$ & $\Gamma_{\mathrm{Raman}}/\Omega_{\mathrm{R}}^2$ \\
    \hline 
    $^{171}\text{Yb}^{+}$ & 355 & $2\times 10^{-13}$ \\
    $^{133, 137}\text{Ba}^{+}$ & 532 \cite{boguslawski_raman_2023} & $1\times 10^{-13}$ \\
    $^{43}\text{Ca}^{+}$ & 397 \cite{ballance_hybrid_2015} & $4\times 10^{-12}$\\
    \hline\hline
    \end{tabular*}
\end{table}

We compare photon scattering from the primary source during sympathetic cooling with that occurring in a typical qubit circuit consisting of digitally amplitude-modulated (AM) MS gates. For the 23-ion mixed Yb$^+$ chain used in our demonstration, the average carrier Raman Rabi frequency for each ion in the gate is approximately $2\pi\times100~\mathrm{kHz}$. We extrapolate from this for the gates implemented on other ion species. To achieve the same gate speed across different ion species, assuming same mode participation and gate detuning, the carrier Rabi frequency should scale as $\lambda_{\mathrm{R}}^2 M^{-1/2}$ when constrained by the trap breakdown voltage and as $\lambda_{\mathrm{R}}^2 M^{-2/3}$ when constrained by Ohmic heating. Accordingly, the estimated carrier Raman Rabi frequencies for Ba$^+$ and Ca$^+$ are between $2\pi\times250~\mathrm{kHz}$ and $2\pi\times310~\mathrm{kHz}$. 

Assuming the same duty cycle as in our demonstration, 
we estimate that each qubit participates in MS gates for 4\% of the circuit duration, with sympathetic cooling occupying 75\%. Based on these duty cycles, the ratios of the primary scattering induced by sympathetic cooling to the Raman scattering during the gates are approximately $1$, $100$–$150$, and $20$–$30$ for Yb$^+$, Ba$^+$, and Ca$^+$, respectively. Here, for the spontaneous-emission-crosstalk estimate, we conservatively considered the qubit state with the higher scattering rate.

%%%%%%%%%%%%%%%%%%%%%%%%%%%%%%%%%%%%%%%%%%%%%%%%%%%%%%%%%%%%%%%%%%%%%%%%%%%%%%%%%%%%%%%%%%%%%%%%%%%%%%%%%%%%%%%%%%%%%%%%%%%%
\pagebreak
\section{Radial Recoil Heating Estimation
\label{app:recoil_cal}}
\noindent
\subsection{Participation Factor in Mixed-Species Chains}
Near the equilibrium configuration, we denote the displacement of ion $i$ from its equilibrium position by $\mathbf{u}_i$. The linearized equations of motion can then be written in matrix form as
\begin{align}
   \mathsf{M} \mathbf{\ddot{u}} = -\mathsf{K}\mathbf{u}. \label{eq:harm_eqn}
\end{align}
Here, $\mathsf{M}$ is the diagonal mass matrix. $\mathsf{K}$ is the Hessian matrix of the total potential energy evaluated at the equilibrium configuration, including contributions from both the external trapping potential and the Coulomb interactions among the ions. $\mathbf{u}$ is the $(3N\times1)$ column vector containing the displacements of all ions. The overdots denote derivatives with respect to time.

A normal-mode solution may be written as
\begin{align}
\mathbf{u}(t)=\mathbf{b}_k e^{-i\omega_k t}.\label{eq:bik_ansatz}
\end{align}
$\mathbf{b}_k$ is the ($3N\times1$) participation-vector associated with mode $k$ normalized as $\mathbf{b}_k^T\mathbf{b}_k=1$. $\omega_k$ is the corresponding mode frequency. Substituting this ansatz into Eq.~\ref{eq:harm_eqn} yields the generalized eigenvalue equation
\begin{align}
   \mathsf{M} \omega_k^2 \mathbf{b}_k = \mathsf{K}\mathbf{b}_k. \label{eq:bik_eqn}
\end{align}
By aligning the coordinate axis with the principal axis of the mode and discarding the vanishing components, we reduce $\mathbf{b}_k$ to an $N\times1$ column vector. We note that, in a mixed-species chain, the participation-factor vectors $\mathbf{b}_k$ are generally not orthogonal for non-degenerate modes. This can be shown using the generalized eigenvalue equation together with the symmetry of the Hessian matrix, $\mathsf{K}^{T}=\mathsf{K}$:
\begin{align}
\omega_k^2 \mathbf{b}_l^T \mathsf{M}\mathbf{b}_k
= \mathbf{b}_l^T \mathsf{K}\mathbf{b}_k
= \mathbf{b}_k^T \mathsf{K}\mathbf{b}_l
= \omega_l^2 \mathbf{b}_k^T \mathsf{M}\mathbf{b}_l.
\label{eq}
\end{align}
Subtracting the rightmost expression from the leftmost one therefore yields
\begin{align}
\left(\omega_k^2-\omega_l^2\right)
\sum_i m_i b_{ik}b_{il}=0.
\label{eq}
\end{align}
Thus, for nondegenerate modes, for which $\omega_k\neq\omega_l$,
\begin{align}
\sum_i m_i b_{ik}b_{il}=0.
\label{eq}
\end{align}

\subsection{Resolved-Sideband Model of Axial Cooling}
Here, we characterize the photon scattering rate and cooling rate during axial sympathetic cooling using a resolved sideband model with no free parameters. The experimental setting is as described in Sec.~\ref{sec:axial}. Because the 935-nm quenching laser rapidly couples the metastable $^2D_{3/2}$ states to the short-lived $^3[3/2]_{1/2}$ manifold, which subsequently decays to the ground state, the three-level system can be approximated as an effective two-level system. We therefore take the measured repumping rates to define the effective excited-state linewidths, $\Gamma_{m_J}$. The two cooling pathways are respectively $^2S_{1/2}, m_J=-1/2$ -- $^2D_{3/2}, m_J=-1/2$ -- $^3[3/2]_{1/2}, m_J=-1/2$, and $^2S_{1/2}, m_J=+1/2$ -- $^2D_{3/2}, m_J=+1/2$ -- $^3[3/2]_{1/2}, m_J=+1/2$. We analyze these two pathways independently and sum their contributions to obtain the total photon scattering rate, assuming equal branching ratios for spontaneous decay from the $^3[3/2]_{1/2}$ manifold.

We perform axial sympathetic cooling in the regime where the effective excited-state linewidths are comparable to the axial mode frequency. In this regime, the internal-state dynamics are not sufficiently fast relative to the ion's secular motion for the adiabatic approximation to hold well. As a result, the internal dynamics remain coherent over multiple oscillation periods. Although numerical simulations can accurately reproduce the experimental observations in this intermediate regime \cite{kulosa_systematic_2023}, our objective here is to develop an analytical model that captures the essential physics while retaining reasonable simplicity. To this end, we employ a resolved-sideband model to estimate the photon scattering and cooling rates measured in the experiment. Within this model, we neglect the overlap between sidebands of different orders. We further assume that the coolant ion is initially prepared in its internal ground state and in a thermal motional state with a mean occupation number $\bar{n}$ of the LA mode. The corresponding phonon-number distribution is given by $P(n,\bar{n})=\bar{n}^n/(\bar{n}+1)^{n+1}$.

Following Ref.~\cite{wineland_experimental_1998}, the sideband Rabi frequency for coolant ion $i$ is given by
\begin{equation}
\begin{aligned}
\Omega_{m_J,i}(n,s)
&= \Omega_{\mathrm{SD},m_J=-1/2}
e^{-\eta_{a,i}^2/2}\eta_{a,i}^s \\
&\qquad
\sqrt{\frac{n_<!}{n_>!}}
L_{n_<}^{|s|}(\eta_{a, i}^2)\label{eq:sd_res_omega}.
\end{aligned}
\end{equation}
Here, $n$ is the phonon number of the LA mode, and $s$ is the sideband order of the transition. Negative and positive values of $s$ correspond to red- and blue-sideband transitions, while $s=0$ corresponds to the carrier transition. $\eta_{a,i}$ is the Lamb--Dicke parameter of coolant ion $i$ for the LA mode. The quantities $n_{<}$ and $n_{>}$ denote the smaller and larger of $n$ and $n+s$ respectively, and $L_{n_{<}}^{|s|}$ is the generalized Laguerre polynomial.

The photon scattering rate for coolant $i$ through $m_J$-path is then  
\begin{equation}
\begin{aligned}
   R_{m_J, i}&=\sum_n P(n, \bar{n})\sum_s R_{i, m_J, n, s}\\
   &=\sum_n P(n, \bar{n}) \frac{\sum_{s}a_{i, m_J, s, n}}{1+\sum_{s\prime}a_{i, m_J, s\prime, n}}\label{eq:axial_total_scat}.
\end{aligned}
\end{equation}
The saturation parameter 
\begin{align}
a_{i, m_J, s, n}=\frac{2\Omega_{m_J,i}^2(n, s)/\Gamma_{m_J}^2}{1+\left(\frac{\Delta_{\mathrm{SD}, m_J}-s\omega_0}{\Gamma_{m_J}/2}\right)^2}.\label{eq:sat_para}
\end{align}
The cooling rate is then 
\begin{equation}
\begin{aligned}
   R_{\mathrm{c}}(\bar{n})&=\sum_i\sum_{m_J}\frac{1}{2}\sum_n P(n, \bar{n}) \sum_s sR_{i, m_J, s, n}\label{eq:axial_total_cooling_rate},
\end{aligned}
\end{equation}
The axial recoil heating rate is 
\begin{equation}
\begin{aligned}
   R_{\mathrm{h, rec}}(\bar{n})&=\sum_i\sum_{m_J}\frac{b_{i}^2}{2\omega_0}R_{m_J, i}\\&(\omega_{\mathrm{rec, 435, axial}}+\omega_{\mathrm{rec, 297, axial}}+\omega_{\mathrm{rec, 935, axial}})\label{eq:axial_recoil_heating_rate}.
\end{aligned}
\end{equation}
In Eq.~\ref{eq:axial_recoil_heating_rate}, $\omega_{\mathrm{rec, 435, axial}}=2\pi\times3.083~\mathrm{kHz}$, $\omega_{\mathrm{rec, 935, axial}}=2\pi\times 1.326~\mathrm{kHz}$, and $\omega_{\mathrm{rec, 297, axial}}=2\pi\times 4.380~\mathrm{kHz}$ are the recoil frequencies associated with the absorption of the 435-nm and 935-nm photons respectively, together with the spontaneous emission of the 297-nm photon along the axial direction. With the idle heating rate $R_{\mathrm{h, idle}}=268(8)~\mathrm{ms^{-1}}$ from \Fig{fig:axial_heating}, we get the net cooling rate as 
\begin{equation}
\begin{aligned}
   R_{\mathrm{c, net}}(\bar{n})=R_{\mathrm{c}}(\bar{n})+R_{\mathrm{h, rec}}(\bar{n})+R_{\mathrm{h, idle}}\label{eq:axial_net_cooling_rate}.
\end{aligned}
\end{equation}

In the experiment shown in \Fig{fig:axial_cooling}, the hot longest-wavelength LR mode introduces an offset in the measured equilibrium occupation of the LA mode. Because radial sympathetic cooling is not applied in this experiment, the longest-wavelength LR mode remains hot after the initial 20-ms delay and contribute to the decay of the Raman carrier Rabi oscillations used for axial thermometry. Consequently, the extracted value of $\bar{n}_0$ contains an offset arising from this coupling to hot LR mode. This effect also explains why the fitted equilibrium phonon number, $n_{\mathrm{eq}}$, differs from the expected value of $\Gamma_{\mathrm{h}}/\Gamma_{\mathrm{c}}$ in Sec~\ref{sec:axial}.

To estimate this contribution, we measure a heating rate of $0.267~\mathrm{ms}^{-1}$ for the LR mode of a single tightly confined ion with a mode frequency close to that of the longest-wavelength LR mode in the chain. We then approximate the heating rate of the corresponding chain mode by scaling with the number of ions, yielding $6.141~\mathrm{ms}^{-1}$. The corresponding contribution to the carrier-decay parameter during the Raman thermometry sequence is 
\begin{align}
\theta_{i, r}\approx b_{i,r}^2\frac{\hbar}{2M_i\omega_{0, r}} (\Delta \mathbf{k})^2 \bar{n}_r. 
\label{eq: radial_theta}
\end{align}
Here, $b_{i,r}$ is the participation factor of qubit ion $i$ in the longest-wavelength LR mode with frequency $\omega_{0,r}$. $\bar{n}_r$ is the mean phonon number of the longest-wavelength LR mode. $\Delta \mathbf{k}$ is the Raman momentum kick as defined in \Fig{fig:exp_setup}. $M_i=171~\mathrm{a.m.u.}$ corresponding to the \qubit qubit ion mass. Following the initial 20-ms delay, the mean occupation of this radial mode during the first 5-ms axial sympathetic cooling is approximately $\bar{n}_r\approx138$. Using Eq.~\ref{eq: radial_theta}, we calculate the corresponding radial contribution to the carrier-decay parameter for each qubit. Averaging over all qubits and converting to the mean phonon number, it yields an effective axial offset of $\bar{n}_0=1180$.

\begin{figure}[H]
  \centering
	\includegraphics[width=0.48\textwidth]{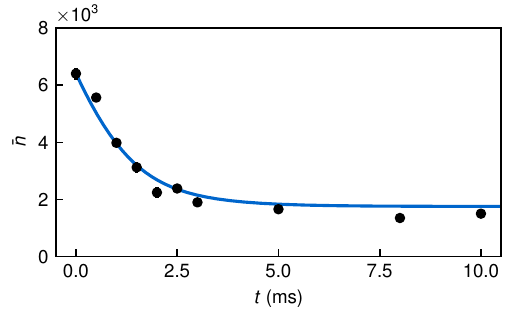}
	\caption{Resolved-sideband model for the mean occupation of the LA mode during axial sympathetic cooling. Error bars denote the standard error of the ion mean with weights inversely proportional to the individual binomial variances. Carrier Raman Rabi oscillations are measured on all qubits following a 20-ms idle delay and subsequent axial sympathetic cooling of duration $t$. The solid curve is calculated using the resolved-sideband model with the net cooling rate $R_{\mathrm{c,net}}$ given by Eq.~\ref{eq:axial_net_cooling_rate}, an initial mean occupation of $\bar{n}_0=1180$, and an idle heating rate of $R_{\mathrm{h,idle}}=268(8)~\mathrm{ms}^{-1}$ measured in \Fig{fig:axial_heating}.}
	\label{fig:axial_cooling_model}
\end{figure}

We characterize the axial cooling shown in \Fig{fig:axial_cooling} using Eq.~\ref{eq:axial_net_cooling_rate} together with the effective axial offset of $\bar{n}_0$ arising from the hot radial mode. The resulting theoretical prediction, shown in \Fig{fig:axial_cooling_model}, is in good agreement with the experimental data.

\subsection{Estimate of Radial Recoil Heating}
We estimate the average recoil heating rate during the 3-ms axial sympathetic cooling interval applied immediately after state preparation. For each coolant ion $i$, the total photon scattering rate during axial cooling is calculated using Eq.~\ref{eq:axial_total_scat}. We evaluate the scattering rate at $\bar{n}=2000$ with zero offset. The recoil frequencies associated with the absorption of 435-nm photons and the spontaneous emission of 297-nm photons projected onto the lower-frequency radial axis are $\omega_{\mathrm{rec,435,lr}}=2\pi\times1.531~\mathrm{kHz}$ and $\omega_{\mathrm{rec,297,lr}}=2\pi\times4.380~\mathrm{kHz}$ respectively. The resulting recoil heating rate of LR mode $k$ is then given by
\begin{align}
   \dot{\bar{n}}_{\mathrm{k, rec}} \approx \frac{\omega_{\mathrm{rec, 435, lr}}+\omega_{\mathrm{rec, 297, lr}}}{\bar{\omega}_{\mathrm{lr}}}\sum_i\sum_{m_J}\frac{1}{2} R_{m_J, i} b_{ik}^2\label{eq:total scattering_rate}.
\end{align}
Here, we approximate the LR-mode frequencies using their average value, $\bar{\omega}_{\mathrm{lr}}=2\pi\times2.900~\mathrm{MHz}$. The predicted recoil heating rates are shown in \Fig{fig:radial_heating}.

%%%%%%%%%%%%%%%%%%%%%%%%%%%%%%%%%%%%%%%%%%%%%%%%%%%%%%%%%%%%%%%%%%%%%%%%%%%%%%%%%%%%%%%%%%%%%%%%%%%%%%%%%%%%%%%%%%%%%%%%%%%%
\section{Cryogenic Implementation Outlook
\label{app:cryo_estimate}}
\noindent
We estimate the performance of our sympathetic cooling scheme in a cryogenic system using the heating-rate measurements reported in Ref.~\cite{brandl_cryogenic_2016}. In that work, the heating rate of a single idle $^{40}$Ca$^{+}$ ion with a mode frequency of $\omega \approx 2\pi\times1.1~\mathrm{MHz}$ was measured to be approximately $2.14~\mathrm{s}^{-1}$ at a distance $d=115~\mathrm{\mu m}$. Following Ref.~\cite{brownnutt_ion-trap_2015, sedlacek_distance_2018}, we assume that the electric-field-noise spectral density scales as $1/(d^4 \omega)$, leading the single-ion heating rate to scale as $(M\omega^2d^4)^{-1}$, where $M$ is the ion mass.

We consider a 60-ion mixed Yb$^{+}$ chain composed of \qubit qubit ions and \coolant coolant ions in a similar cryogenic system where the ions are trapped at a distance of $d=68~\mathrm{\mu m}$ above the trap surface, consistent with our current setup. In equidistant chains, the LA mode scales with the ion number $N$ as $N^{-0.856}$ \cite{cetina_control_2022}. Assuming the same ion spacing of $3.8~\mathrm{\mu m}$, the LA mode of the 60-ion chain is estimated to be $\omega_a \approx 2\pi\times75~\mathrm{kHz}$. The corresponding idle heating rate is approximately $50~\mathrm{ms^{-1}}$. For a radial mode frequency of approximately $2\pi\times3~\mathrm{MHz}$, the corresponding idle radial heating rate for the longest-wavelength mode is approximately $30~\mathrm{s^{-1}}$.

Following Ref.~\cite{cetina_control_2022}, the infidelity of a single maximally entangling XX AM MS gate between ions $i$ and $j$ due to axial heating is given by Eq.~\ref{eq:xx_inf_axial}.
\begin{align}
1-F_{ij}=\frac{1}{2}\left(1-\frac{1}{\sqrt{1+(\pi/2)^2(\theta_i+\theta_j)^2}}\right).\label{eq:xx_inf_axial}
\end{align}
We assume equal mode participation for the two ions with each participation factor equal to $1/\sqrt{N}$. Assuming the same individual beam waist as in our system, we use Eq.~\ref{eq:theta} to obtain $\theta=\theta_i=\theta_j=0.026$ for the LA mode with a mean occupation of $\bar{n}=1500$, corresponding to a gate infidelity of approximately $10^{-3}$.

In a quantum circuit, we consider applying axial sympathetic cooling whenever the axial mean occupation reaches $\bar{n}=1500$. For a chain containing 10 coolant ions, and assuming the same axial cooling parameters and equal mode participation factors, a 3-ms cooling pulse should reduce the mean occupation to $\bar{n}\approx300$. At $\bar{n}=1500$, according to the model developed in Appendix~\ref{app:recoil_cal}, the total photon scattering rate is approximately $0.3~\mathrm{\mu s^{-1}}$ per coolant ion. We estimate that the resulting recoil heating increases the mean occupation of the long-wavelength radial modes by no more than $0.3$ during the axial cooling process.

We consider sympathetically cooling the entire LR mode manifold using the 10 coolant ions, which we assume are positioned to achieve good participation in the selected radial modes. In this case, the coolant sideband Rabi frequency is reduced by a factor of $r=\sqrt{23/60}$ compared with that of the 23-ion mixed chain. To compensate, we increase the radial cooling duration by a factor of $1/r$, resulting a total radial cooling duration of approximately $11~\mathrm{ms}$.

We consider applying one round of 3-ms axial sympathetic cooling every 25 ms during circuit execution. After every three rounds of axial cooling, we perform one round of 11-ms radial sympathetic cooling to recool the radial modes from recoil and idle heating. Under this cooling schedule, we predict the mean occupation of the LA mode to remain below $1500$, while that of the longest-wavelength radial mode remains below $3.5$. Consequently, over every $95~\mathrm{ms}$ of circuit execution, only $20~\mathrm{ms}$ is used for sympathetic cooling, corresponding to a cooling overhead of $21\%$ of the total runtime.

%%%%%%%%%%%%%%%%%%%%%%%%%%%%%%%%%%%%%%%%%%%%%%%%%%%%%%%%%%%%%%%%%%%%%%%%%%%%%%%%%%%%%%%%%%%%%%%%%%%%%%%%%%%%%%%%%%%%%%%%%%%%
\newpage
\section{Supplemental data
\label{app:cooling_supp_data}}
\noindent
\begin{figure}[H]
  \centering
	\includegraphics[width=0.44\textwidth]{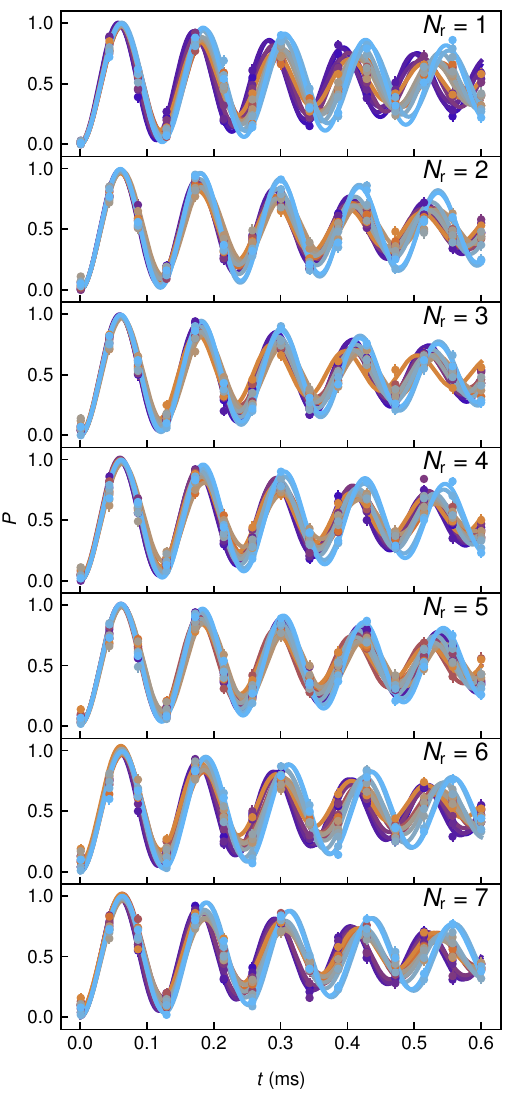}
	\caption{Collective carrier Raman Rabi oscillations. The probability of population transfer to the excited state $P$ is plotted as a function of the Rabi pulse duration $t$. Error bars denote $1\sigma$ binomial uncertainties. Rabi pulses are performed after $N_r$ rounds of prototypical sequence with complete sympathetic cooling (labeled in the upper right corner). Qubits (indexed from $-8$ to $+8$) are distinguished by a linearly segmented colormap ranging from dark blue ($-8$), through orange ($0$), to light blue ($+8$). The oscillations are fitted to the model of Eq.~\ref{eq:axialmodel} to extract the decay parameters $\theta_i$, which are subsequently converted to the mean occupation of the LA mode $\bar{n}$ using Eq.~\ref{eq:theta}.
    }
	\label{fig:acrc_carrier_flop_supp}
\end{figure}
\begin{figure}[H]
  \centering
	\includegraphics[width=0.43\textwidth]{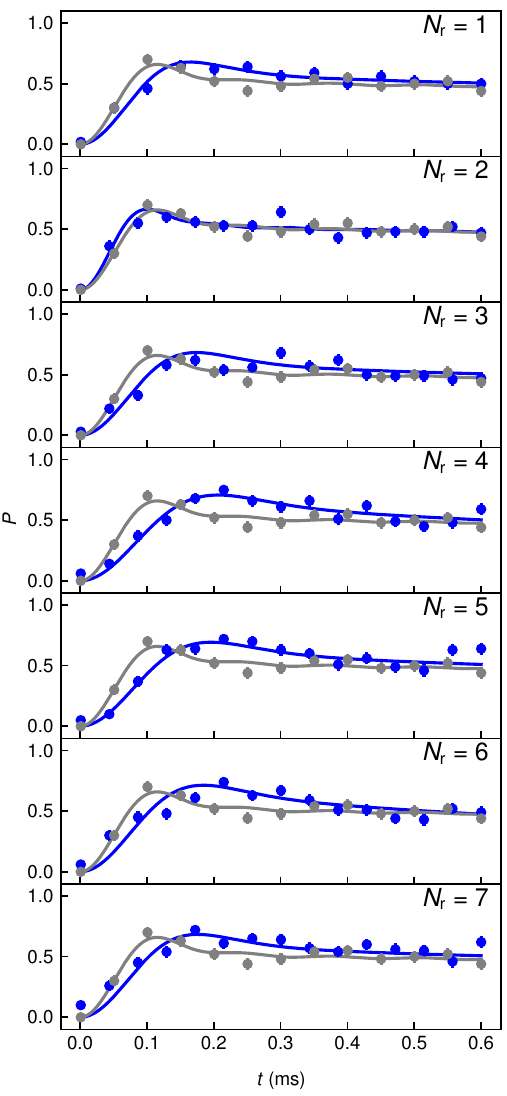}
	\caption{First-order blue-sideband Raman Rabi oscillations of the longest-wavelength LR mode. The probability of population transfer to the excited state $P$ is plotted as a function of the Rabi pulse duration $t$. Error bars denote $1\sigma$ binomial uncertainties. The sideband is measured on qubit ion $-4$ with a carrier Rabi frequency of $\Omega_{\mathrm{c,hr}}=2\pi\times70~\mathrm{kHz}$. The mean occupation of the highest-frequency radial mode, $\bar{n}{\mathrm{hr}}$, is measured either immediately after state preparation (gray) or following $N_r$ rounds of the prototypical sequence (labeled in the upper right corner) with complete sympathetic cooling (blue). The data are fitted with $\bar{n}{\mathrm{hr}}$ as a free parameter. The reference measurement immediately after state preparation yields $\bar{n}_{\mathrm{hr}}=16(2)$. For $N_r=1$–$7$, the fitted values are $8(4)$, $13(3)$, $7(5)$, $4(2)$, $6(3)$, $4(2)$, $7(5)$, demonstrating that the longest-wavelength LR mode remains consistently recooled throughout the sequence. }
	\label{fig:acrc_bsb_com_flop_supp}
\end{figure}
\begin{figure*}[t]
  \centering
	\includegraphics[width=\textwidth]{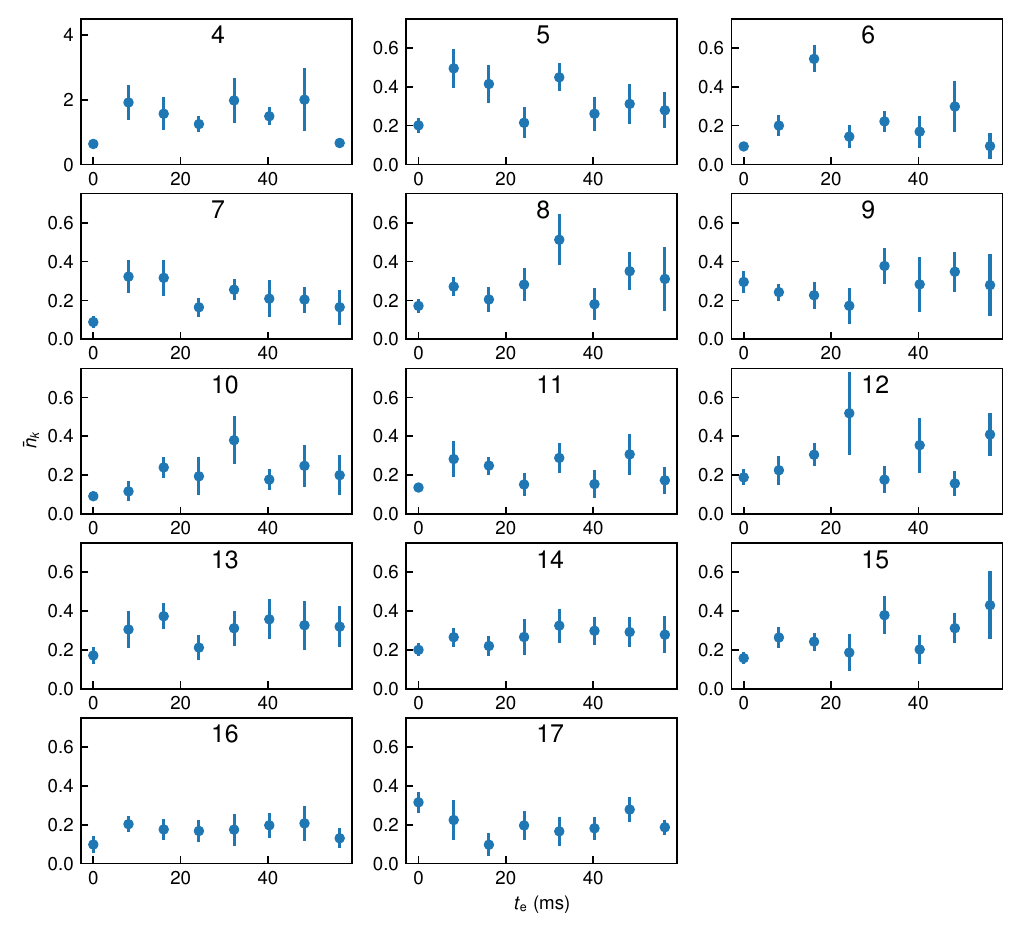}
	\caption{Mean occupations of the middle LR modes extracted from blue-sideband data. The mean occupations $\bar{n}$ of radial modes 4–17 are measured after repeated rounds of a prototypical sequence with complete sympathetic cooling of total circuit duration $t_e$. The values of $\bar{n}$ are extracted as fit parameters from first-order blue-sideband Raman Rabi oscillations with error bars as the parameter fit errors. The mode index is labeled in the top of the graph.}
	\label{fig:acrc_radial_supp}
\end{figure*}

\begin{figure*}[!t]
    \centering

    \includegraphics[width=0.95\textwidth]{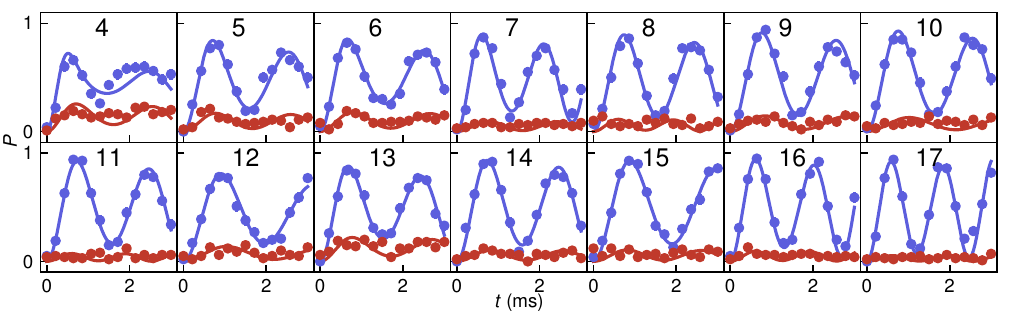}
	\caption{First-order blue(blue)/red(red) sideband Raman Rabi oscillations following $N_r=2$ round of the prototypical sequence. Error bars denote $1\sigma$ binomial uncertainties. The probability of population transfer to the excited state $P$ is plotted as a function of the Rabi pulse duration $t$. Raman sideband pulses are applied in parallel after the sympathetic cooling cycle to qubits {-5, 7, 8, -6, 3, 1, 5, -1, 6, -3, -4, 4, -8, -7} respectively for modes 4 to 17 (labeled at top). Blue-sideband fitted $\bar{n}_k$ are 1.57(51), 0.41(10), 0.54(7), 0.32(9), 0.20(6), 0.23(7), 0.24(6), 0.25(5), 0.31(6), 0.37(7), 0.22(5), 0.24(5), 0.18(6), 0.10(6). Red-sideband fitted $\bar{n}_k$ are 0.57(9), 0.30(3), 0.13(2), 0.13(2), 0.15(2), 0.08(1), 0.09(1), 0.07(1), 0.08(1), 0.10(2), 0.13(2), 0.07(1), 0.09(2), 0.16(2).}
	\label{fig:acrc_radial_bsb_rsb_N_2}

     \includegraphics[width=0.95\textwidth]{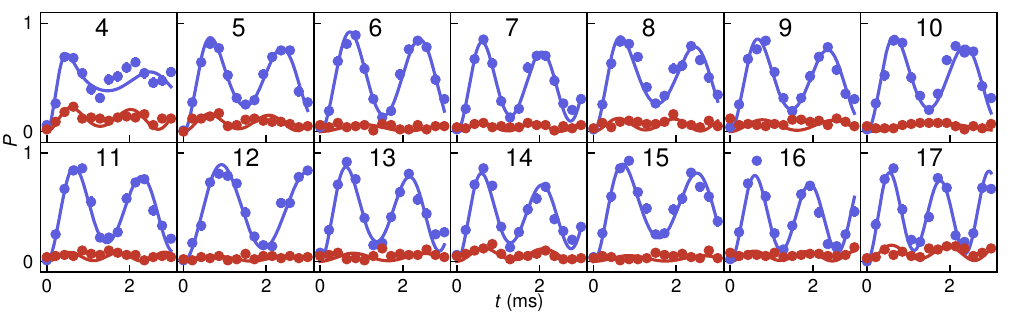}
	\caption{First-order blue(blue)/red(red) sideband Raman Rabi oscillations following $N_r=4$ rounds of the prototypical sequence. Error bars denote $1\sigma$ binomial uncertainties. The probability of population transfer to the excited state $P$ is plotted as a function of the Rabi pulse duration $t$. Raman sideband pulses are applied in parallel after the sympathetic cooling cycle to qubits {-5, 7, 8, -6, 3, 1, 5, -1, 6, -3, -4, 4, -8, -7} respectively for modes 4 to 17 (labeled at top). Blue-sideband fitted $\bar{n}_k$ are 1.98(68), 0.45(7), 0.22(6), 0.26(6), 0.51(13), 0.38(9), 0.38(12), 0.29(8), 0.18(7), 0.31(9), 0.32(9), 0.38(10), 0.17(8), 0.17(8). Red-sideband fitted $\bar{n}_k$ are 0.57(9), 0.30(3), 0.13(2), 0.13(2), 0.15(2), 0.08(1), 0.09(1), 0.07(1), 0.08(1), 0.10(2), 0.13(2), 0.07(1), 0.09(2), 0.16(2).}
	\label{fig:acrc_radial_bsb_rsb_N_4}
    
     \includegraphics[width=0.95\textwidth]{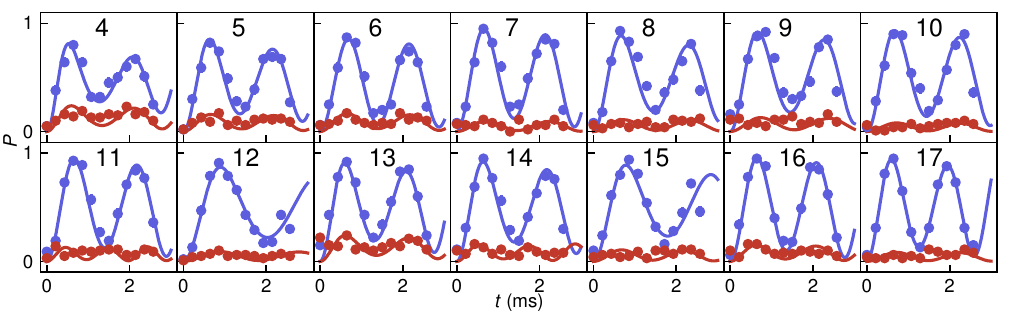}
	\caption{First-order blue(blue)/red(red) sideband Raman Rabi oscillations following $N_r=7$ rounds of the prototypical sequence. Error bars denote $1\sigma$ binomial uncertainties. The probability of population transfer to the excited state $P$ is plotted as a function of the Rabi pulse duration $t$. Raman sideband pulses are applied in parallel after the sympathetic cooling cycle to qubits {-5, 7, 8, -6, 3, 1, 5, -1, 6, -3, -4, 4, -8, -7} respectively for modes 4 to 17 (labeled at top). Ion loss occurs during the final two measurements, thus those data points are excluded from the fit. Blue-sideband fitted $\bar{n}_k$ are 0.66(16), 0.28(9), 0.09(7), 0.16(9), 0.31(17), 0.28(16), 0.20(10), 0.17(7), 0.41(11), 0.32(10), 0.28(10), 0.43(18), 0.13(5), 0.19(4). Red-sideband fitted $\bar{n}_k$ are 0.57(9), 0.30(3), 0.13(2), 0.13(2), 0.15(2), 0.08(1), 0.09(1), 0.07(1), 0.08(1), 0.10(2), 0.13(2), 0.07(1), 0.09(2), 0.16(2).
}
	\label{fig:acrc_radial_bsb_rsb_N_7}
    
\end{figure*}

\begin{figure*}
    \centering

    \includegraphics[width=0.95\textwidth]{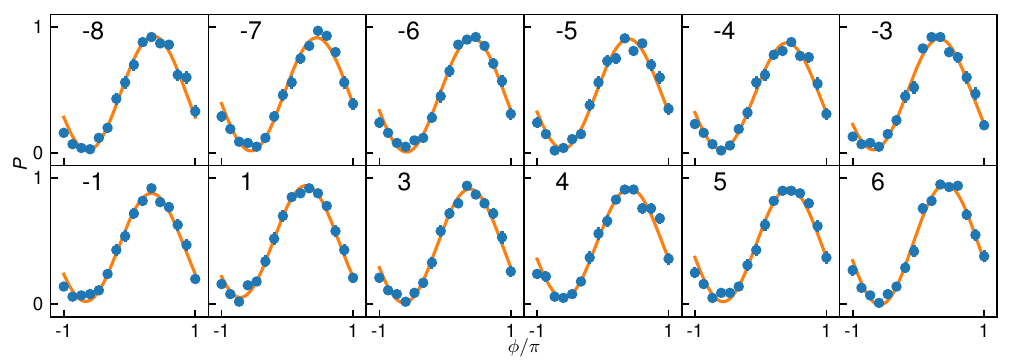}
	\caption{Ramsey phase-scan fringes following $N_r=1$ round of the prototypical sequence. Error bars denote $1\sigma$ binomial uncertainties. The probability of population transfer to the excited state $P$ is plotted as a function of the final pulse phase $\phi$. The qubit indices are labeled in the top left. The Ramsey fringe data are fitted to $y(\phi) = 1/2 [A \cos(\phi - \phi_0) + 1] + y_0$, where the fringe amplitude $A$, phase offset $\phi_0$, and fringe offset $y_0$ are extracted as fit parameters. We take the fringe contrast as the absolute contrast of the fringe $|A|$.}
	\label{fig:acrc_ramsey_fringe_N_1}

     \includegraphics[width=0.95\textwidth]{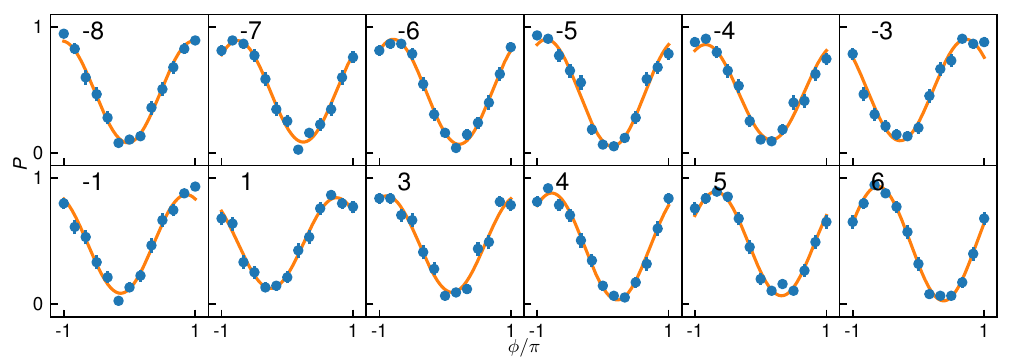}
	\caption{Ramsey phase-scan fringes following $N_r=4$ round of the prototypical sequence. Error bars denote $1\sigma$ binomial uncertainties. The probability of population transfer to the excited state $P$ is plotted as a function of the final pulse phase $\phi$. The qubit indices are labeled in the top left. The Ramsey fringe data are fitted to $y(\phi) = 1/2 [A \cos(\phi - \phi_0) + 1] + y_0$, where the fringe amplitude $A$, phase offset $\phi_0$, and fringe offset $y_0$ are extracted as fit parameters. We take the fringe contrast as the absolute contrast of the fringe $|A|$.}
	\label{fig:acrc_ramsey_fringe_N_4}

    \includegraphics[width=0.95\textwidth]{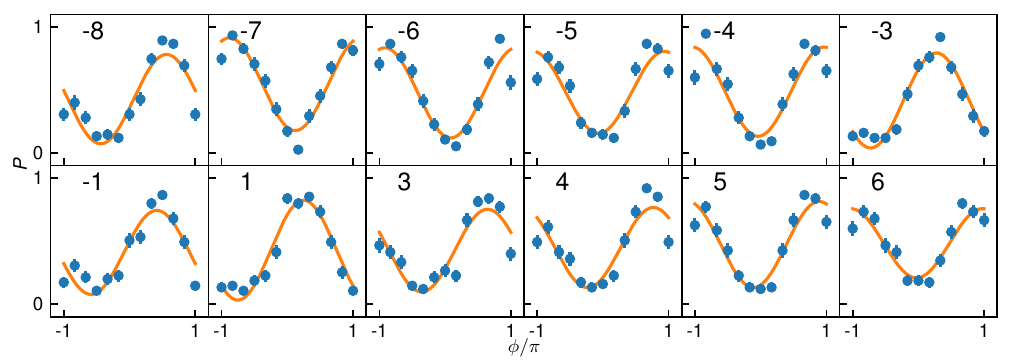}
	\caption{Ramsey phase-scan fringes following $N_r=6$ round of the prototypical sequence. Error bars denote $1\sigma$ binomial uncertainties. The probability of population transfer to the excited state $P$ is plotted as a function of the final pulse phase $\phi$. The qubit indices are labeled in the top left. The Ramsey fringe data are fitted to $y(\phi) = 1/2 [A \cos(\phi - \phi_0) + 1] + y_0$, where the fringe amplitude $A$, phase offset $\phi_0$, and fringe offset $y_0$ are extracted as fit parameters. We take the fringe contrast as the absolute contrast of the fringe $|A|$.}
	\label{fig:acrc_ramsey_fringe_N_6}

\end{figure*}

\begin{figure*}[!t]
    \centering
    
    \includegraphics[width=\textwidth]{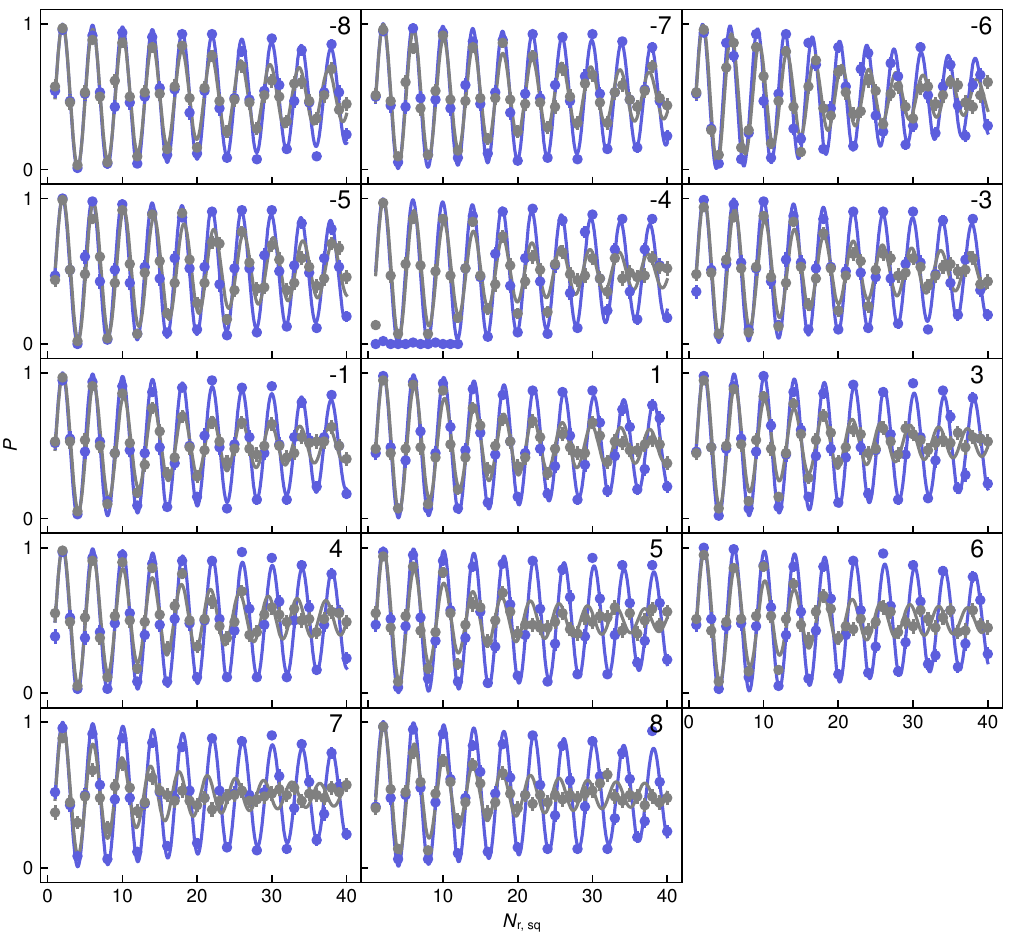}
	\caption{Performance of single-qubit gates. A total of $N_{\mathrm{r,sq}}$ blocks of SK1 $\pi/2$ gates applied sequentially to each qubit are applied either consecutively (gray) or interleaved with $100~\mathrm{\mu s}$ of axial sympathetic cooling (blue). The measured state populations of qubits with selected indices (indicated in the upper-right corner of each panel) are shown as function of $N_{\mathrm{r,sq}}$. Error bars denote $1\sigma$ binomial uncertainties. Ion -4 briefly goes dark in the beginning of the experiments when axial cooling is applied, thus shows low population transfer.}
	\label{fig:acrc_sk1_compare_complete}

\end{figure*}

%%%%%%%%%%%%%%%%%%%%%%%%%%%%%%%%%%%%%%%%%%%%%%%%%%%%%%%%%%%%%%%%%%%%%%%%%%%%%%%%%%%%%%%%%%%%%%%%%%%%%%%%%%%%%%%%%%%%%%%%%%%%
\clearpage

\bibliography{bib}
\end{document}